\documentclass[pre,onecolumn,showpacs,preprintnumbers,amsmath,%
amssymb,superscriptaddress,eqsecnum,showpacs,
   preprintnumbers,floatfix]{revtex4-1}[12pt]
\usepackage{color}
\usepackage{morefloats}
\usepackage{graphicx}
\usepackage{subfigure}

\begin{document}

\newcommand{\vphi}{\varphi}
\newcommand{\bq}{\begin{equation}}
\newcommand{\be}{\begin{equation}}
\newcommand{\ba}{\begin{eqnarray}}
\newcommand{\eq}{\end{equation}}
\newcommand{\ee}{\end{equation}}
\newcommand{\ea}{\end{eqnarray}}
\newcommand{\tchi} {{\tilde \chi}}
\newcommand{\tA} {{\tilde A}}
\newcommand{\sech} { {\rm sech}}
\newcommand{\pstar}{\mbox{$\psi^{\ast}$}}
\newcommand {\bPsi}{{\bar \Psi}}
\newcommand {\bpsi}{{\bar \psi}}
\newcommand {\barf}{{\bar f}}
\newcommand {\tu} {{\tilde u}}
\newcommand {\tv} {{\tilde v}}
\newcommand{\dq}{{\dot q}}
\newcommand {\tdelta} {{\tilde \delta}}
 \newcommand{\shao}[1]{\textcolor{red}{#1}}
 \newcommand{\etc}{\textit{etc}{}}
 \newcommand{\ie}{\textit{i.e.}{~}}
 \newcommand{\videpost}{\textit{vide post}{}}
 \newcommand{\eg}{\textit{e.g.}{~}}
 \newcommand{\ansatz}{\textit{ansatz}{ }}
 
\maxdeadcycles=1000

\preprint{LA-UR 12- (Draft fnlde.tex) : \today}
\newpage

\title{Nonlinear Dirac equation solitary waves in the presence of external driving forces }
\author{Franz G.  Mertens}
\email{franzgmertens@gmail.com}
\affiliation{Physikalisches Institut, Universit\"at Bayreuth, D-95440 Bayreuth, Germany} 
\author{Fred Cooper}
\email{cooper@santafe.edu}
\affiliation{Santa Fe Institute, Santa Fe, NM 87501, USA}
\affiliation{Theoretical Division and Center for Nonlinear Studies, 
Los Alamos National Laboratory, Los Alamos, New Mexico 87545, USA}
\author{Niurka R. Quintero} 
\email{niurka@us.es} 
\affiliation{IMUS and Departamento de Fisica Aplicada I, E.P.S. Universidad de Sevilla, 41011 Sevilla, Spain}
\author{Sihong Shao}\email{sihong@math.pku.edu.cn}
\affiliation{LMAM and School of Mathematical Sciences, Peking University, Beijing 100871, China}
\author{Avinash Khare} 
\email{khare@iiserpune.ac.in}
\affiliation{ Indian Institute of Science Education and Research, Pune 
411021, India}
\author{Avadh Saxena}
\email{avadh@lanl.gov}
\affiliation{Theoretical Division and Center for Nonlinear Studies, 
Los Alamos National Laboratory, Los Alamos, New Mexico 87545, USA}
\begin{abstract}
We consider the nonlinear Dirac (NLD) equation in 1+1 dimension with 
scalar-scalar  self-interaction 
 in the presence of  external forces as well as damping of the form  $ f(x,t) - i \mu  \gamma^0 \Psi$, where both $f$ and $\Psi$ are two-component spinors. We develop an approximate variational approach using collective coordinates (CC) for studying the time dependent response of the solitary waves to these external forces. This approach predicts intrinsic oscillations of the solitary waves, i.e. the amplitude, width and phase all oscillate with the same frequency. The translational motion is also affected, because the soliton position oscillates around a mean trajectory. We then compare the results of the variational approximation with numerical simulations of the NLD equation, and find a good agreement, if we take into account a certain linear excitation with specific wavenumber that is excited 
 together with the intrinsic oscillations such that the momentum in a transformed NLD equation is conserved.  We also  solve explicitly the CC equations of the variational approximation in the non-relativistic regime for a homogeneous external force and obtain excellent agreement with the numerical solution of the CC equations.
\end{abstract}

\maketitle

\section{Introduction}

Since the first nonlinear generalization of the Dirac equation by Ivanenko \cite{Ivanenko1938}, the nonlinear Dirac (NLD) equation  
has emerged naturally as a practical model in many physical systems, 
such as extended particles \cite{FinkelsteinLelevierRuderman1951,
FinkelsteinFronsdalKaus1956,Heisenberg1957},
the gap solitons in nonlinear optics \cite{Barashenkov1998},
light solitons in waveguide arrays and experimental realization of an optical analog for relativistic quantum mechanics \cite{Longhi2010,DreisowHeinrichKeil2010,TranLonghiBiancalana2014},
Bose-Einstein condensates in honeycomb optical lattices \cite{Haddad2009},
phenomenological models of quantum chromodynamics
\cite{Fillion-Gourdeau2013},
as well as matter influencing
the evolution of the Universe in cosmology \cite{Saha2012}.
To maintain  the Lorentz invariance of the NLD equation,
the self-interaction Lagrangian can be built up from the bilinear covariants. Different self-interactions give rise to different 
NLD equations. Several interesting models have been proposed and investigated
based on the scalar bilinear covariant \cite{Gursey1956,Soler1970,GrossNeveu1974,Mathieu1984},
the vector bilinear covariant \cite{Thirring1958},
the axial vector bilinear covariant \cite{Mathieu1983},
both scalar and pseudoscalar bilinear covariants \cite{RanadaRanada1984},
both scalar and vector bilinear covariants \cite{Stubbe1986-jmp,NogamiToyama1992} among others.


A key feature of these NLD equations is that
they allow solitary wave solutions or particle-like solutions -- localized solutions
with finite energy and charge \cite{Ranada1983}.
That is, the particles appear as intense localized regions of field
which can be recognized as the basic ingredient
in the description of extended objects in quantum field theory \cite{Weyl1950}. For the NLD equation in (1+1) dimensions (\ie one time dimension plus one space dimension),
several analytical solitary wave solutions were derived 
for the cubic nonlinearity \cite{LeeKuoGavrielides1975,ChangEllisLee1975}, 
for fractional nonlinearity \cite{Mathieu1985-prd} as well as
 for general nonlinearity \cite{Stubbe1986-jmp,CooperKhareMihailaSaxena2010,XuShaoTangWei2013}
by using explicitly the constraints resulting from energy-momentum conservation; 
and this is well summarized by Mathieu \cite{Mathieu1985-jpa-mg}.
With the help of the analytical expressions of the NLD solitary wave
solutions, the interaction dynamics among them 
 has been studied and rich nonlinear phenomena have been revealed
in a series of works 
\cite{AlvarezCarreras1981,ShaoTang2005,ShaoTang2006,ShaoTang2008,
XuShaoTang2013,TranNguyenDuong2014}.


An interesting topic for the NLD equation solitary waves is the stability issue, which has been the central topic in works
spread out over several decades that is still ongoing.
Analytical studies of the NLD solitary wave stability face serious obstacles \cite{StraussVazquez1986,AlvarezSoler1986,BlanchardStubbeVazquez1987},
while results of computer simulations are contradictory \cite{Bogolubsky1979-pla,AlvarezSoler1983,Mathieu1983,Alvarez1985}.
Numerical results inferred that both
the multi-hump  profile and high-order nonlinearity could
undermine the stability during the scattering of the NLD solitary
waves \cite{ShaoTang2005,XuShaoTang2013}.  For the NLD equation with scalar-scalar interactions (i.e. the Soler model)  the solitary wave solutions can have either one hump or two humps.
Quite recently, for the Soler model,  we found that 
all stable NLD solitary waves have a one-hump profile, but not all one-hump waves are stable, while all waves with two humps are unstable \cite{ShaoQuinteroMertens2014}. This result is in agreement with the rigorous analysis in the nonrelativistic limit \cite{ComechGuanGustafson2014}. 
In order to further understand the behavior and the stability of NLD solitary waves, the NLD equation in the presence of external potentials has been investigated \cite{NogamiToyamaZhao1995,Toyama1998,ToyamaNogami1998,
MertensQuinteroCooper2012}
and a sufficient dynamical condition for instability to arise was postulated through a collective coordinates (CC) theory \cite{MertensQuinteroCooper2012}. In this work, we will continue to study the NLD solitary waves under external forces.


For the forced nonlinear Schr{\"o}dinger (NLS) equation  when subject to 
an external force of the form $f(x)= r \exp(-i K x)$, the authors found \cite{MertensQuinteroBarashenkovBishop2011,QuinteroMertensBishop2015,MertensQuinteroBishop2010} 
that intrinsic soliton oscillations are excited, i.e., the soliton amplitude, width, phase, momentum, and velocity all oscillate with the same frequency. This behavior was predicted by a collective coordinates 
theory and was confirmed by numerical simulations. Moreover, one specific plane wave phonon (short for a linear excitation) with wavenumber $k=-K$ is also excited such that the total momentum in a transformed NLD equation is conserved. This phonon mode was not included in the CC theory and had 
to be calculated separately \cite{MertensQuinteroBishop2013}.


In the present paper we consider the relativistic generalization of our previous work on the forced NLS equation, namely the behavior of solitary wave solutions to the NLD equation when subjected to an external force which is now a two-component spinor. In Sec. \ref{sec2} 
we review exact analytical solutions for the unperturbed NLD equation. In Sec. \ref{sec3} 
we present the NLD equation with external force $f_{j}(x,t)=r_{j} \exp[i (\nu_j t-K_j x)]$, $j=1,2$, and the corresponding Lagrangian density. Using the energy-momentum tensor we show that the total energy is conserved if the force is time independent ($\nu_j=0$).

For the case $K_1=K_2=K$, $\nu_j=0$ and zero dissipation we perform in Sec. \ref{sec4} a transformation such that the 
transformed NLD equation is invariant under space translations and thus the momentum is conserved. 
In Sec. \ref{sec5} we make a variational ansatz with three collective coordinates. All integrals that appear in the Lagrangian can be performed exactly and we finally have a set of 3 ODEs as CC equations. For a special case these equations can be simplified and an approximate analytical solution can be obtained. 
Solutions are also obtained in the non-relativistic regime, when $K=0$, 
by an expansion up to order $v^2$, where $v$ is the velocity. 

In Sec. \ref{sec6} the spectrum of the linear excitations (phonons) is calculated and together with the numerical solutions of the CC equations this is compared with the results from our numerical simulations (Sec. \ref{sec7}). We always obtain periodic solutions and the spectra of these solutions exhibit two dominant peaks: The phonon frequency $\Omega_K=\sqrt{m^2+K^2}$ and the intrinsic oscillation frequency $\Omega_{sim}$. The frequency $\Omega_{sim}$ agrees nearly perfectly with the prediction $\Omega_{cc}$ from solving numerically the CC equations (Table \ref{tab1}). The soliton position $q(t)$ performs small oscillations around a mean trajectory $\bar{v}_{sim} \ t$. 
This translational motion is only weakly affected by the intrinsic soliton oscillations. Further $\bar{v}_{cc}$ agrees with 
$\bar{v}_{sim}$ within an error of about $14\%$ (Table \ref{tab1}). The reason is that the plane wave phonons with $k=-K$ are not taken into account in the CC theory. The summary of our main results is contained in Sec. \ref{sec8}. 


\section{Review of exact solutions to the  NLD equation} \label{sec2}
In this section we review the exact solitary wave solutions to the NLD equation,
\bq
(i \gamma^{\mu} \partial_{\mu} - m) \Psi +g^2 (\bPsi  \Psi)^{\kappa} \Psi 
= 0 \>, \label{nlde1}
\eq
where we use the representation for the 1+1 dimensional Dirac Gamma matrices:
$\gamma^0 = \sigma_3$;  ~~~$\gamma^1= i \sigma_2,$ which we also used in  \cite{MertensQuinteroCooper2012}. 
The solitary wave solution in the rest frame is represented by 
\bq \label{eq2.3}
\Psi(x,t) = e^{-i\omega t} \psi(x)= e^{-i\omega t} \left(\begin{array} {cc}
      A(x) \\
      i ~B(x) \\ 
   \end{array}\right), 
\eq
where $A$ and $B$ satisfy
\ba
&& \frac{dA}{dx} + (m+\omega ) B - g^2(A^2-B^2)^{ \kappa} B=0\,, \nonumber \\
&&\frac{dB}{dx} + (m-\omega ) A - g^2(A^2-B^2)^{ \kappa} A=0\,. \nonumber \\
\ea
The solutions of these equations vanishing at infinity are
\ba \label{eq2.33}
A & = & \sqrt{ \frac{(m+\omega)  \cosh ^2(\kappa \beta x)}{m+\omega \cosh(2\kappa \beta x)}} 
\bigg [\frac{(\kappa+1) \beta ^2}{g^2 (m+\omega  \cosh (2\kappa \beta x))} 
\bigg ]^{\frac{1}{2\kappa}}, \nonumber   \\ 
B & =  & \sqrt{ \frac{(m-\omega) \sinh^2(\kappa \beta x)}{m+\omega \cosh(2\kappa \beta x)}} 
\bigg [\frac{(\kappa +1) \beta ^2}{g^2 (m+\omega  \cosh (2\kappa \beta x))} 
\bigg ]^{\frac{1}{2\kappa}}, \nonumber   \\ 
\ea
where $\beta = \sqrt{m^2-\omega^2}$.
We are interested in bound state solutions that correspond to positive frequency $\omega > 0$ and which have energies in the rest frame less than the mass parameter $m$, i.e. $\omega < m$. 

  Because of Lorentz invariance we can find the solution in a frame moving with velocity $v$ with respect to the rest frame.
  The Lorentz boost  is given in terms of  the rapidity variable $ \eta$ as follows  (here $c=1$): 
  \bq
  v = \tanh \eta;~~   \gamma = \frac{1}{\sqrt{1-v^2}} = \cosh \eta; ~~ \sinh \eta =  \frac{v}{\sqrt{1-v^2}} . 
  \eq
  
  In the moving frame, the transformation law for spinors tells us that:
  \bq
  \Psi(x,t) =    \left(\begin{array}{cc}
        \cosh(\eta/2) & \sinh(\eta/2) \\
        \sinh(\eta/2) &  \cosh(\eta/2\\
     \end{array} \right)  \left(  \begin{array} {cc}
      \Psi_1^0[\gamma(x-vt), \gamma(t-vx)] \\
  \Psi_2^0[\gamma(x-vt), \gamma(t-vx)]\\ 
   \end{array} \right) , 
\eq
since
\bq
\cosh (\eta/2) = \sqrt{(1+\gamma)/2};~~  \sinh (\eta/2) = \sqrt{(\gamma-1)/2} . 
 \eq
 This in component form reads:
 \ba \label{eq2.37}
 &&\Psi_1(x,t) = \left( \cosh(\eta/2) A(x') + i \sinh(\eta/2) B(x') \right) e^{-i\omega t'}  , \nonumber \\
&&\Psi_2 (x,t) = \left( \sinh(\eta/2) A(x') + i \cosh(\eta/2) B(x') \right) e^{-i\omega t'}  ,
\ea
where
\bq
x' = \gamma(x-vt); ~~ t' = \gamma(t-vx) . 
\eq
Note that  $\cosh^2(\eta/2)+\sinh^2(\eta/2) = \cosh \eta = \gamma$.

\section{Externally Driven NLD equation} \label{sec3}

In  previous papers \cite{MertensQuinteroBishop2010,MertensQuinteroBarashenkovBishop2011} we 
investigated the externally driven NLS equation 

\bq
 i \frac{\partial}{\partial t} \psi + \frac{\partial^2}{\partial x^2} \psi + {g ^2}(\psi^\star \psi)^{ \kappa} \psi+\delta \psi  = r e^{-i K x}-i\mu \psi  \label{psieq} , 
\eq
where $\mu$ is the dissipation coefficient, and $r$ and $K$ are constants.  
This equation can be derived by means of a generalization of the Euler-Lagrange 
equation  
\bq \label{df2}
\frac{d}{dt} \frac{\partial {\cal L}}{\partial \psi_{t}^{*}} + 
\frac{d}{dx} \frac{\partial {\cal L}}{\partial \psi_{x}^{*}}-
\frac{\partial {\cal L}}{\partial \psi^{*}}= 
\frac{\partial {\cal F} }{\partial \psi^{*}_{t}},
\eq
where the Lagrangian density reads  
\bq \label{df3}
{\cal L} = \frac{i}{2} (\psi_{t} \psi^{*}-\psi_{t}^{*} \psi)-|\psi_{x}|^{2}+ 
\frac{g^2} {\kappa+1} (\psi^\star \psi)^{\kappa+1}  +\delta |\psi|^{2}-r e^{-i K x} \psi^{*}- r e^{i K x}\psi,
\eq
and the dissipation function density  is given by 
\bq \label{df4}
{\cal F} = -i \mu (\psi_{t} \psi^{*}-\psi_{t}^{*} \psi).
\eq

For the NLD case we instead consider a two-component spinor forcing term
\bq \label{force0}
f = \left({\begin{array}{c} 
  f_1(x,t) \\ 
  f_2(x,t) \\ 
\end{array}} \right)
\eq
with the NLD equation
\bq
(i \gamma^{\mu} \partial_{\mu} - m) \Psi +g^2 (\bPsi  \Psi)^{\kappa} \Psi 
=   f(x,t) - i \mu \gamma^0 \Psi  \>. \label{nlde1a}
\eq
In what follows we will generalize our choice for the NLS equation by choosing 
\bq \label{force}
f_j(x,t) =  r_j e^{i( \nu_j t - K_j x)}, \quad j=1,2, 
\eq
with real parameters $r_j$, $\nu_j$ and $K_j$. 
Note that the phase of $f$ is invariant under Lorentz transformations. 
As the second component of the spinor $\Psi$ is the so-called ``small 
component", which is smaller than the first component by the factor 
$\alpha=\sqrt{(m-\omega)/(m+\omega)}$, 
we will only consider cases with $r_2=\epsilon r_1$, where $\epsilon=O(\alpha)$ 
or smaller. 

Equation \ (\ref{nlde1a}) can be derived in a standard fashion from the Lagrangian density
\bq \label{eq3.8}
\mathcal{L} =  \left(\frac{i}{2}\right) [\bPsi \gamma^{\mu} \partial_{\mu} \Psi 
-\partial_{\mu} \bPsi \gamma^{\mu} \Psi] - m \bPsi \Psi 
+ \frac{g^2}{\kappa+1} (\bPsi \Psi)^{\kappa+1} - \bPsi  f - \bar{f}  \Psi + \mathcal{L}_0(b),
\eq
where $\mathcal{L}_0(b)$ is determined later on and $b=\lim_{x \to \pm \infty} \Psi(x,t)$. 
The term in the Lagrangian density which pertains to  forcing  can be written as 
\bq
{\cal L} _3 = - 2 Re (\barf \Psi),
\eq
and the full interaction part of the Lagrangian density is now
\bq
{\cal L}_I = \frac{g^2}{\kappa+1} (\bPsi \Psi)^{\kappa+1} - \bPsi  f - \bar{f}  \Psi  . \ 
\eq
The generalized Euler-Lagrange equation can be written as
\bq \label{E-L} 
\partial_\mu \frac{\partial \mathcal{L}}{ \partial( \partial_\mu \bPsi)} -  \frac{\partial \mathcal{L}}{\partial \bPsi} = \frac{\partial \mathcal{F}}{\partial(\partial_t \bPsi)} ,
\eq
where the dissipation function density is now
\bq \label{df5}
{\cal F} =-i \mu ( \bPsi \gamma^0 \partial_t \Psi - \partial_t \bPsi \gamma^0 \Psi ) .
\eq

The adjoint equation comes from the Euler-Lagrange equation:
\bq 
\partial_\mu \frac{\partial \mathcal{L}}{ \partial( \partial_\mu \Psi)} -  \frac{\partial \mathcal{L}}{\partial \Psi} = \frac{\partial \mathcal{F}}{\partial(\partial_t \Psi)},
\eq
from this we get the adjoint driven NLD equation
\bq
-i \partial_{\mu}  \bPsi \gamma^\mu - m \bPsi  +g^2 (\bPsi  \Psi)^{\kappa}  \bPsi 
= \barf  + i \mu \bPsi \gamma^0  \>.  \label{nldradj}
\eq
To generalize our discussion of external forces from the NLS equation to the NLD equation we have included a dissipation term in our general formulations. However, in most sections that follow we will concentrate on the case where the dissipation coefficient $\mu=0$, so that the energy is conserved. 

When $K_1=K_2=K$, $\nu_i =0$ and $\mu=0$ we choose $K L= \pi n$ ($n$ is an integer, 
$2 L$ is the total length of the system) and periodic boundary conditions such that 
$f_j(x,t)$ goes to $r_j$ and 
the wave function goes to a constant spinor $b$  for $L \to \infty$.   In that case a constant solution
of the forced NLD equation   with $\Psi = b$ satisfies the nonlinear equation:
\bq \label{eqb}
 -m b + g^2 (\bar b b)^\kappa b  = r, 
 \eq
where $r$ represents a spinor with components $r_1$ and $r_2$.

\subsection{Energy flow equations and the conservation of energy}

From the NLD equation with external sources and the definition of the energy-momentum tensor:
\bq  \label{emc1}
 T^{\mu \nu} = \frac{i}{2} \left[ \bPsi \gamma^\mu \partial^ \nu \Psi  -  \partial^\nu \bPsi  \gamma^\mu  \Psi \right]  - g^{\mu \nu} \cal{L}, 
 \eq
we have that 
\bq
\partial_\mu T^{\mu \nu} = F^\nu, 
\eq
where
\bq \label{fdens}
F^\nu = \bPsi (\partial^\nu  f)+ (\partial^\nu  \barf )\Psi . 
\eq
The energy density is given by 
\bq 
T^{00} = -\frac{i}{2} \left[  \bPsi  \gamma^1 \partial_x \Psi-\partial_x \bPsi \gamma^1 \Psi \right]+ m \bPsi \Psi - \mathcal{L}_I - 
\mathcal{L}_0,    \label{eq:hdensity2}
\eq
where  now 
\bq
 \mathcal{L}_I  =   \frac{g^2}{ \kappa+1} (\bPsi \Psi)^{ \kappa+1} -\barf \Psi -\bPsi f, \ 
\eq
and $\mathcal{L}_0$ is chosen so that $T^{00}$ vanishes at $x= \pm L$, when $L \to \infty$. Therefore, 
from Eqs. (\ref{eqb}) and (\ref{eq:hdensity2}) we obtain  
\bq
 \mathcal{L}_0  = m \bar b  b - \frac{g^2}{ \kappa+1} (\bar b  b)^{ \kappa+1} + \bar b r  + \bar{r} b =  -m \bar b  b + \frac{g^2(2 \kappa+1)}{ \kappa+1} (\bar b  b)^{ \kappa+1}, \ 
\eq
Now we will assume that in the lab frame $f(x,t)$ is independent of $t$ and of the form:
\bq \label{force2}
f_j(x) =  r_j e^{-iK_j x}, \quad j=1,2, 
\eq
with real parameters $r_j$ and $K_j$. 
In that case  from Eq. \eqref{fdens}, we have that $F^0 = 0$ and
\bq
\partial_{t} T^{00} + \partial_{x} T^{10}= 0, \label{c1}
\eq 
where
\begin{eqnarray} \label{de}
T^{00} &=& -\frac{i}{2} \left[  \bPsi  \gamma^1 \partial_x \Psi-\partial_x \bPsi \gamma^1 \Psi \right]+ m \bPsi \Psi - \frac{g^2}{ \kappa+1} (\bPsi \Psi)^{ \kappa+1} + \bPsi f+  \bar{f}  \Psi, \\  \label{t10}
 T^{10} &=& -\frac{i}{2} \left[\bPsi_{t}  \gamma^1 \Psi- \bPsi \gamma^1 \Psi_{t} \right]. 
\end{eqnarray}

Integrating Eq.\ (\ref{c1}), and under the assumption that $T^{10}(+\infty,t)-T^{10}(-\infty,t)=0$, then the energy of the driven NLD equation,  
\begin{eqnarray} \label{energy}
 E^{total}  &= &\int_{-\infty}^{+\infty}\, dx \,T^{00},  
\end{eqnarray}
is conserved.   

\section{Transformed NLD equation with external force} \label{sec4}

Let us consider the case of external force (\ref{force0}) and (\ref{force}) with $\nu_j=0$ and $K_1=K_2=K$, 
\bq
f(x) = 
\left(
\begin{array}{ccc}
    r_1   \\
   r_2   \\
   \end{array}
\right) e^{-i Kx} \equiv r e^{-i Kx} . 
\eq
After the following transformation 
\ba \label{psichi}
\Psi(x,t) &=& \chi(x,t)  e^{-i Kx},  
\ea
Eq.\ (\ref{eq3.8}) becomes
\bq \label{eq3.8a}
\mathcal{L} =  \left(\frac{i}{2}\right) [\bar{\chi} \gamma^{\mu} \partial_{\mu} 
\chi 
-\partial_{\mu} \bar{\chi} \gamma^{\mu} \chi] - m \bar{\chi} \chi  
+ \frac{g^2}{\kappa+1} (\bar{\chi} \chi)^{\kappa+1} - \bar{\chi} r - \bar{r} \chi+ 
 K \bar{\chi} \gamma^1 \chi+\mathcal{L}_0(a), 
\eq
where the constant $\mathcal{L}_0(a)$ now depends on a constant vector $a=\lim_{x \to \pm \infty} \chi(x,t)$, determined by 
the algebraic equations:
\bq
-m a + g^2 (\bar{a}  a)^{\kappa} a
  = r -K \gamma^1 a.  \label{chi1}
\eq
From  the Euler-Lagrange equation:
\bq 
\partial_\mu \frac{\partial \mathcal{L}}{ \partial( \partial_\mu \bar{\chi})} -  \frac{\partial \mathcal{L}}{\partial \bar{\chi}} = 0,
\eq
we obtain the following perturbed NLD equation
\bq
(i \gamma^{\mu} \partial_{\mu} - m) \chi +g^2 (\bar{\chi}  \chi)^{\kappa} \chi
  = r -K \gamma^1 \chi.  \label{chi1a}
\eq
The corresponding adjoint equation reads 
\bq
-i  \partial_{\mu} \bar{\chi} \gamma^{\mu} - m \bar{\chi} +g^2 (\bar{\chi}  \chi)^{\kappa} \bar{\chi}
  = \bar{r} -K \bar{\chi} \gamma^1.  \label{adchi1}
\eq
 Equation  \ (\ref{chi1a}) is not only invariant under time-translation, but also under space-translation. Therefore, the total energy and the total momentum 
 should be conserved quantities. Indeed, multiplying Eq. (\ref{chi1a}) from the left by $\bar{\chi}_{t}$ and Eq. (\ref{adchi1}) from the right by $\chi_{t}$, and then adding  both expressions,  we again obtain the continuity equation (\ref{c1}), now with   
\bq \label{dechi}
T^{00}_{\chi} = -\frac{i}{2} \left[  \bar{\chi}  \gamma^1 \partial_x \chi-\partial_x \bar{\chi} \gamma^1 \chi \right]+ m \bar{\chi} \chi - \frac{g^2}{ \kappa+1} (\bar{\chi} \chi)^{ \kappa+1} + \bar{\chi} r+ \bar{r} \chi -K \bar{\chi} \gamma^1 \chi-\mathcal{L}_0(a),    
\eq
where $\mathcal{L}_0(a)$ is chosen as 
\bq
 \mathcal{L}_0(a)  = m \bar a  a - \frac{g^2}{ \kappa+1} (\bar a  a)^{ \kappa+1} + \bar a r  + \bar{r} a - K  \bar a \gamma^1 a,  
\eq
so that $T^{00}_{\chi}$ vanishes at $x=\pm L$, when $L \to \infty$. 
Moreover, 
\bq \label{t10chi}
 T^{10}_{\chi} = -\frac{i}{2} \left[\bar{\chi}_{t}  \gamma^1 \chi- \bar{\chi} \gamma^1 \chi_{t} \right]. 
\eq
Integrating Eq. \eqref{c1}, and again assuming  $T^{10}_{\chi}(+\infty,t)-T^{10}_{\chi}(-\infty,t)=0$, the  energy of the driven NLD equation (\ref{chi1})  given by
\bq \label{energychi}
 E^{total}_{\chi} = \int_{-\infty}^{+\infty}\, dx \, T^{00}_{\chi},   
\eq
is conserved. 
Inserting the transformation (\ref{psichi}) into the equation (\ref{energy}), it can be verified that 
\ba
E^{total}_{\Psi}=E^{total}_{\chi}.
\ea 
Now multiplying Eq. (\ref{chi1a}) from the left by $\bar{\chi}_{x}$ and Eq. (\ref{adchi1}) from the right by $\chi_{x}$, and then adding  both expressions,  we obtain the continuity equation for momentum flow
\ba \label{c2}
\frac{\partial}{\partial t} T^{01}_{\chi} + \frac{\partial}{\partial x} T^{11}_{\chi}=0 , 
\ea
with  
\bq \label{dePchi}
T^{01}_{\chi} = \frac{i}{2} \left[\bar{\chi}_x  \gamma^0 \chi-\bar{\chi} \gamma^0 \chi_x \right], 
\eq
and 
\bq \label{t11chi}
 T^{11}_{\chi} = \frac{i}{2} \left[\bar{\chi}  \gamma^0 \chi_t -\bar{\chi}_t \gamma^0 \chi \right] - m \bar{\chi} \chi + 
K \bar{\chi}  \gamma^1 \chi 
 +\frac{g^2}{ \kappa+1} (\bar{\chi} \chi)^{ \kappa+1} - \bar{\chi}  r- \bar{r}  \chi+\mathcal{L}_0(a).   
\eq
Integrating Eq.\ (\ref{c2}), if $T^{11}_{\chi}(+\infty,t)-T^{11}_{\chi}(-\infty,t)=0$, then the momentum of the transformed 
driven NLD equation  is given by
\bq \label{Pchi}
 P_{\chi} = \int_{-\infty}^{+\infty}\, dx \frac{i}{2} \left[  
 \bar{\chi}_x  \gamma^0 \chi- \bar{\chi} \gamma^0 \chi_x \right],  
\eq 
which is also conserved. Now inserting the transformation (\ref{psichi}) into (\ref{Pchi})
\bq \label{PPsi}
 P_{\chi} = \int_{-\infty}^{+\infty}\, dx \left \{ K \Psi^{\dag} \Psi+\frac{i}{2} \left[  
 \bar{\Psi}_x  \gamma^0 \Psi- \bar{\Psi} \gamma^0 \Psi_x \right] \right \}.  
\eq
The right hand side of Eq.\ (\ref{PPsi}) cannot be separated into two integrals due to the fact that $\Psi^{\dag} \Psi$ does not 
vanish at $x \to \pm \infty$. However, in simulations when we deal with a finite domain, we have 
\bq \label{relationP}
P_{\chi} = K Q+P_{\Psi},  
\eq
where now 
\begin{eqnarray} \label{PPsimu}
 P_{\chi} &=& \int_{-L}^{+L}\, dx \frac{i}{2} \left[
 \bar{\chi}_x  \gamma^0 \chi- \bar{\chi} \gamma^0 \chi_x \right], \\
 Q & =&  \int_{-L}^{+L}\, dx \Psi^{\dag} \Psi, \\
 P_{\Psi} &=& \int_{-L}^{+L}\, dx \frac{i}{2} \left[
 \bar{\Psi}_x  \gamma^0 \Psi- \bar{\Psi} \gamma^0 \Psi_x \right].   
\end{eqnarray}

\section{Variational (collective coordinate)  Ansatz for the NLD equation with external driving forces} \label{sec5}

Our ansatz for the trial variational wave function is to assume that because of the smallness of the perturbation the main modification to our exact solutions to the NLD equation  (without driving forces)  is that the parameters describing the position $q(t)$, width parameter $\beta(t)$ and phase $\phi(t)$ become time dependent  functions.  We assume that the driving term is specified in the lab frame, and that the initial condition on the solitary wave is that it is a Lorentz boosted
exact solution moving with velocity $v$.  To describe the position of the solitary wave we introduce the parameter $q(t)$ which replaces $vt$ for the unforced case. We then let the width parameter $\beta$ and thus $\omega = \sqrt{m^2- \beta^2}$  become functions of time.  We next rewrite the phase of the exact solution as 
\bq
\omega t' = \gamma \omega( t-vx)  \rightarrow \phi(t) - p(t)(x-q(t))
\eq
to mimic our parametrization of the collective coordinates in the nonlinear Schr{\"o}dinger equation.  Next, we let 
$p(t) \equiv  \omega(t) \gamma(\dot q) \dot q  $  be determined from $\omega(t)$ and $q(t)$ and let the phase $\phi(t)$ be an independent collective variable.
That is, in Eq.\ (\ref{eq2.37}) we replace
\bq
vt \rightarrow q(t);~~ \beta \rightarrow \beta(t);  ~~~  \omega t'=  \gamma\omega( t - vx) \rightarrow \phi(t) -p(t)(x-q(t)) , 
\eq
where  $ p(t)= \gamma(t)   \omega(t) \dot q (t)$.   

Thus our trial wave function in component form is given by: 

\ba \label{eq4.2}
&&
\Psi_1(x,t) = \left( \cosh{\frac{\eta}{2}} A(z) 
+ i \sinh{\frac{\eta}{2}} B(z) \right) e^{-i \phi + i p(x-q)} , \nonumber \\
&&\Psi_2(x,t) = \left( \sinh{\frac{\eta}{2}} A(z) 
+ i \cosh{\frac{\eta}{2}} B(z) \right) e^{-i \phi + i p(x-q)} , 
\ea
where $z = \cosh \eta~( x-q(t))$. 
Note that $\omega$, which was a parameter in Eq. (\ref{eq2.33}), now is time dependent because of $\omega=\sqrt{m^2-\beta^2(t)}$.  
Using the trial wave function Eq.\ (\ref{eq4.2})  we can determine the effective Lagrangian for the variational parameters.
Writing the Lagrangian density as 
\bq
\mathcal{L} = \mathcal{L}_1 + \mathcal{L}_2+\mathcal{L}_3 , 
\eq
where 
\ba \mathcal{L}_1&&= \frac{i}{2} \left( \bPsi \gamma^\mu \partial_\mu \Psi - \partial_\mu \bPsi \gamma^\mu \Psi \right)  , \nonumber \\
{\cal L}_2 &&= - m \bPsi \Psi + \frac{g^2}{\kappa+1} (\bPsi \Psi)^{\kappa+1};~~ {\cal L}_3 = - \bPsi f -\barf  \Psi . 
\ea
Integrating over $x$ and changing integration variables to $z$ one obtains
\bq
L_1 = \int_{-\infty}^\infty  dx  \mathcal{L}_1 = Q \left( p {\dot q}+ {\dot \phi} - p \tanh \eta \right) - I_0  \cosh \eta - J_0 \tanh \eta , 
\eq
where the charge   
\bq
Q= \int dz[A^2(z)+B^2(z)],  
\eq
and the rest frame kinetic energy $I_0=H_1$
\bq
I_0 = \int dz \left( B' A-A'B  \right) = {H_1} , \label{izero}
\eq 
are given by Eqs. (\ref{A1}) and (\ref{A2}), respectively, in the Appendix.
 Here $ B'(x') = \frac{ dB(x')}{dx'}$, and  
\bq
J_0 = \int dz \left( {\dot  B} A- {\dot A}  B  \right)  ,  
\eq
and 
\bq {\dot  A} = \frac{ d A}{dt} = \frac{dA}{dz} \frac{dz}{dt}  , 
\eq
with a similar relation holding for $\dot B$.
Since $z=(x-q(t)) \cosh \eta$, we have 

\bq
\frac{dz}{dt}  = - {\dot q} \cosh \eta - z \tanh \eta { \dot \eta}  
\eq
and 

\bq
J_0 = - \cosh \eta  \dot{q} I_0 -{ \dot \eta}   \tanh{\eta}     \int dz 
` z  \left(  A  B' - B   A'   \right) . 
\eq 
The integrand in the second term is odd in $z$, so the integral vanishes and we are left with;
\begin{eqnarray} \label{L1}
L_1 &=& \int dx  \mathcal{L}_1 = Q \left( p {\dot q}+ {\dot \phi} - p \tanh \eta \right) - I_0 \left ( \cosh \eta -{\dot q} \sinh \eta \right), \\
L_2 &=& \int dx {\cal L}_2= -\frac{m}{\cosh \eta} I_1 
+ \frac{g^2}{(\kappa+1) \cosh \eta} I_2, \label{L2}
\end{eqnarray}
where 
\ba
I_1&& = \int dz \left( A^2(z)-B^2(z)\right)= \frac{H_2}{m}; \qquad I_2=\int dz \left( A^2(z)-B^2(z) \right)^{\kappa+1}=\frac{\kappa+1}{g^2} H_3=\frac{\kappa+1}{g^2 \kappa} H_1.  \label{i2} 
\ea
For $L_3$ we have
\ba
L_3 &&= -  2 \int dx Re \left[f^\star_1 \Psi_1 - f_2 \Psi_2 \right]= \frac{1}{\gamma} \int dz {\mathcal L}_3. \nonumber \\
\ea
We obtain for the integrand
\begin{eqnarray}
{\mathcal L}_3 &=& - 2 
r_1 \cos(\phi-K_1 q) \left\{ \cosh \frac{\eta}{2}  A(z) \cos \frac{p+K_1}{\gamma} z
-  \sinh \frac{\eta}{2}  B(z) \sin \frac{p+K_1}{\gamma} z \right\}
 \nonumber \\ 
&+& 2 r_2 \cos(\phi-K_2 q) \left\{ \sinh \frac{\eta}{2}  A(z) \cos \frac{p+K_2}{\gamma} z
-  \cosh \frac{\eta}{2}  B(z) \sin \frac{p+K_2}{\gamma} z \right\} , 
\end{eqnarray}
where  we have not included terms that are  odd in $z$.
Performing the integration we get
\begin{eqnarray} \label{4.15}
L_3 &=& - 2 
\frac{r_1}{\gamma} \cos(\phi-K_1 q) \left( \cosh \frac{\eta}{2} J_1
-  \sinh \frac{\eta}{2}  N_1 \right) 
- 2 \frac{r_2}{\gamma} \cos(\phi-K_2 q)  \left( \sinh \frac{\eta}{2}  J_2
-  \cosh \frac{\eta}{2}  N_2 \right), \label{4.15a} \\
J_j(\omega,\dot{q}) &=& \int dz A(z)  \cos \frac{p+K_j}{\gamma} z = 
\frac{\pi \cos b_j}{g \sqrt{\omega} \cosh a_j \pi}, \quad a_j=\frac{p+K_j}{2 \beta \gamma}, 
\quad b_j= a_j \cosh^{-1} m/\omega, \label{4.15b} \\
N_j(\omega,\dot{q}) &=& \int dz B(z)  \sin \frac{p+K_j}{\gamma} z = 
\frac{\pi \sin b_j}{g \sqrt{\omega} \cosh a_j \pi}.
\end{eqnarray}
The integrals $I_1$, $I_2$, $J_j$ and $N_j$ are done exactly in the 
Appendix. 
Putting all terms together and using the fact that $ \dot q = v = \tanh \eta$ we obtain:
\ba
L&& = Q {\dot \phi}  - I_0 ~ \sech \eta 
  -{m} I_1  \, {\sech \eta} +\frac{g^2}{\kappa+1}  I_2 \, \sech \eta + L_3, \nonumber \\
 L_3 && = - 
  \frac{2\pi}{g \gamma \sqrt{\omega}}  \left\{
  \frac{r_1 \cos(\phi-K_1 q)}{ \cosh a_1 \pi} 
 C_1  - \frac{r_2 \cos(\phi-K_2 q)}{ \cosh a_2 \pi} 
 S_2 
  \right\}, \nonumber \\ \label{eq4.16}
C_j && = \cosh \frac{\eta}{2} \cos b_j 
  -\sinh \frac{\eta}{2} \sin b_j, \quad   
S_j  = \sinh \frac{\eta}{2} \cos b_j 
 -\cosh \frac{\eta}{2} \sin b_j, \quad j=1,2.
\ea
Since we are using the exact solutions  of the NLD equation as our trial wave functions for the forced problem, the integrals
$I_0, I_1$ and $I_2$ are related since
 for the NLD equation without the presence of external forces, the solitary wave  with $v=0$ obeys the relationship 
\cite{LeeKuoGavrielides1975}  
\bq
\omega \psi^\dag \psi - m \bpsi \psi + \frac{g^2}{\kappa+1}
(\bPsi \Psi)^{\kappa+1} =0.
\eq
For our problem this converts into 
\bq
\omega(A^2+B^2) -m(A^2-B^2) + \frac{g^2}{\kappa+1} (A^2-B^2)^{\kappa+1} =0.
\eq
Integrating this relationship we obtain:
\bq
mI_1 - \frac{g^2}{\kappa+1} I_2 - \omega Q=H_2 -\frac{H_1}{\kappa}-\omega Q=0.  \label{relation1}
\eq
Using this relation to replace $I_1$ and $I_2$ in $L$ we obtain 
\bq
L = Q {\dot \phi}  - \frac{1}{\gamma}  (I_0 + \omega Q)  - { U }(q ,\dot q, \beta,\phi) = Q {\dot \phi}  - \frac{M_0}{\gamma}   - { U }(q ,\dot q, \beta, \phi)  ,~~ \label{Lag}
\eq
where $U = - L_3$, and $M_0=I_{0}+\omega Q$ is the rest frame energy of the solitary wave for $\kappa=1$. 
 
From Eq. \eqref{df5} we can calculate the dissipation function $F$ for the CC equations.
We find
\ba
F&& = 2 \mu \int_{-\infty}^{+\infty} dx Im (\Psi^\dag \partial_t \Psi) \nonumber \\
&&= 2 \mu  \int_{-\infty}^{+\infty} \frac {dz}{\cosh \eta} \left[ \sinh \eta \,(A \dot B - B \dot A) - \cosh \eta \,(p \dot q + \dot \phi) (A^2+B^2) \right]. 
\ea
We recognize the integrals as being related to $J_0 = - \cosh \eta  \dot q I_0$ and $Q$, so we obtain
\bq
F = -2 \mu  \left[ I_0 \sinh \eta {\dot q} + Q (p \dot q + \dot \phi) \right].
\eq

We can simplify this by introducing the boosted rest frame mass: 
\bq \label{eqM}
M =   \gamma M_0 \equiv \gamma (I_0 + \omega Q) 
\eq
and use the definition of $p(t) = \gamma \omega \dot q$ so that 
\bq
F=- 2 \mu (M {\dot q}^2 + Q \dot \phi).
\eq

This is the relativistic generalization of our expression that we found for the forced NLS equation \cite{MertensQuinteroBishop2010}. 
Now we are ready to derive Lagrange's equations for the collective coordinates using Eq. (\ref{Lag}).
From
\bq \label{eq4.24}
\frac{d}{dt} \frac {\partial L}{\partial \dot q} -  \frac{\partial L} {\partial q} = \frac {\partial { F}}{\partial \dot q}, 
\eq
we obtain
\bq \label{eq4.25}
\frac{d}{dt} \left( M \dot q \right)  = {F} _{eff}, \eq
where
\ba \label{eq4.25a}
 { F} _{eff}& =& \frac{d}{dt} \frac {\partial {U} }{\partial \dot q} -  \frac{\partial {U} } {\partial q}+ \frac {\partial  F}{\partial \dot q} .  
\end{eqnarray}
We also have a contribution from dissipation from the equation
\bq
\frac{d}{dt} \frac {\partial L}{\partial \dot \phi} -  \frac{\partial L} {\partial \phi} = \frac {\partial {F}}{\partial \dot \phi},
\eq 
which gives us a first order differential equation for $\omega$
\bq \label{omegadot}
\dot Q = Q'(\omega) \dot \omega = - 2 \mu  Q-  \frac{\partial U}{\partial \phi}, 
\eq
where the prime denotes the derivative with respect to $\omega$.

As $L$ does not depend on $\dot{\beta}$, the final Lagrange equation is 
$\partial L/\partial \beta=0$. After changing to the variable $\omega=\sqrt{m^2-\beta^2}$ we have  
\bq
\frac{\partial L} {\partial \omega} = 0. \label{final} 
\eq

This leads to  a first order differential equation for $\phi$

\bq \label{relation3}
Q'(\omega) \dot \phi = \frac{1}{\gamma}  M_0'(\omega) +\frac{\partial U}{\partial \omega} . 
\eq

In what follows we will make the simplification $\mu =0$ in our comparison of the CC equations with the numerical solution of the forced NLD equation, so that we have energy conservation as a check for our simulations. 

\subsection{Simplification at $\kappa=1$, $K_1=K$, $r_1=r$, $r_2=\mu =0$} 

In that case $U=-L_3$ simplifies to be:
\bq \label{eqU}
U(q, \dot q, \omega, \phi) = 2 r \cos (\phi- K q) \left[ \frac{\cosh \eta/2}{\cosh \eta} J_1(\omega,\dot{q}) - \frac{\sinh \eta/2}{\cosh \eta} N_1(\omega,\dot{q}) \right], 
\eq
where 
\ba
&&J_1(\omega, \dot{q}) = \frac{\pi \cos b}{g \sqrt{\omega} \cosh a \pi};~~N_1(\omega,\dot{q}) = \frac{\pi \sin b}{g \sqrt{\omega} \cosh a \pi} \nonumber \\
\ea
and
\ba
&& a(\omega,\dot{q}) = \frac{p+K}{2 \beta \gamma} = \frac{\omega \gamma \dot q +K}{2 \beta \gamma}, \qquad \beta= \sqrt{m^2-\omega^2},\nonumber \\
&& b(\omega,\dot{q}) = a \cosh^{-1} (m/\omega), \qquad \gamma(\dot q) = \cosh \eta(\dot q) = (1- (\dot q)^2)^{-1/2}. 
\ea
From the expressions (\ref{A1}) and (\ref{A3}) in the Appendix  we obtain
\bq
M_0'(\omega) = \omega Q'(\omega). \label{eqM0p} 
\eq  
Inserting Eqs.\ (\ref{eqM0p}), (\ref{A1}), and (\ref{eqU}) in Eqs.\ (\ref{relation3}) and (\ref{omegadot}) we obtain, respectively,  
\begin{eqnarray}
\dot{\phi} &=&  \frac{\omega} {\gamma}  -\frac{ g^2 r \beta(\omega)    \omega^{2} \cos ( \phi-K q) }{ m^2}   \left[\frac{\cosh \eta/2}{\cosh \eta}  \frac{\partial J_1(\dot q, \omega)}{\partial \omega} -\frac{\sinh \eta/2}{\cosh \eta}  \frac{\partial N_1(\dot q, \omega)}{\partial \omega} \right], 
\label{dotphi1} \\
 \dot \omega &=& -\frac{ g^2 r \beta   \omega^{2} \sin (\phi-K q) }{ m^2}  \left[ \frac{\cosh \eta/2}{\cosh \eta} J_1(\dot q, \omega) - \frac{\sinh \eta/2}{\cosh \eta}
  N_1(\dot q, \omega) \right]. \label{omdot1}
\end{eqnarray}
In the special case of $\mu=0$, from Eqs.\ (\ref{eq4.25}) and (\ref{eq4.25a}) we obtain 
\bq \label{final}
\frac{d}{dt} (M_0 \gamma(t) \dot q) = \frac{d}{dt} \frac {\partial {U} }{\partial \dot q} -  \frac{\partial {U} } {\partial q}. 
\eq

\subsection{ Solutions when $K=v_0 = q_0 = 0$}
When we look in the rest frame where $v=0$ and look for solutions, we notice when $K=v_0 =0$ that
$a=b=0$. Thus $N_1=0$ and 
\bq
J_1 = \frac{\pi}{g \sqrt \omega}; ~~ U =  2 r \cos(\phi)  \frac{\pi}{g \sqrt \omega} ,~~\frac {dJ_1}{d \omega} |_{ (K=0)} =-\frac{\pi }{2 g \omega^{3/2}} . 
\eq
Therefore, the equation of motion given by (\ref{final}) is always satisfied. The Eqs.\ (\ref{dotphi1}) and (\ref{omdot1}), respectively,  become:
\begin{eqnarray}
\dot{\phi} &= & {\omega}  +\frac{ g \pi  r \beta   \omega^{1/2} \cos  \phi}{ 2m^2}  \label{eqc0nr}, \\
 \dot \omega &=& -\frac{ \pi g r \beta   \omega^{3/2} \sin \phi}{ m^2}. 
\end{eqnarray}
Note that if $\phi = \bar \phi + \phi_0$ and we start with $\phi_0 = n \pi$ ($n$ is an integer) we have 
\begin{eqnarray}
\dot{ \bar \phi} &=&  {\omega}  +\frac{ g \pi  r \beta   \omega^{1/2} (-1)^n  \cos  \bar \phi}{ 2m^2}, \\
\dot \omega &=& -\frac{ \pi g r \beta   \omega^{3/2} (-1)^n \sin  \bar \phi}{ m^2} .  
\end{eqnarray}
This has an unstable stationary solution for $n=1$,  namely
\bq
 \dot {\bar \phi}= 0, ~~ \bar \phi =0, ~ \omega_0=  \frac{g \pi r}{2 m^2}  \beta_0 \sqrt{\omega_0} .
\eq
For small $ g \pi r $
\bq
\omega_0= \frac{g^2  \pi^2 r^2}{4 m^2 }.
\eq
This solution is unstable to small perturbations.
\subsubsection{Analytic solutions for $\phi_0=\pi/2$, $K=v=0$ and  $g \pi r/m^2 \ll 1$}
 
\begin{figure} 
\begin{tabular} {cc}
\includegraphics[width=3in]{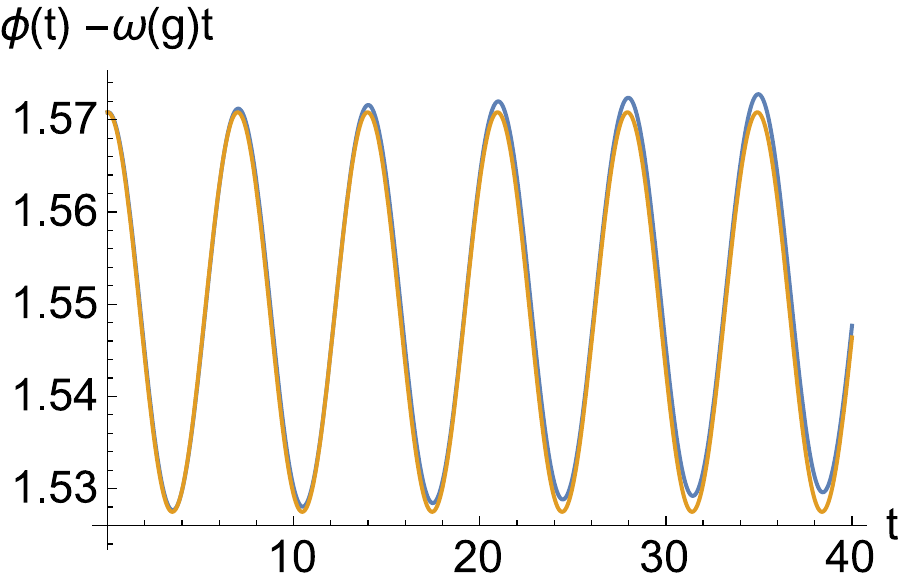}
& \includegraphics[width=3in]{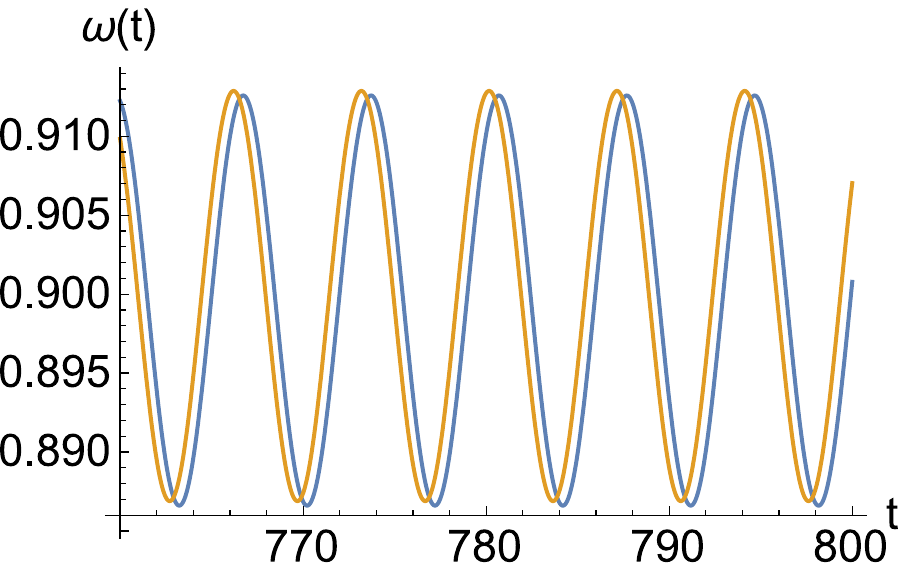}
\end{tabular}
\caption{Left panel: $\phi(t) - \omega(g)  t$ vs $t$, 
Right panel: $\omega(t)$ vs $t$. Blue curves are the analytic result. Red curves are the result of solving numerically the CC equations. 
Parameters are: $\omega_0=0.9$, $\dot{q}(0)=0$, $K=0$,  $r_1=0.01$, $r_2=0$, $m=1$ and $g=1$.}
\label{figcompare}
\end{figure}

If we choose  $\phi_0=\pi/2$, we get the simple equations:
 \begin{eqnarray}
\dot{ \bar \phi} &=&  {\omega}  - \frac{ g \pi  r \beta   \omega^{1/2}   \sin  \bar \phi}{ 2m^2}, \\
 \dot \omega &=& -\frac{ \pi g r \beta   \omega^{3/2} \cos  \bar \phi}{ m^2}. 
\end{eqnarray}
One can expand these equations as a power series in $\tilde g = g \pi r/m^2$. To order ${\tilde g}^2$ we obtain:
\begin{eqnarray} \label{phian}
\phi(t) &=& \frac{\pi}{2} +  \left[\omega_0+ \frac{\tilde g^2}{2} \left(m^2-2 \omega_0 ^2\right)  \right] t 
+\frac{3 \tilde g \beta_0 [\cos (\omega_0 t) -1]}{2 \sqrt{\omega_0}}+\tilde g^2 \frac{\sin (2 \omega_0 t)}{8 \omega_0} \left(
7 \omega_0^2-5 m^2 \right)+O\left(\tilde g^3\right), \label{omanalytic} \\
   \omega(t) &=& 
  \omega_0-\tilde g \sqrt{\omega_0} \beta_0 \sin ( \omega_0 t)+\tilde g^2 \sin ^2( \omega_0 t) \left(\frac{3}{2} m^2 -
  2 \omega_0^2 \right)+O\left(\tilde g^3\right),  \label{analytic}
   \end{eqnarray}
where $\beta_0= \sqrt{m^2-\omega_0^2}$.    
   We notice that the term in $\phi(t)$ that is linear in $t$  shows a constant shift from its initial value $\omega_0$, namely
   $\omega_0   \rightarrow   \omega (g) =  \omega_0 + \frac{1}{2}  {\tilde g} ^2 \left(m^2-2 \omega_0 ^2 \right)$.
   For $g=m=1$, $r= 0.01$ and $\omega_0=0.9$ we find  $\omega(g) = 0.899694$.  For these initial data we have compared our analytic solutions with the numerical solutions of the CC equations.  At  all times  the $\omega(t)$ found analytically from Eq. \eqref{analytic}  tracks the numerical solution almost perfectly slightly getting out of phase at late times. The phase $\phi$, after we eliminate the linear growth as determined numerically, gradually starts diverging from the analytical result 
   Eq. \eqref{phian}  even at modest times.  These results are shown in Fig. \ref{figcompare}.


   \subsection{Variational method for the transformed NLD equation with external force, $K_1=K_2=K$}

Let us now  consider the case of an external force (\ref{force}) with $\nu_j=0$ and $K_1=K_2=K$. From 
(\ref{psichi}) and the ansatz (\ref{eq4.2})  it follows  that
\ba \label{eq4.2a}
\chi_1(x,t) &=& \left( \cosh{\frac{\eta}{2}} A(z) 
+ i \sinh{\frac{\eta}{2}} B(z) \right) e^{-i \tilde{\phi} + i \tilde{p}(x-q)},  \\
\chi_2(x,t) &=&  \left( \sinh{\frac{\eta}{2}} A(z) 
+ i \cosh{\frac{\eta}{2}} B(z) \right) e^{-i \tilde{\phi} + i \tilde{p}(x-q)},\nonumber 
\ea
where
\ba \label{tildes}
\tilde{p} &=& \gamma \omega \dot{q}+K, \\ 
\tilde{\phi} &=& \phi- K q. \nonumber
\ea  
Inserting (\ref{eq4.2a}) into the Lagrangian density (\ref{eq3.8a}) and integrating over $x$ we obtain
\begin{equation}	\label{lagrangianchi}
\tilde{L} = Q (\dot{\tilde{\phi}}+K \dot{q})  - \frac{M_0}{\gamma}   - \tilde{U}(\dot q, \beta, \tilde{\phi}), 
\end{equation}
where
\begin{equation}	\label{utilde}
 \tilde{U}(\dot q, \beta,  \tilde{\phi})= \frac{2 \pi \cos \tilde{\phi} (r_1 \tilde{C}-r_2 \tilde{S})}{g \gamma \sqrt{\omega} \cosh \tilde{a} \pi}, 
\end{equation}
with  $\tilde{a}=\tilde{p}/(2 \beta \gamma)$, $\tilde{b}=\tilde{a} \cosh^{-1} m/\omega$, and $\tilde{C}$ and $\tilde{S}$ are defined by 
Eq. (\ref{eq4.16}) with $\tilde{b}$ instead of $b$. Remarkably, $\tilde{L}$ does not depend on $q$ because the potential 
$\tilde{U}$ in Eq. (\ref{utilde}) no longer depends on $q$. 

From the Lagrange equation 
\bq \label{lq}
\frac{d}{dt} \frac {\partial \tilde{L}}{\partial \dot Y} -  \frac{\partial \tilde{L}} {\partial Y} = 0,
\eq
with $Y=q$, 
it is clear that the canonical momentum
\begin{equation} \label{eq4.42z}
\tilde{P}_q=\frac{\partial \tilde{L}}{\partial \dot{q}}=\gamma M_0 \dot{q}+K Q-\frac{\partial \tilde{U}}
{\partial \dot{q}}, 
\end{equation}
is conserved. Moreover, we obtain 
\begin{eqnarray} \label{reltildep}
(\dot{\tilde{\phi}}+K \dot{q}) \frac{dQ}{d\omega}-\frac{1}{\gamma} \frac{dM_0}{d\omega} -\frac{\partial \tilde{U}}{\partial \omega}&=&0, \label{relQ} \\
\frac{d Q}{dt} +\frac{\partial \tilde{U}}{\partial \tilde{\phi}}&=&0, \label{relphi}
\end{eqnarray}
from (\ref{lq}) with  $Y=\omega$, $Y=\tilde{\phi}$, respectively.
Multiplying Eq.\ (\ref{eq4.42z}) by $\ddot{q}$, using (\ref{reltildep})-(\ref{relphi}), and integrating over $t$ we obtain 
that the energy 
\begin{equation} \label{energycc}
\tilde{E}_{cc}=\frac{M_0}{\gamma}+(\tilde{P}_q-K Q) \dot{q}+\tilde{U}, 
\end{equation}
is conserved. 

Inserting the ansatz (\ref{eq4.2a}) into the expression for the field momentum (\ref{Pchi}) we obtain
\begin{equation} \label{eqPtilde}
\tilde{P}_{\chi} = \tilde{p} Q + \dot{q} \gamma H_1.
\end{equation}
This means that in the solutions of collective coordinates equations, 
$\tilde{P}_q$ must be a conserved quantity. In simulations of the transformed NLD equation (\ref{chi1}), 
 the momentum (\ref{Pchi}) 
 must be conserved and equal to $\tilde{P}_{\chi}(0)$.  

A similar remark holds for the energy. Indeed, inserting the ansatz (\ref{eq4.2a}) into (\ref{energychi}) and 
integrating over $x$ we obtain 
\ba \label{energy1}
\tilde{E}_{\chi}=\gamma H_2 + \tilde{U}(\dot{q},\beta,\tilde{\phi}). 
\ea
This energy is equal to the energy of the driven NLD equation  $E_{\Psi}$ obtained by inserting 
the ansatz (\ref{eq4.2}) into (\ref{energy}). 
The energy $\tilde{E}_{cc}$ (\ref{energycc}) obtained from the solutions of the collective coordinates equations must be constant, while the energies 
$E^{total}_{\Psi}$ and 
$E^{total}_{\chi}$, computed from the simulations of the driven NLD equation and transformed 
NLD equation, respectively, not only 
must be  equal to each other, but also they must be constant if the energy density satisfies certain condition at the boundaries.  

As $\tilde{U}$ in Eq.\ (\ref{utilde}) does not depend on $q(t)$; using $v(t)=\dot{q}$ as a new collective coordinate we obtain one    algebraic equation (\ref{eq4.42z}) and the following two differential equations of first order:  
\begin{eqnarray} \label{odew1}
 \dot \omega&=& -\frac{\pi g \beta   \omega^{3/2}}{\gamma m^2} \frac{\sin \tilde \phi}{\cosh \tilde a \pi} \left 
 (r_1  C  - r_2  S \right) + \frac{2}{m^2} \mu \omega \beta^2, \\
\label{odephit1}
\dot{\tilde{\phi}}&=& -K \dot{q}+ \frac{\omega}{\gamma} 
-  \frac{g^2 \omega^2 \beta}{2 m^2} \frac{\partial U}{\partial \omega}, 
\end{eqnarray}
where in $\gamma$ we should also replace $\dot{q}=v(t)$ and 
\begin{eqnarray} \label{cc1}
\frac{\partial U}{\partial \omega} &=& -\frac{U}{2 \omega} - \pi U \tilde{a}_\omega \tanh \tilde{a} \pi + \frac{ 
2 \pi \cos \tilde{\phi} (r_1  C_\omega  - r_2  S_\omega )}{g \gamma \sqrt{\omega} \cosh \tilde{a} \pi}, \\
\label{cc2}
\tilde{a}_\omega&=& \frac{K \omega+\gamma m^2 \dot{q}}{2 \gamma \beta^3}, \qquad \tilde{b}_\omega=\tilde{a}_\omega \cosh^{-1} \frac{m}{\omega} -\frac{\tilde{a} m}{\omega \beta , }\\
 \label{cc3}
C_\omega &=& -\tilde{b}_\omega  \left[\sinh \frac{\eta}{2} \cos \tilde{b} 
 +\cosh \frac{\eta}{2} \sin \tilde{b} \right], \\
 \label{cc4}
S_\omega &=& -\tilde{b}_\omega \left[\cosh \frac{\eta}{2} \cos \tilde{b}+\sinh \frac{\eta}{2} \sin \tilde{b}\right],\\
\label{cc5}
\frac{\partial U}{\partial \dot{q}} &=& \frac{2 \pi \cos \tilde{\phi}}{g \gamma \sqrt{\omega} \cosh \tilde{a} \pi} \left(
-(\dot{q} \gamma^2 + \pi \tilde{a}_{\dot{q}} \tanh \pi \tilde{a} )(r_1 C - r_2 S) + r_1 C_{\dot{q}} -r_2 S_{\dot{q}}  \right), \\
\label{cc6}
\tilde{a}_{\dot{q}} &=& \frac{\omega -K \dot{q} \gamma}{2 \beta}, \\
 \label{cc7}
C_{\dot{q}} &=& \frac{\gamma^2 S}{2} - \tilde{a}_{\dot{q}}  \cosh^{-1} \frac{m}{\omega} \left[\sinh \frac{\eta}{2} \cos \tilde{b} 
 +\cosh \frac{\eta}{2} \sin \tilde{b} \right], \\
 \label{cc8}
S_{\dot{q}} &=& \frac{\gamma^2 C}{2} - \tilde{a}_{\dot{q}} \cosh^{-1} \frac{m}{\omega} \left[\cosh \frac{\eta}{2} \cos \tilde{b} 
 +\sinh \frac{\eta}{2} \sin \tilde{b} \right].
\end{eqnarray} 
The Eqs.\ (\ref{odew1})-(\ref{odephit1}) can be obtained using the Lagrange equation (\ref{lq}) and setting $Y=\tilde{\phi}, \omega$, respectively. Now we need to solve two first order differential 
equations (\ref{odew1}) and (\ref{odephit1}) and one algebraic equation 
(\ref{eq4.42z}), where the three unknowns are $\omega(t)$, $v(t)$ and $\tilde{\phi}(t)$. Then, using the substitution  (\ref{tildes}) one can obtain  $\phi(t)$. 

In the above equations the subscripts on the variables $ \tilde a, \tilde b, C, S$ refer to (partial)  derivatives with respect to the subscript variable.

\subsection{Non-relativistic (NR)  regime with $K_1=K$, $r_1=r$, $r_2=0$.}

In the non-relativistic regime, $v=\dot q \ll 1$. Therefore,  
\bq
\sinh { \delta_1 \eta } \rightarrow  \delta_1 \dot q ;~~ \cosh  {\delta_1 \eta}  \rightarrow 1;  ~~ p \rightarrow \omega \dot q ;  ~~ \gamma \rightarrow 1, 
\eq
where in our case $\delta_1=1/2$ or $\delta_1=1$.
Thus in determining $U$, 
\bq
\tilde a (\dot q, \omega) = \frac{\omega \dot q+K}{2 \beta}; ~~ \tilde b (\dot q, \omega) =  \tilde a \cosh^{-1} (m/\omega). 
\eq
 For determining the NR equations for $\phi$ and $\omega$ we need an expression for $\tilde U$  valid up to terms linear in $\dot q$. We find
\bq
\tilde U = 2   r \cos \tilde \phi \left[J_1(\dot q, \omega) - \frac{\dot q}{2} N_1(\dot q, \omega) \right]. 
\eq
Thus the equations of motion (\ref{odew1}) and (\ref{odephit1}) become, respectively,
\begin{eqnarray}  \label{odew1nr}
 \dot \omega &=& -\frac{ g^2 r \beta   \omega^{2} \sin \tilde \phi}{ m^2}  \left[ \ J_1(\dot q, \omega) - \frac{\dot q}{2} 
  N_1(\dot q, \omega) \right], \\ 
\dot{\tilde{\phi}} &=&  -K \dot{q}+ {\omega}  -\frac{ g^2 r \beta   \omega^{2} \cos \tilde \phi}{ m^2}   \left[  \frac{\partial J_1(\dot q, \omega)}{\partial \omega} - \frac{\dot q}{2}  \frac{\partial N_1(\dot q, \omega)}{\partial \omega} \right] ,  
\label{eqc1nr}
\end{eqnarray}  
and the canonical momentum (\ref{eq4.42z}) becomes:
\bq
\tilde{P}_q = \frac{ 2 K \beta}{g^2 \omega} + \frac{4 m}{g^2}  \tanh ^{-1} \sqrt{ \frac{m-\omega}{m+  \omega}} ~\dot q   - \frac {\partial \tilde U} {\partial \dot q} . 
 \eq
To determine the conserved  $\tilde{P}_q $ up to terms linear in $v$,  we need an expression for  $\tilde U$ up to terms of order $v^2$. We find
\bq
\tilde U = -\frac{\pi  r \cos ( \tilde \phi ) \text{sech}\left(\frac{\pi  K}{2 \sqrt{m^2-\omega^2}}\right)}{4 g \sqrt{\omega} \left(m^2-\omega^2\right)^{3/2}} ( a_0+a_1 v + a_2 v^2 + ...) , 
\eq
where we have
\bq
a_0=-8 \left(m^2-\omega^2\right)^{3/2} \cos \left(\frac{K \cosh ^{-1}\left(\frac{m}{\omega}\right)}{2 \sqrt{m^2-\omega^2}}\right)
\eq
and 
\ba
&&a_1=4 \left(m^2-\omega^2\right)^{3/2} \sin \left(\frac{K \cosh ^{-1}\left(\frac{m}{\omega}\right)}{2 \sqrt{m^2-\omega^2}}\right)+4 \omega \left(m^2-\omega^2\right) \cosh
   ^{-1}\left(\frac{m}{\omega}\right) \sin \left(\frac{K \cosh ^{-1}\left(\frac{m}{\omega}\right)}{2 \sqrt{m^2-\omega^2}}\right) \nonumber \\
 &&  +4 \pi  \omega \left(m^2-\omega^2\right) \tanh
   \left(\frac{\pi  K}{2 \sqrt{m^2-\omega^2}}\right) \cos \left(\frac{K \cosh ^{-1}\left(\frac{m}{\omega}\right)}{2 \sqrt{m^2-\omega^2}}\right) . \nonumber \\
   \ea
   Note that  when $K=0$, $a_1 \rightarrow 0$. 
   We will display $a_2$ for $K=0$ below. 

  \subsubsection{Nonrelativistic solutions with $K=0$}
\begin{figure}
\includegraphics[width=4in]{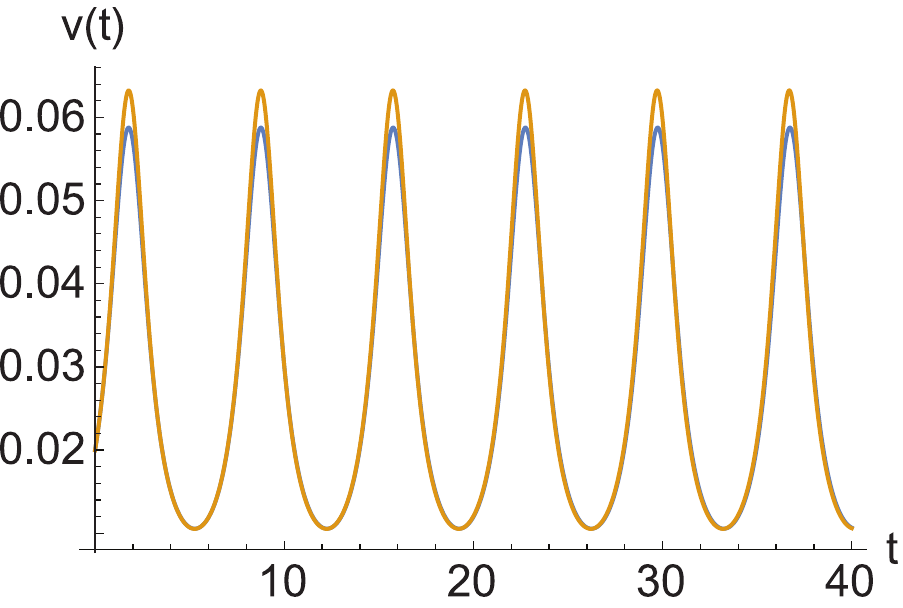}
\caption{$v$ as a function of time $t$ for an initial non-relativistic velocity $v_0 = 1/50$. 
Parameters are: $\omega_0=0.9$, $K=0$,  $r_1=r=0.01$, $r_2=0$, $m=1$, $g=1$. The upper curve is the analytic solution.}
 \label{vanalytic} 
\end{figure}

  When $K=0$, it is easy to expand $ \tilde U$ up to order $v^2$,  and obtain the  two equations of motion and one constraint equation up to order $v$. 
  One has:
  \ba
  \tilde U_{nr}  &&  = \frac{2 \pi  r \cos (\phi ) } { g \sqrt{\omega }}   \nonumber \\
 &&  \times \left( 1-\frac{v^2 \left(\sqrt{m^2-\omega ^2} \left(3 m^2+\left(\pi ^2-3\right) \omega ^2\right)+\omega ^2 \sqrt{m^2-\omega ^2} \cosh ^{-1}\left(\frac{m}{\omega
   }\right)^2+2 \omega  \left(m^2-\omega ^2\right) \cosh ^{-1}\left(\frac{m}{\omega }\right)\right)}{8 \left(m^2-\omega ^2\right)^{3/2}} \right) . 
 \nonumber \\
   \ea
   We have
   \ba
   - \frac{\partial \tilde U_{nr}} {\partial v}&& = \frac{2 \pi  r \cos (\phi ) } { g \sqrt{\omega }}  \nonumber \\
&& \times \frac{v \left(\sqrt{m^2-\omega ^2} \left(3 m^2+\left(\pi ^2-3\right) \omega ^2\right)+\omega ^2 \sqrt{m^2-\omega ^2} \cosh ^{-1}\left(\frac{m}{\omega
   }\right)^2+2 \omega  \left(m^2-\omega ^2\right) \cosh ^{-1}\left(\frac{m}{\omega }\right)\right)}{4 \left(m^2-\omega ^2\right)^{3/2}} . \nonumber \\
   \ea
   The non-relativistic conservation law for $ \tilde P_q$ when $K=0$, valid  to linear order in $v$ is then
   \bq
    \tilde P_q = M_0(\omega) v -  \frac{\partial \tilde U_{nr}} {\partial v} . 
   \eq
   This allows us to explicitly solve for $v$  as a function of $\omega(t)$, $\phi(t)$   and the initial values of $v,  \phi$ and $\omega$.  Note that because the velocity  corrections to $ \tilde U$ start at $v^2$,  the leading terms to solve for $\phi$ and $\omega$ are the {\it same} equations that we needed to solve in the $v=0$ case, 
   namely:
   \begin{equation}  \label{odnr1}
 \dot \omega = -\frac{ \pi g r \beta   \omega^{3/2} \sin \phi}{ m^2}   , 
\end{equation}
\begin{equation}
\dot{\phi} =  {\omega}  +\frac{ g \pi  r \beta   \omega^{1/2} \cos  \phi}{ 2m^2}  \label{eqc0nr1} , 
\end{equation} 
which we have already solved explicitly  as a power series in $\tilde g$.  These solutions are given in Eqs.  \eqref{phian}, \eqref{analytic}. We have that 
\ba
&&  \frac{v}{v_0} =   \tilde P_q( v_0, \omega_0, \phi_0)   \nonumber   \\
&&  \times \left[ M_0(\omega) +  \frac{2 \pi  r \cos (\phi ) } { g \sqrt{\omega }}   \frac{\left(\beta(\omega) \left(3 m^2+\left(\pi ^2-3\right) \omega ^2\right)+\omega ^2 \beta(\omega) \cosh ^{-1}\left(\frac{m}{\omega} \right)^2+2 \omega \beta^2(\omega)  \cosh ^{-1}\left(\frac{m}{\omega }\right)\right)}{4 \beta^3(\omega)}  \right] ^{-1} . \nonumber \\
\ea
For the case $\phi_0 = \frac{ \pi}{2}$, one has $ \tilde P_q( v_0, \omega_0, \phi_0) = M_0(\omega_0)$.  For the initial conditions we considered earlier when $v_0=0$, namely $K=0, r=1/100, m=1, g=1$, and initial velocity $v_0=1/50$ we get the result shown in Fig.  \ref{vanalytic}.  In that figure we compare the analytic result with the solution of the CC equations.  The analytic result is slightly higher at the maxima of $v(t)$. 


%
%

\section{Spectrum of the Linear Excitations (Phonons)} \label{sec6}
 Similar to the case of the forced NLS equation \cite{MertensQuinteroBishop2013}, the external force $f_j=r_j e^{-iK_j x}$ [see Eq. \eqref{force2}] excites not only soliton oscillations, but also a plane wave phonon (short for a linear excitation) such that the total momentum is conserved.
 
 The general solution of the linearized NLD equation without damping [Eq. \eqref{nlde1a} with $\mu=0$] and $K_1=K_2=K$, reads
 \bq
 \Psi_{ph} = b ~e^{i( kx-\omega_{ph} t + \theta)} + c~ e^{-iKx}, \label{phonon}
 \eq
 with the phonon dispersion curve
 \bq
 \omega_{ph} (k) = \sqrt{m^2+k^2},
 \eq
 and arbitrary, but small $b$.  In the case $r_1=r$ and $r_2=0$, which was considered in Sec. V, the spinor $c$ has the simple form
 \bq
 c= \frac{r}{\Omega_K^2} \left(  \begin{array}{c}
    -m \\ 
    K\\ 
  \end{array} \right),~~ \Omega_K=  \sqrt{m^2+K^2}. \label{omK}
 \eq
We choose $|K|  \ll m$, then the second component of $c$ is much smaller than the first one [see below Eq. \eqref{force}].

In the presence of a soliton we need to know the phonons only far away from the soliton. For example, when both the soliton and the phonon move to the right,  at the far left the phonon wave function is given by Eq. \eqref{phonon} with $\theta = 0$, while at the far right there is a phase shift $\theta \neq 0.$

We now calculate the charge  $Q_{ph}$ and the momentum $P_{ph}$ of the phonon by integrating $\Psi^\dag \Psi$ and $T^{01}$ over the interval $-L \leq x \leq L$.  Here we must distinguish two cases:  $k \neq -K$ and  $k= -K$.  
In the latter case the $x$-dependent parts in the integrands of $P_{ph}$ and $Q_{ph}$ drop out.  We then obtain 
\bq
Q_{ph} = 2 L (b^\dag b + c^\dag c)+ 2 L ~ {\rm Re} \{ c^\dag b e^{- i (\omega_K t- \theta)} - b^\dag c e^{ i \omega_K t } \} . 
\eq
We also find $P_{ph} = - K Q_{ph}$ so that
\bq \label{mom1}
P_{ph} + K Q_{ph} = 0.
\eq
The left hand side of  Eq. \eqref{mom1} is the total momentum in the transformed system discussed in Sec. IV.  It is zero because the wave function describing the phonon,
\bq
\chi_{ph} = \Psi_{ph} e^{iKx} = b ~e^{-i(\omega_K t - \theta)} + c,
\eq
is simply a homogeneous oscillation and does not travel in the transformed system.

In our simulations which will be presented in the next section, we use a finite system of length $2L$ and periodic boundary conditions.  The predicted phonon mode with wavenumber $k= -K$ is clearly identified in the spectrum of $Q(t)$, see panel (e) of Figs.\  
\ref{fig1}, \ref{fig3} and \ref{fig5}.  The observed frequencies $\omega_2$ agree very well with $\Omega_K$ given by Eq. \eqref{omK} for $K=0$ and $K = \pm \frac{3 \pi}{L}$ with $L=100$.

The phonon mode is also seen indirectly in the spectrum of the maximum of the charge density $\rho(x,t)$ shown in panel (f) of Figs. \ \ref{fig1}, \ref{fig3} and \ref{fig5}. This is a {\it local } 
quantity which is used for the computation of the soliton position $q(t)$, in contrast to the  {\it global} quantity $Q(t)$ which is obtained by integration over the whole system. The phonon frequency $\omega_2$ is observed in the difference $\omega_3 = \omega_2 - \omega_1$, where $\omega_1$ is identified as the frequency of the intrinsic soliton oscillations discussed in the next section.


\section{ Simulations vs numerical solutions of collective coordinates equations} \label{sec7}

\begin{figure}[th!]
\begin{tabular}{c}
 \includegraphics[width=3in]{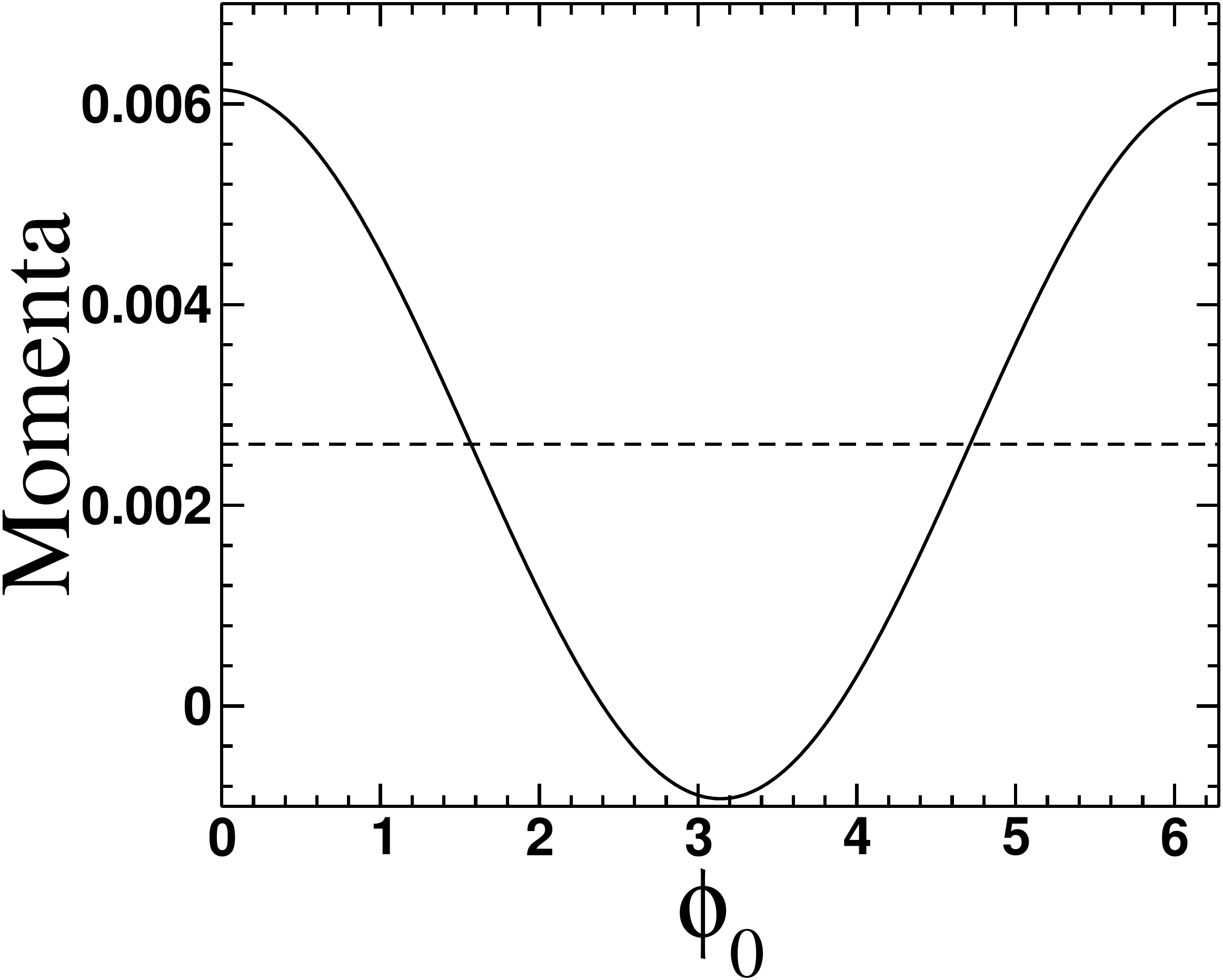}\qquad
 \includegraphics[width=3in]{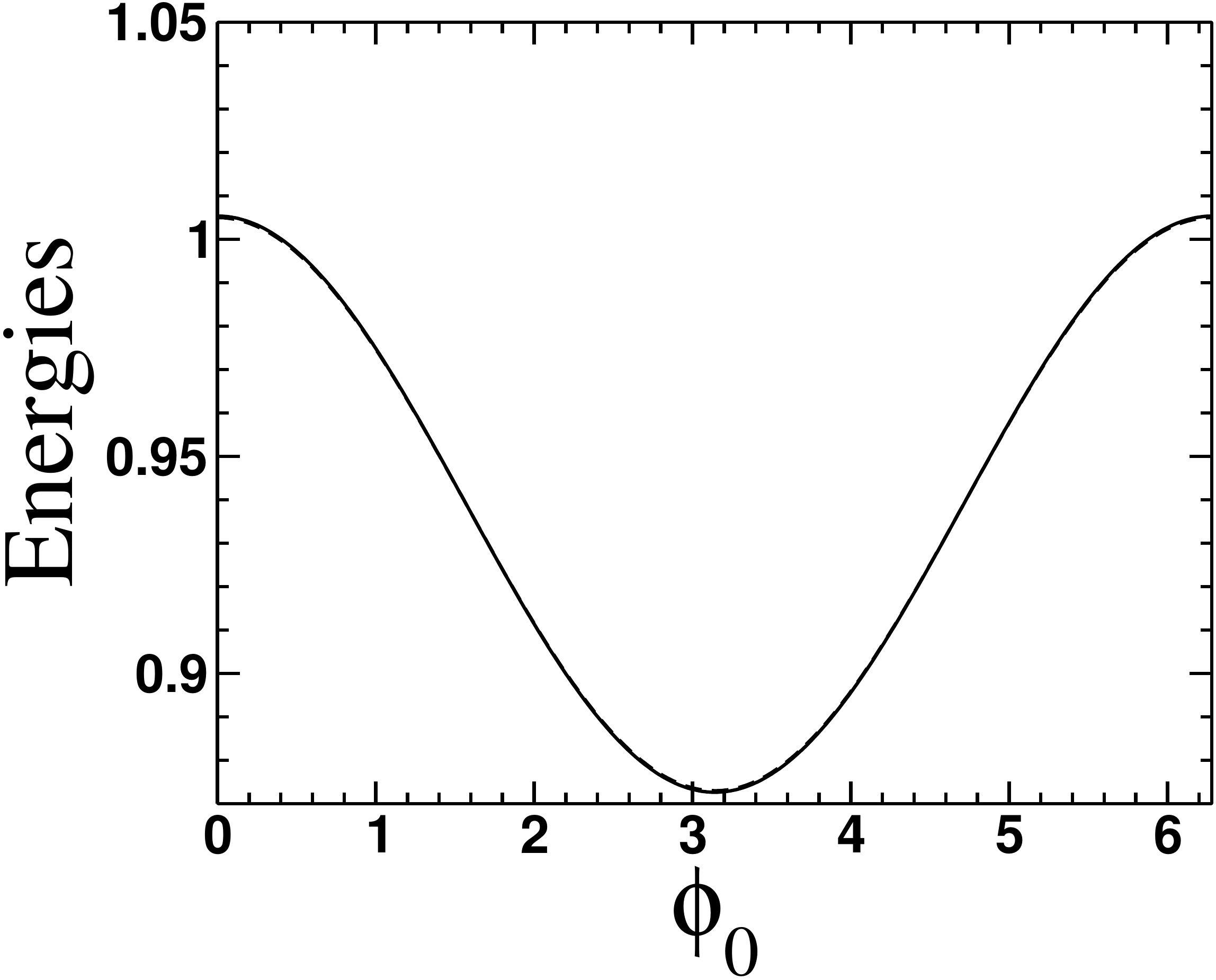}
\end{tabular}
\caption{Left panel: $P_q$ (solid line) and $\tilde{P}_{\chi}$ (dashed line) vs initial phase. 
Right panel: $E_{cc}$ (solid line) and $\tilde{E}_{\chi}$ (dashed line superimposed) vs initial phase. 
The maximum difference between $E_{cc}$ and $\tilde{E}_{\chi}$ is of order of $10^{-4}$. 
Parameters are: $\omega_0=0.9$, $\dot{q}(0)=0.1$, $K=-3 \pi/100$, $r_1=r=0.01$, $r_2=0$, $m=1$, $g=1$.}
 \label{figPnew}
\end{figure}

\begin{figure}[th!]
\begin{tabular}{c}
\includegraphics[width=3in]{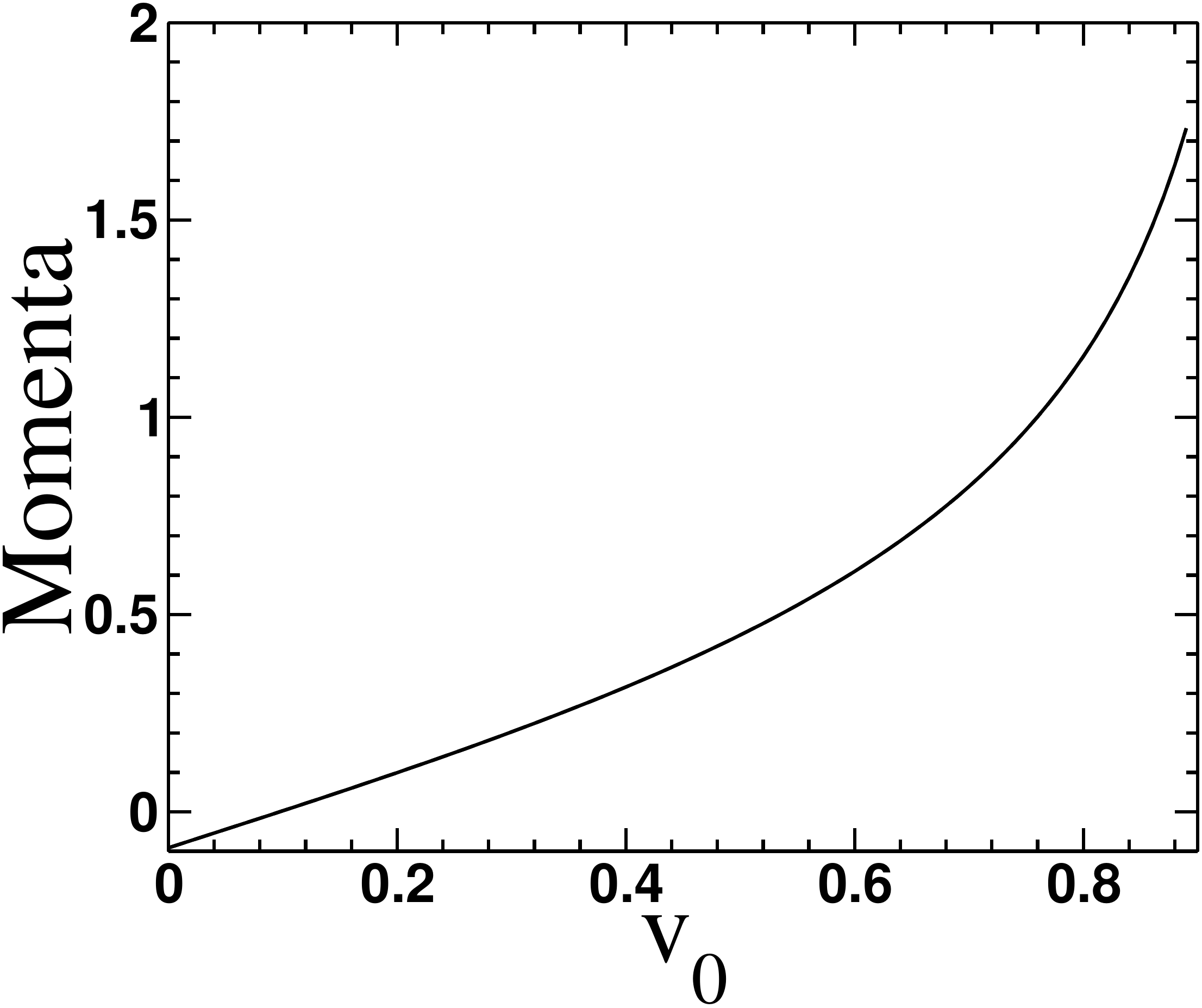}\qquad 
\includegraphics[width=3in]{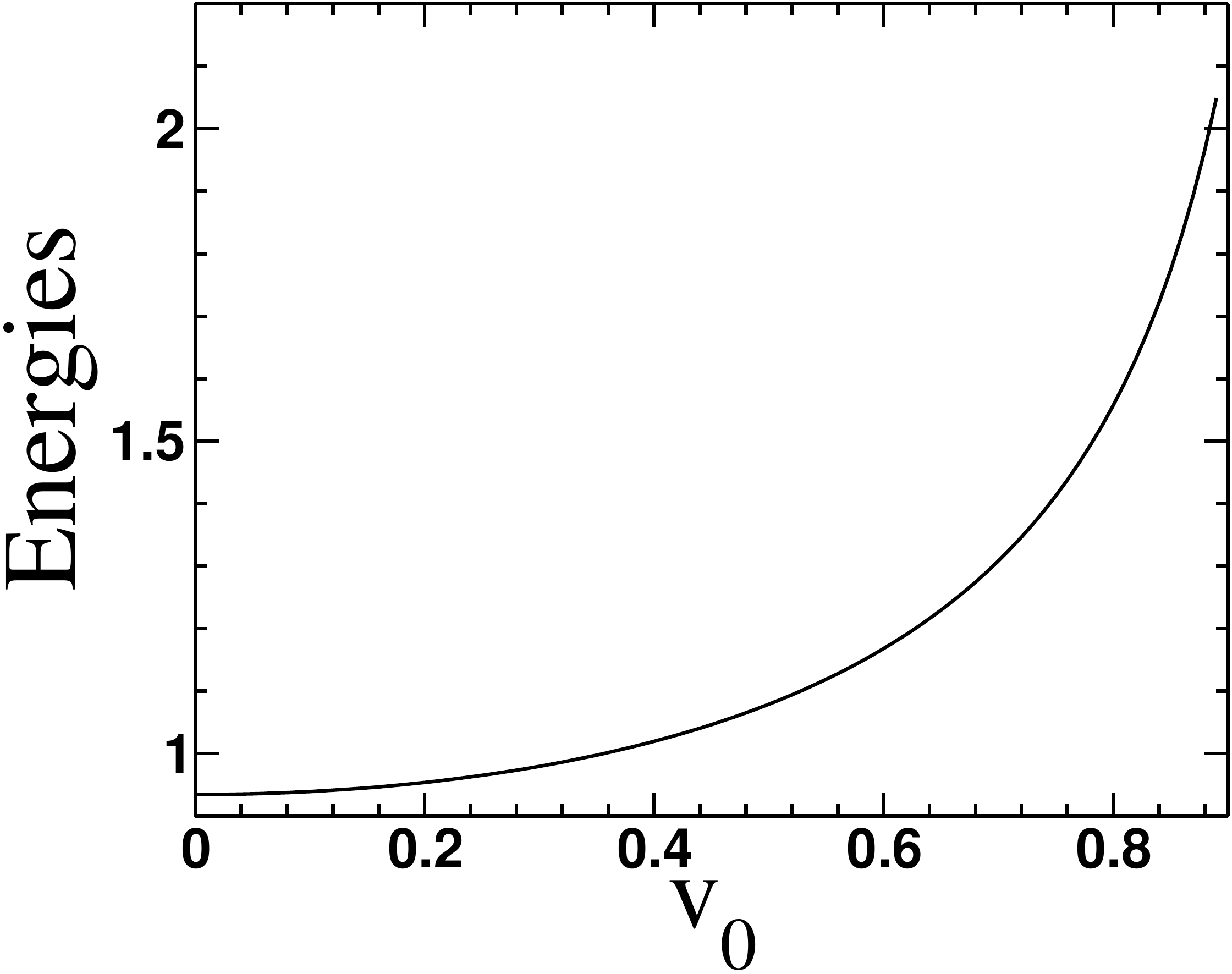}
\end{tabular}
\caption{Left panel: $P_q=\tilde{P}_{\chi}$ vs initial velocity. 
Right panel: $E_{cc}=\tilde{E}_{\chi}$ vs initial velocity. 
Parameters are: $\omega_0=0.9$, $\phi(0)=\pi/2$, $K=-3 \pi/100$, $r_1=r=0.01$, $r_2=0$, $m=1$, $g=1$.}
\label{figEnew}
\end{figure}

In order to obtain the numerical solutions of the collective coordinates equations we need to set initial conditions for $\dot{q}(0)$, 
$\omega(0)$ and $\tilde{\phi}(0)$. In simulations we use the soliton solution of the unperturbed NLD equation with the same initial conditions. 
We would like to stress here that arbitrary sets of these initial conditions produce different quantities for $P_q$ and 
$\tilde{P}_{\chi}$ (see Fig. \ref{figPnew}). Numerical solutions of collective coordinates must conserve $P_q$ and $E_{cc}$, 
whereas in simulations  $\tilde{P}_{\chi}$ 
and  $\tilde{E}_{\chi}$ are conserved. Therefore,  good agreement between simulations and numerical solutions  is expected only 
for the initial conditions that guarantee $P_q=\tilde{P}_{\chi}$ and $E_{cc}=\tilde{E}_{\chi}$. The simplest case is to choose 
initially $\tilde{\phi} (0)= \pm \pi/2$, $q(0)=0$ and arbitrary values for $\dot{q}(0)$ and 
$\omega(0)$ (see Fig.\ \ref{figEnew}). 

The CC theory leads to the algebraic equation  (\ref{eq4.42z}) 
and the two ODEs   \eqref{odew1}, and  \eqref{odephit1} which are solved by a MATHEMATICA program. The driven NLD Eq. \eqref{nlde1a} is a PDE for which various numerical schemes have been proposed that are reviewed in Ref. \cite{XuShaoTang2013}. It is also reported there that the operator splitting (OS) method performs better than the other schemes in terms of accuracy and efficiency. Therefore we have applied a fourth-order OS method in a recent paper on the stability of solitons in the undriven NLD equation \cite{ShaoQuinteroMertens2014} and the readers are referred to Refs. \cite{XuShaoTang2013} and \cite{ShaoQuinteroMertens2014} for a detailed description of the method.
 For the driven NLD equation we again employ the same scheme used in Ref. \cite{ShaoQuinteroMertens2014}, but instead of nonreflecting boundary conditions, we take periodic boundary conditions. This has the advantage that tests of the conservation laws for momentum and energy (see Secs. IV and III) yield an accuracy of the order of $10^{-10} $and $10^{-14}$, respectively. Here we adopt  the computational domain $[-100,100]$, (i.e. $L=100$),  the time step $\Delta t=0.025$, and the final time $t_{fin}=800$.

\begin{figure}
\subfigure[]{\includegraphics[width=0.49\textwidth,height=0.39\textwidth]{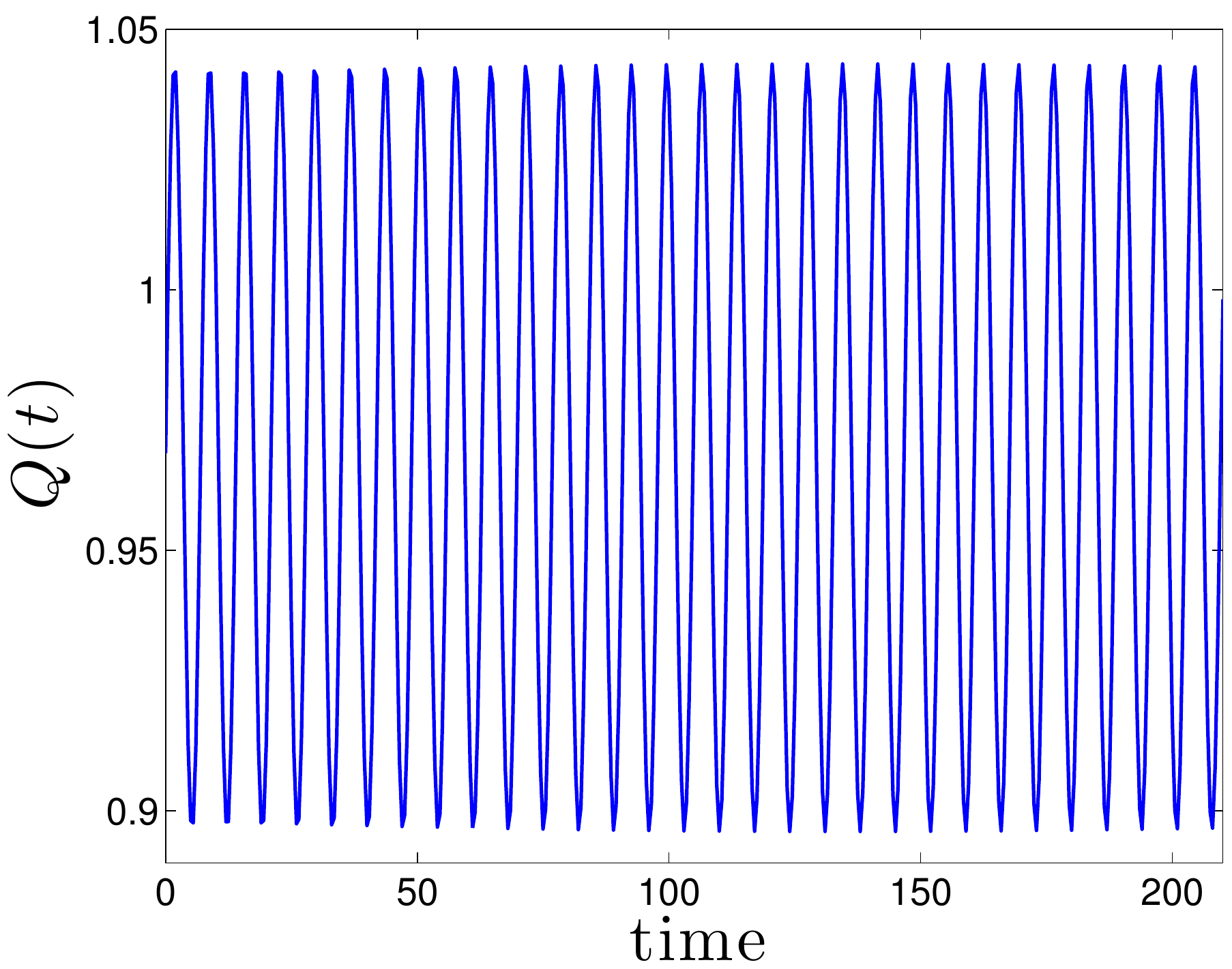}}
\subfigure[]{\includegraphics[width=0.49\textwidth,height=0.39\textwidth]{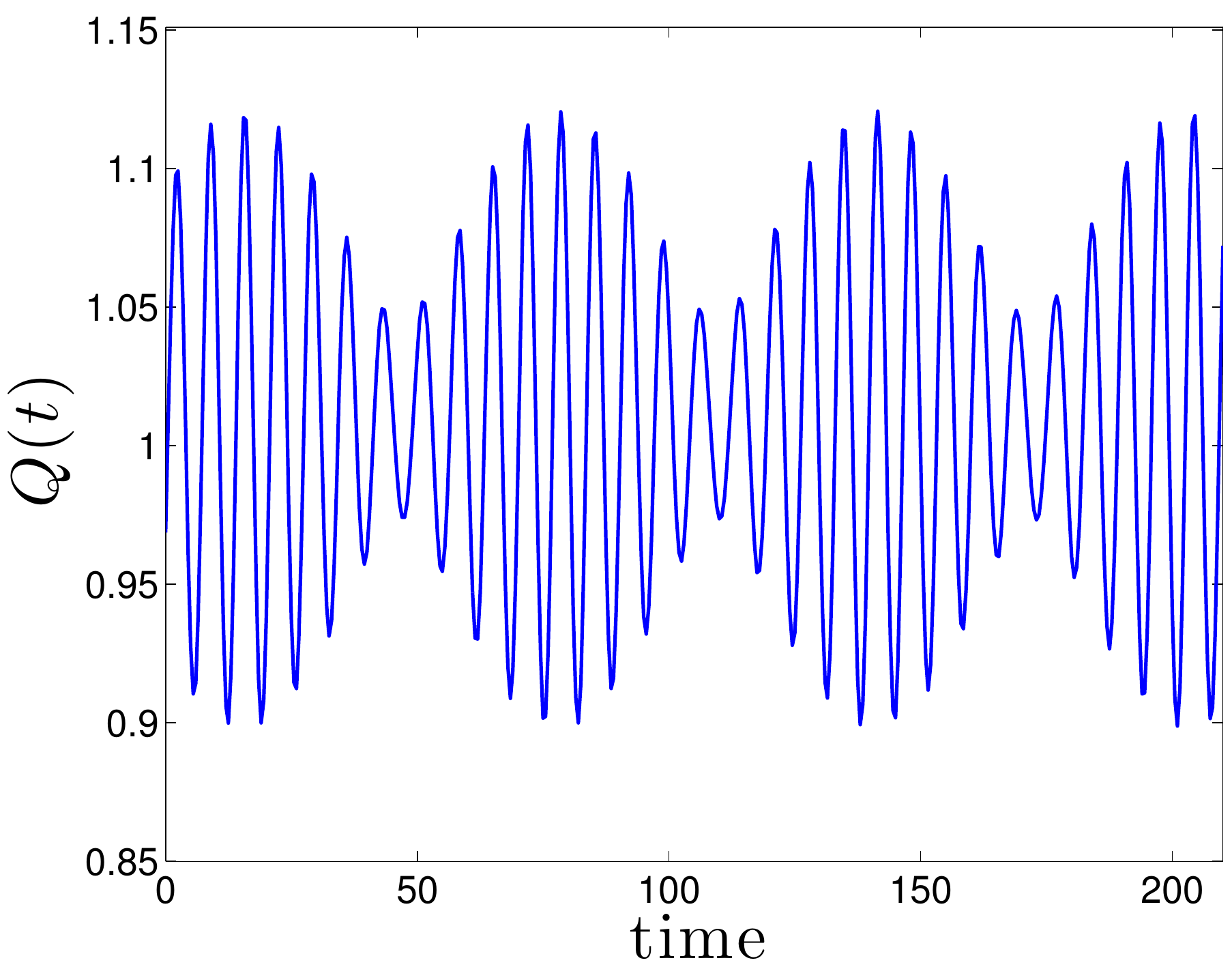}}
\subfigure[]{\includegraphics[width=0.49\textwidth,height=0.39\textwidth]{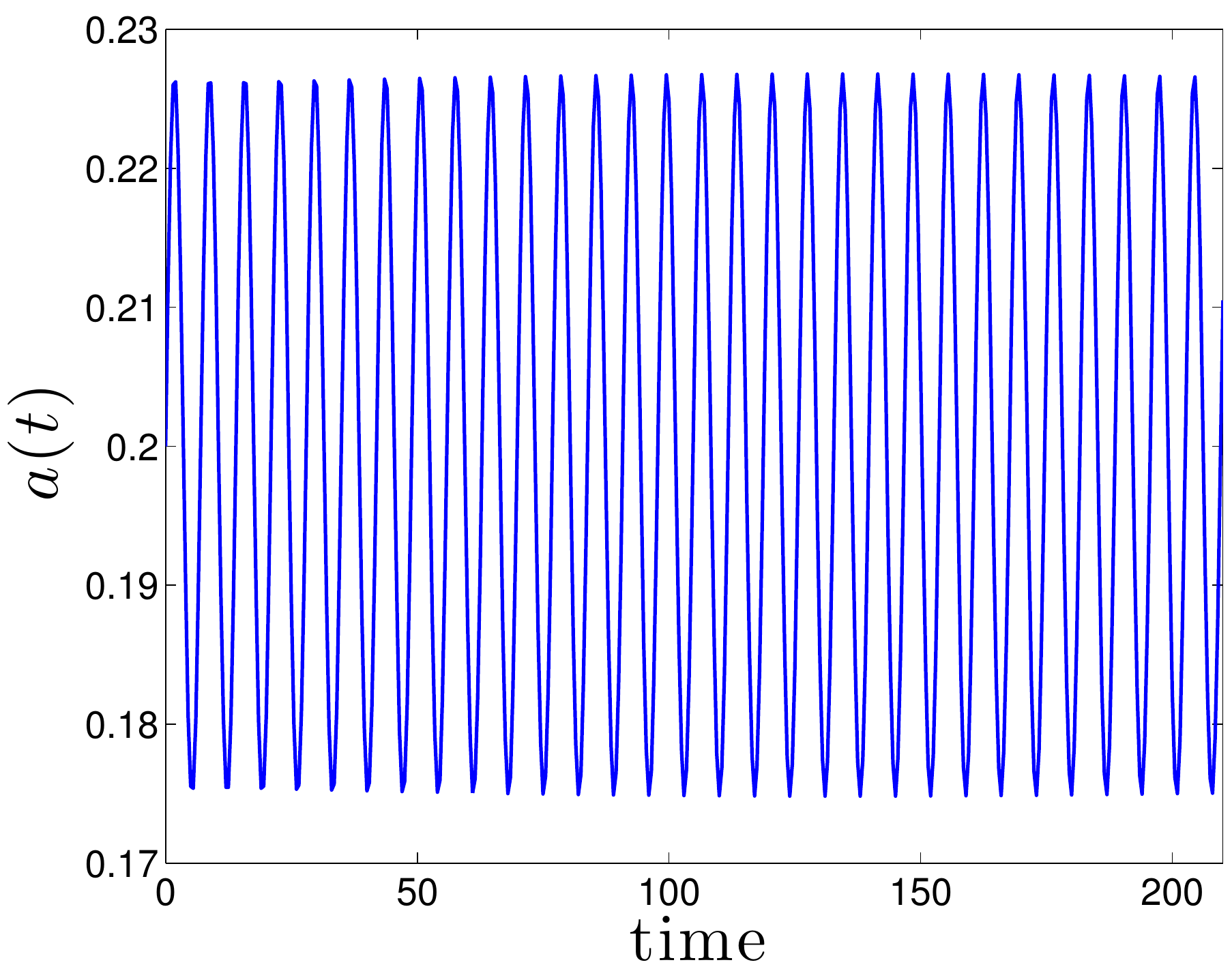}}
\subfigure[]{\includegraphics[width=0.49\textwidth,height=0.39\textwidth]{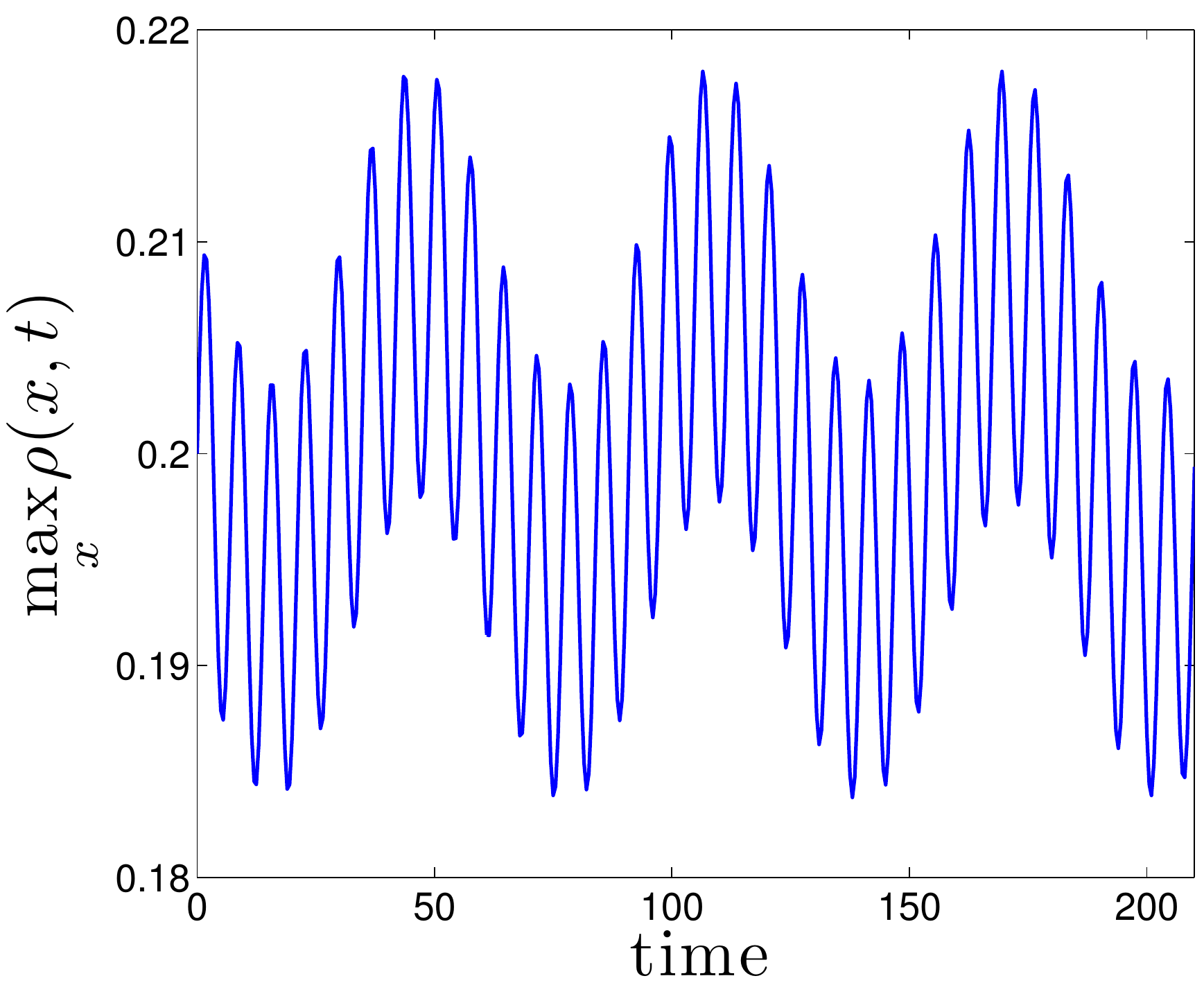}}
\subfigure[]{\includegraphics[width=0.49\textwidth,height=0.39\textwidth]{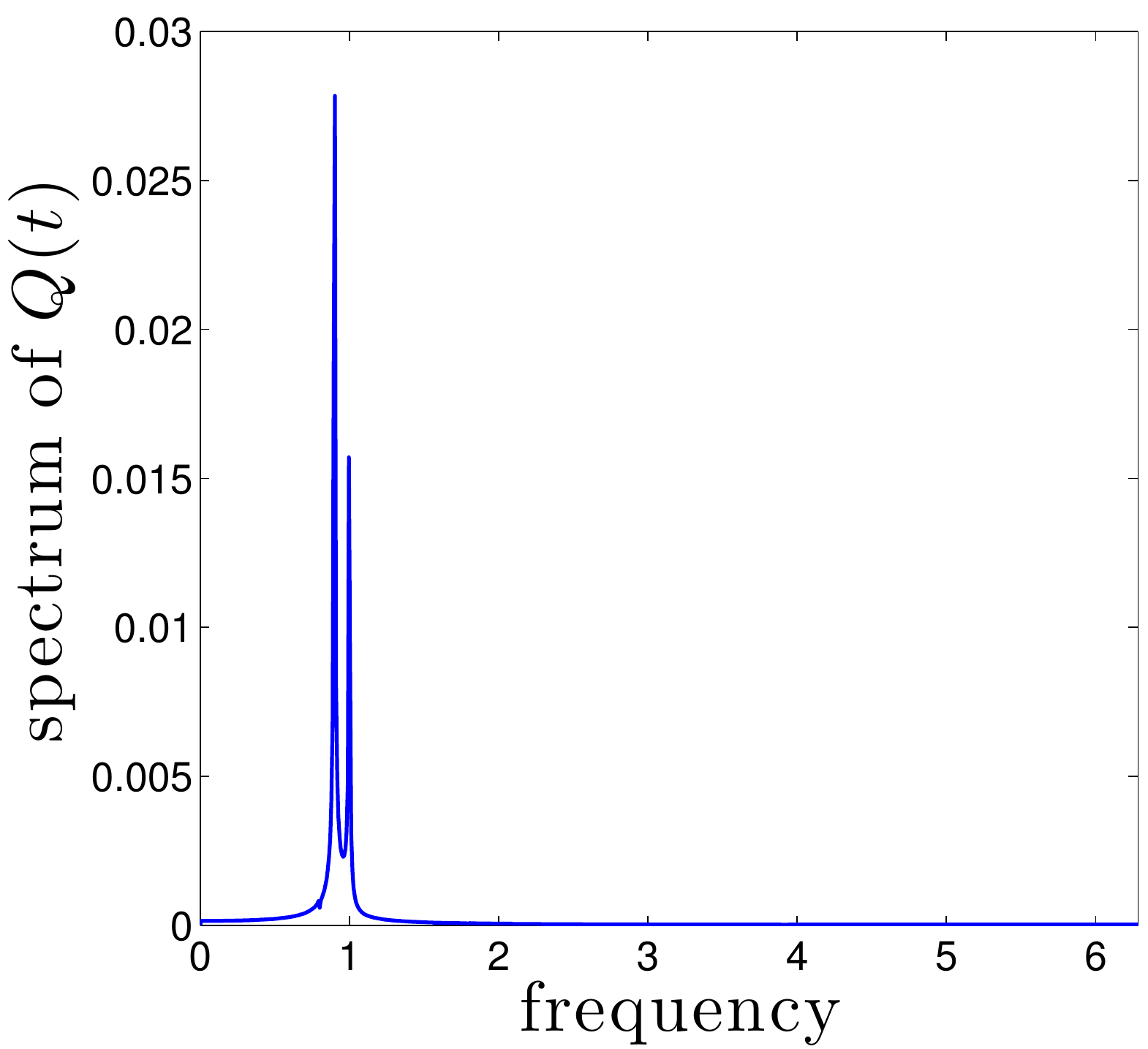}}
\subfigure[]{\includegraphics[width=0.49\textwidth,height=0.39\textwidth]{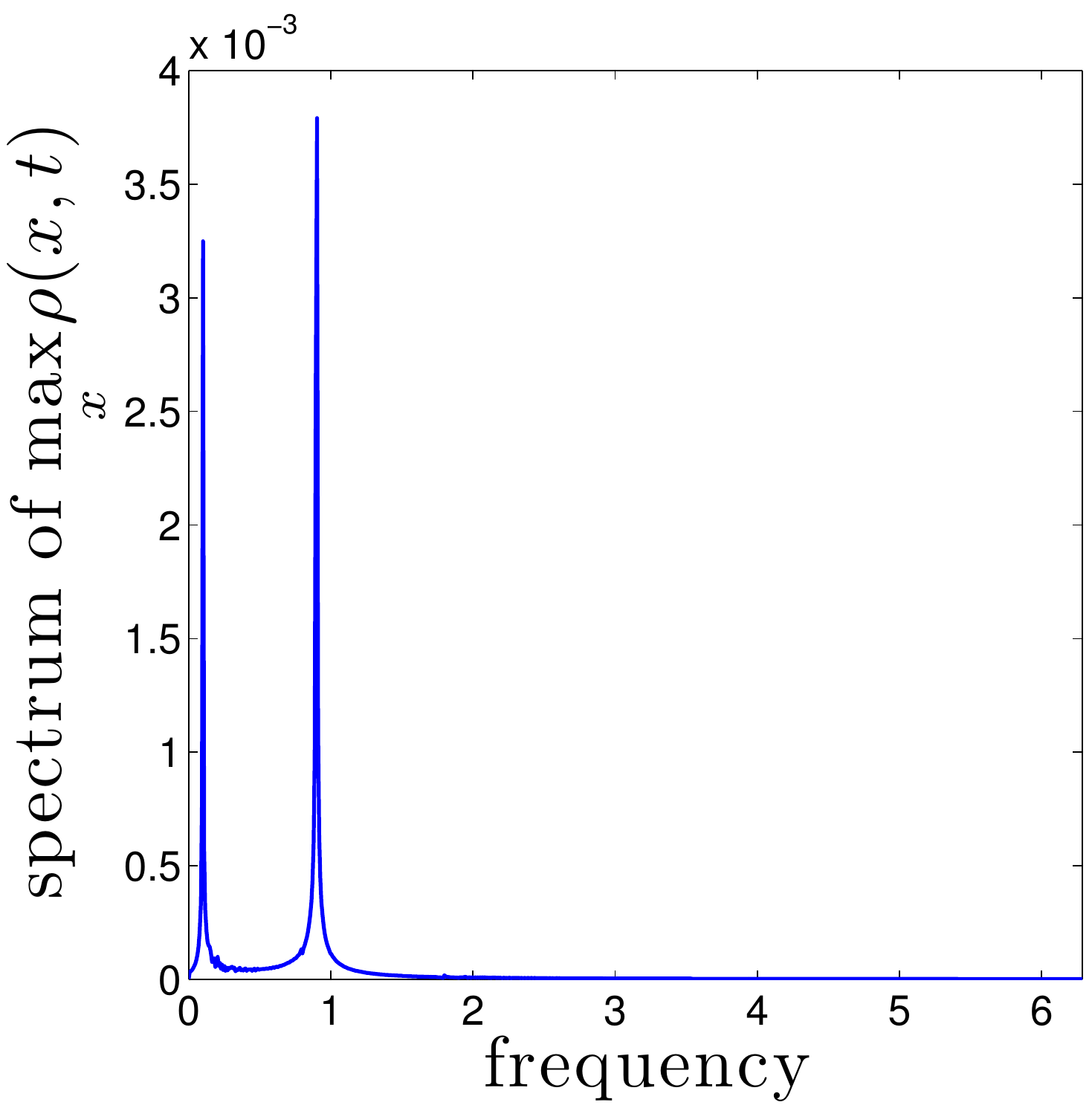}}
\caption{Intrinsic soliton oscillations for the case of constant, homogeneous force ($K=0$) and zero initial velocity 
($v_0=0$). Other parameters and initial condition (IC): $g=1$, $m=1$, $r=0.01$, $\omega_0=0.9$, $\phi_0=\pi/2$, integration time $t_f=800$. 
Panels (a) and (b): charge from CC theory and simulation, respectively. 
Panels (c) and (d): amplitude $a=2 [m -\omega(t)]/g^2$ and $\max_{x} \rho(x,t)$, respectively.
Panel (e): discrete Fourier transform (DFT) of $Q(t)$, soliton peak at $\omega_1=0.9032$ and phonon peak at $\omega_2=0.997 \approx \sqrt{1+k^2}$ with $k=-K=0$. 
Panel (f): DFT of $\max_{x} \rho(x,t)$, peaks at $\omega_1=0.9032$, $\omega_3=\omega_2-\omega_1=0.0942$. 
}
\label{fig1}
\end{figure}

In the previous section we have identified plane wave phonons in the spectra of the soliton charge and amplitude (maximum of the charge density). Now we discuss the peaks due to the intrinsic oscillations of the soliton shape and velocity and compare with our CC results. 
The highest peak $ \omega_1 = \Omega_{sim}$  in the spectra of $Q(t)$ is always close to the initial value $\omega_0 = 0.9$ and always agrees nearly perfectly with the predicted frequency $\Omega_{cc}$  of the CC theory, see Figs. \ref{fig1} and  \ref{fig3}-\ref{fig6} and Table \ref{tab1}. The dependency on the parameter $ K$ is weak which means that the periodicity of the force $ f_1 = r \exp(-i K x)$  has little influence on the intrinsic oscillations; this includes the case $K = 0$ where the force is homogeneous.
Exactly the same frequencies are observed in the spectra of the soliton amplitude [in the simulations this is the maximum of the charge density, and in the CC theory this is  $a = 2 \frac{(m -  \omega(t)) } {g^2} $]. 
Table \ref{tab2}  contains the parameters of harmonic and biharmonic functions which have been fitted to the data for the soliton charge and amplitude for three cases of the parameter K. Comparing CC theory with simulations, we see that the results for the mean values and the amplitudes of the first harmonics agree qualitatively.
Finally we discuss the results for the translational motion of the solitons. There are very small oscillations of the soliton position $q(t)$ around a mean trajectory $\bar v  ~t $, see Figs.\  \ref{fig4} and  \ref{fig6}. We compute the discrete Fourier transform (DFT) of  $q(t) $ - $\bar v t $ for the CC theory and simulations and observe the same frequencies as above for the soliton charge and amplitude. 

Table \ref{tab1} shows that $\bar v$ is always close to the initial value $v_0 = 0.1$ which means that the translational motion is only weakly affected by the intrinsic soliton oscillations. The agreement between $ {\bar v}_{cc}$ and   $ {\bar v}_{sim} $ is not so good (the maximal error in Table \ref{tab1} is about  $14 \%$). The reason is that the plane wave phonons with $ k = -K$ are not taken into account in the CC theory. The monotonic behavior of $ {\bar v}_{sim} $ as a function of $K$ is explained qualitatively in the following way: For positive $K$ the phonon phase velocity is negative, which results in head-on collisions with the soliton. Here $ {\bar v}_{sim} < v_0$, which can be explained by assuming negative spatial shifts of the soliton due to the collisions. For negative K the plane wave phonon overtakes the soliton which results in positive shifts explaining that $ {\bar v}_{sim}  > v_0 $.  
\begin{table}
\begin{tabular}{|c|c|c||c|c|}
\hline 
$K  (\pi/100) $ &  $\Omega_{cc}$ &  $\Omega_{sim}$ &  $\bar v _{cc}$   & $\bar v _{sim}$ \\
\hline
-6	& 0.911062	& 0.91106	  & 0.0855975	& 0.10012	  \\
\hline
-5	& 0.911062	& 0.91106	& 0.0865441	& 0.10008	   \\
\hline
-4	& 0.911062	& 0.91106	& 0.0914469	& 0.10005	  \\
\hline
-3	& 0.903208	& 0.90321	& 0.101902	& 0.10005	  \\
\hline
-2	& 0.903208	& 0.90321	& 0.111025	& 0.10004	  \\
\hline
-1	& 0.895354	& 0.89535	& 0.114524	& 0.10001	 \\
\hline
0	& 0.8875	& 0.89535	& 0.114569	& 0.099968	 \\
\hline
1	& 0.8875	& 0.89535	& 0.112703	& 0.099932   \\
\hline
2	& 0.879646	& 0.8875	& 0.109807	& 0.09987	 \\
\hline
3	& 0.879646	& 0.8875	& 0.10646	& 0.099805  \\
\hline
4	& 0.879646	& 0.87965	& 0.103058	& 0.099745	\\
\hline
5	& 0.871792	& 0.87965	& 0.0999051	& 0.099677	 \\
\hline
6	& 0.871792	& 0.87965	& 0.0972293	& 0.099607	\\
\hline
\end{tabular}
\caption{ Frequency of intrinsic soliton oscillations and average soliton velocity as a function of the parameter $K$. We compare the results of the CC method and numerical simulation at various values of $K$ in multiples of $\pi/100$ . Here we choose the initial conditions $v_0 =0 .1, \omega_0=0.9, \phi_0=\pi/2, r_1=r=0.01, r_2=K_2=0, t_{fin}=800$, and $t_{fin}$ denotes the integration time. }
\label{tab1}
\end{table}

\begin{table}
\begin{tabular}{|c|c|c|}
\hline 
Parameters &  Theory & Simulations \\
\hline
$K=0$, $v_0=0$ & \begin{tabular}{c} $Q(t)\approx 0.97 -0.072 \cos(\omega_1 t+1.15)$ \\  $a(t) \approx 0.2 -0.025 \cos(\omega_1 t+1.15)$ \end{tabular} & 
\begin{tabular}{c} $Q(t)\approx 1.0097 -0.049 \cos(\omega_1 t+0.058)-0.034 \cos(\omega_2 t+0.88)$ \\   
$\max_{x} \rho \approx 0.2 -0.0067 \cos(\omega_1 t+0.036)+0.0065 \cos(\omega_3 t+0.98)$ \end{tabular} \\
\hline
$K=-3 \pi/100$, $v_0=0.1$ & \begin{tabular}{c}$Q(t)\approx 0.97 -0.073 \cos(\omega_1 t+1.67)$ \\  $a(t) \approx 0.2 -0.026 \cos(\omega_1 t+1.67)$ \end{tabular} & 
\begin{tabular}{c} $Q(t)\approx 1.0092 -0.071 \cos(\omega_1 t+2.0239)-0.038 \cos(\omega_2 t-6.74)$ \\   
$\max_{x} \rho \approx 0.2 -0.0096 \cos(\omega_1 t+2.02)+0.0061 \cos(\omega_3 t+0.797)$ \end{tabular} \\
\hline
$K=3 \pi/100$, $v_0=0.1$ & \begin{tabular}{c}$Q(t)\approx 0.97 -0.0594 \cos(\omega_1 t+1.31)$ \\  $a(t) \approx 0.2 -0.021 \cos(\omega_1 t+1.31)$ \end{tabular} & 
\begin{tabular}{c} $Q(t)\approx 1.0093 -0.054 \cos(\omega_1 t+0.72)-0.029 \cos(\omega_2 t-0.2)$ \\   
$\max_{x} \rho \approx 0.2 -0.0089 \cos(\omega_1 t+0.72)-0.00797 \cos(\omega_3 t-0.26)$ \end{tabular}
 \\
\hline
\end{tabular}
\caption{Least-squares fits to $Q(t)$, amplitude $a=2 [m-\omega(t)]/g^2$, and $\max_{x} \rho$ from theory and simulation. 
Other parameters as in Figs.\ \ref{fig1}-\ref{fig6}. Note that $\omega_1$, $\omega_2$ and $\omega_3$ are given in the captions of Figs.\ \ref{fig1}, \ref{fig3}-\ref{fig6}.}
\label{tab2}
\end{table}

\begin{figure}
\includegraphics[width=0.6\textwidth]{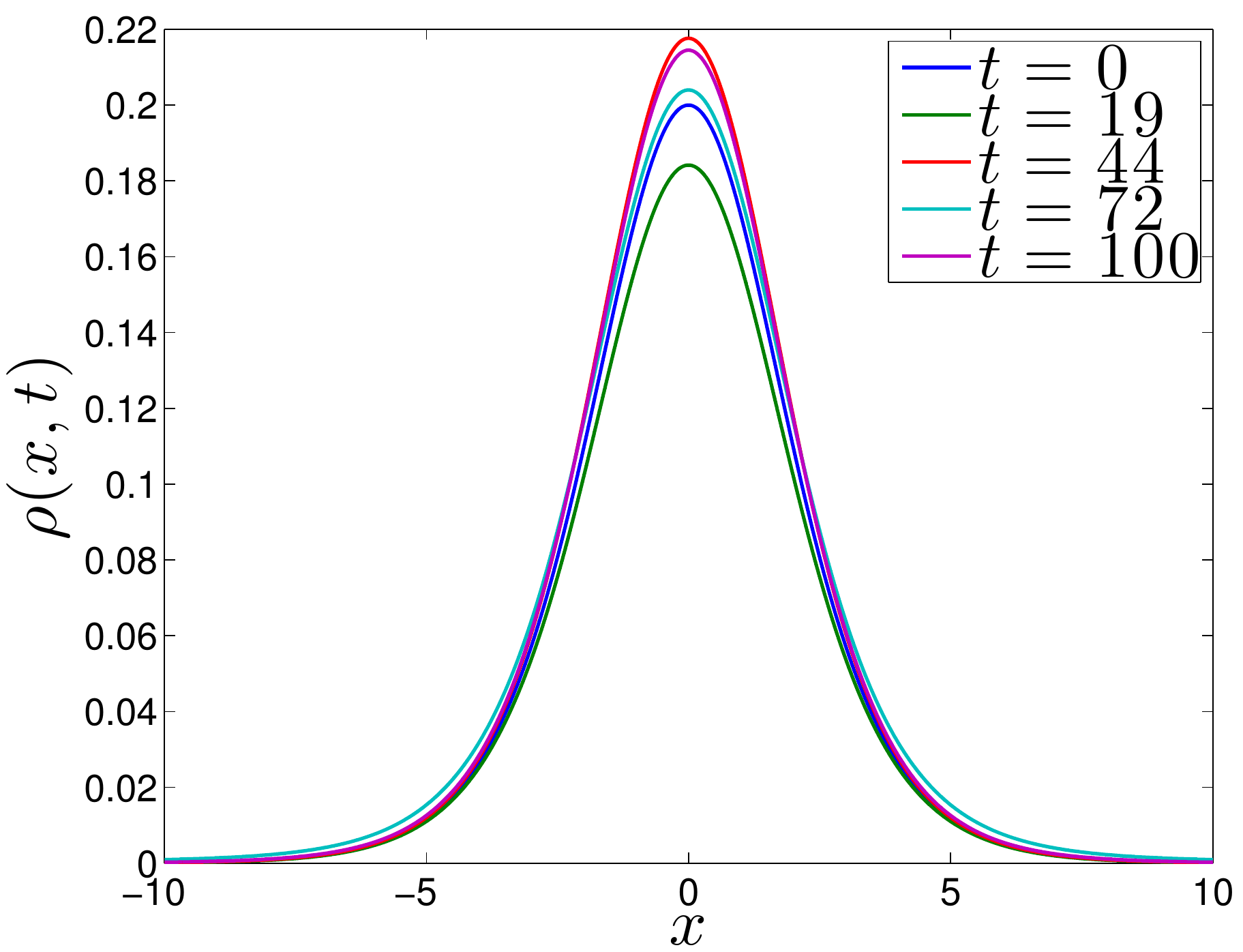}
\caption{Snapshots of the soliton profile at different times. Same parameters and initial conditions as in Fig.\ \ref{fig1}.}
\label{fig2}
\end{figure}

\begin{figure}
\subfigure[]{\includegraphics[width=0.49\textwidth,height=0.39\textwidth]{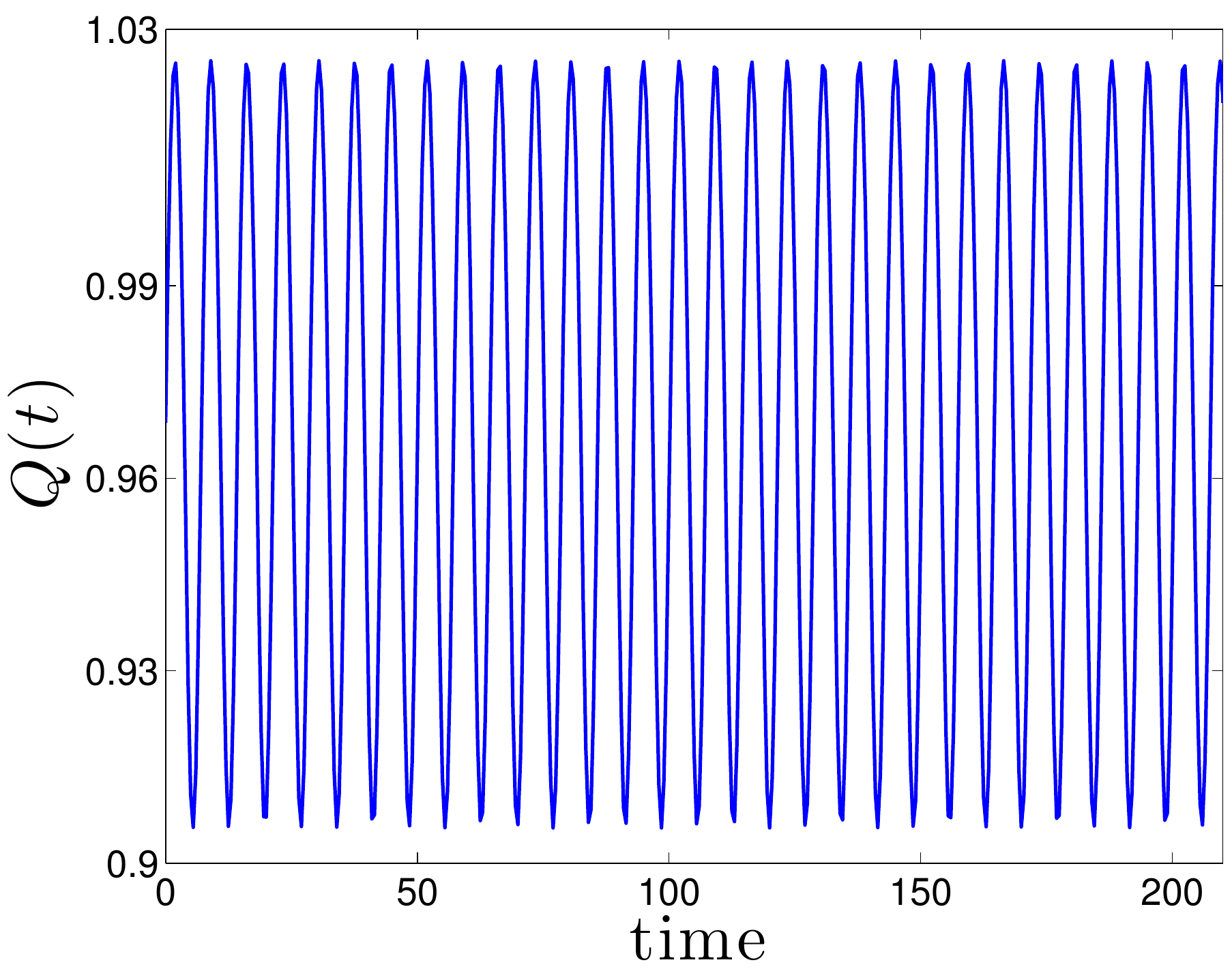}}
\subfigure[]{\includegraphics[width=0.49\textwidth,height=0.39\textwidth]{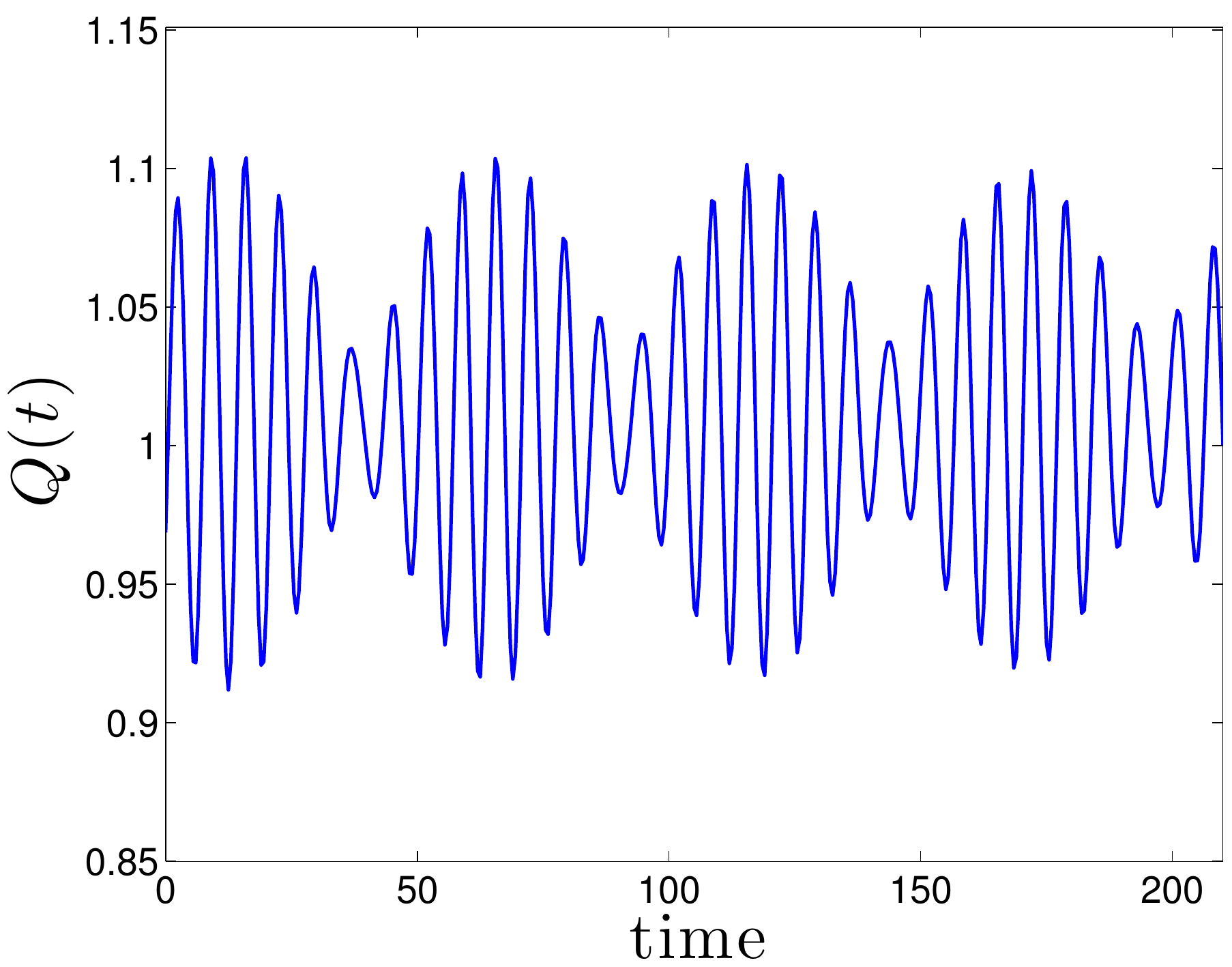}}
\subfigure[]{\includegraphics[width=0.49\textwidth,height=0.39\textwidth]{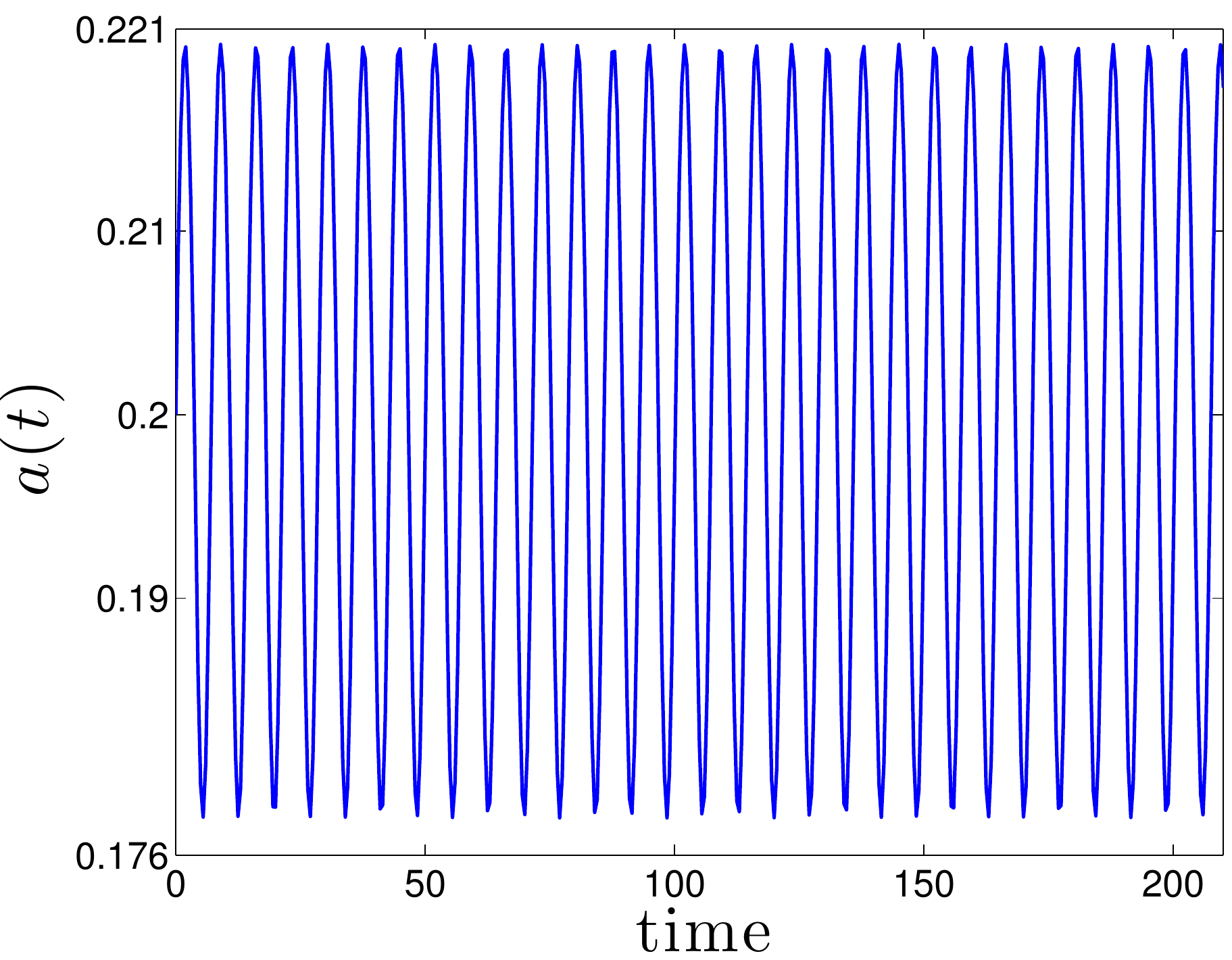}}
\subfigure[]{\includegraphics[width=0.49\textwidth,height=0.39\textwidth]{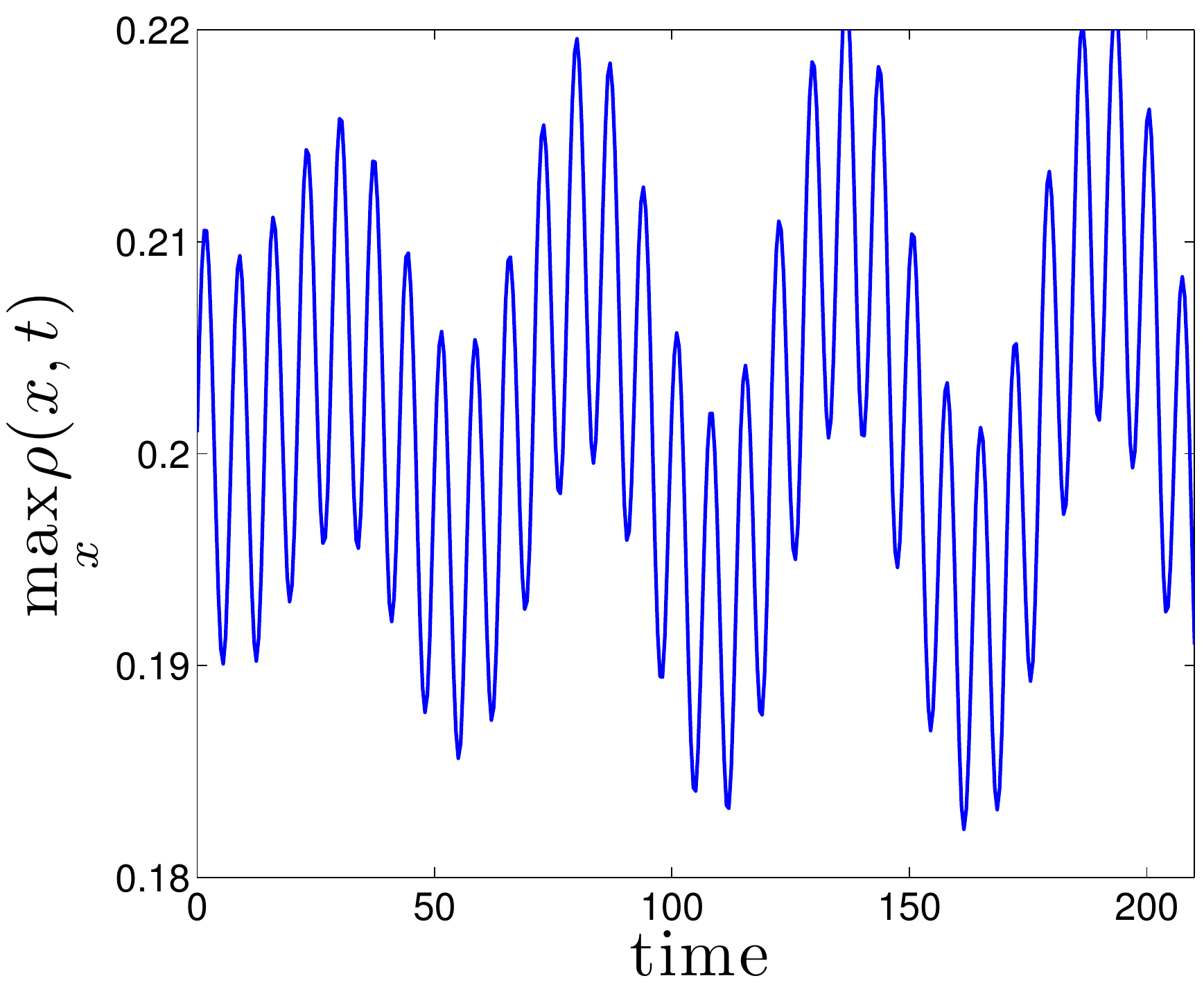}}
\subfigure[]{\includegraphics[width=0.49\textwidth,height=0.39\textwidth]{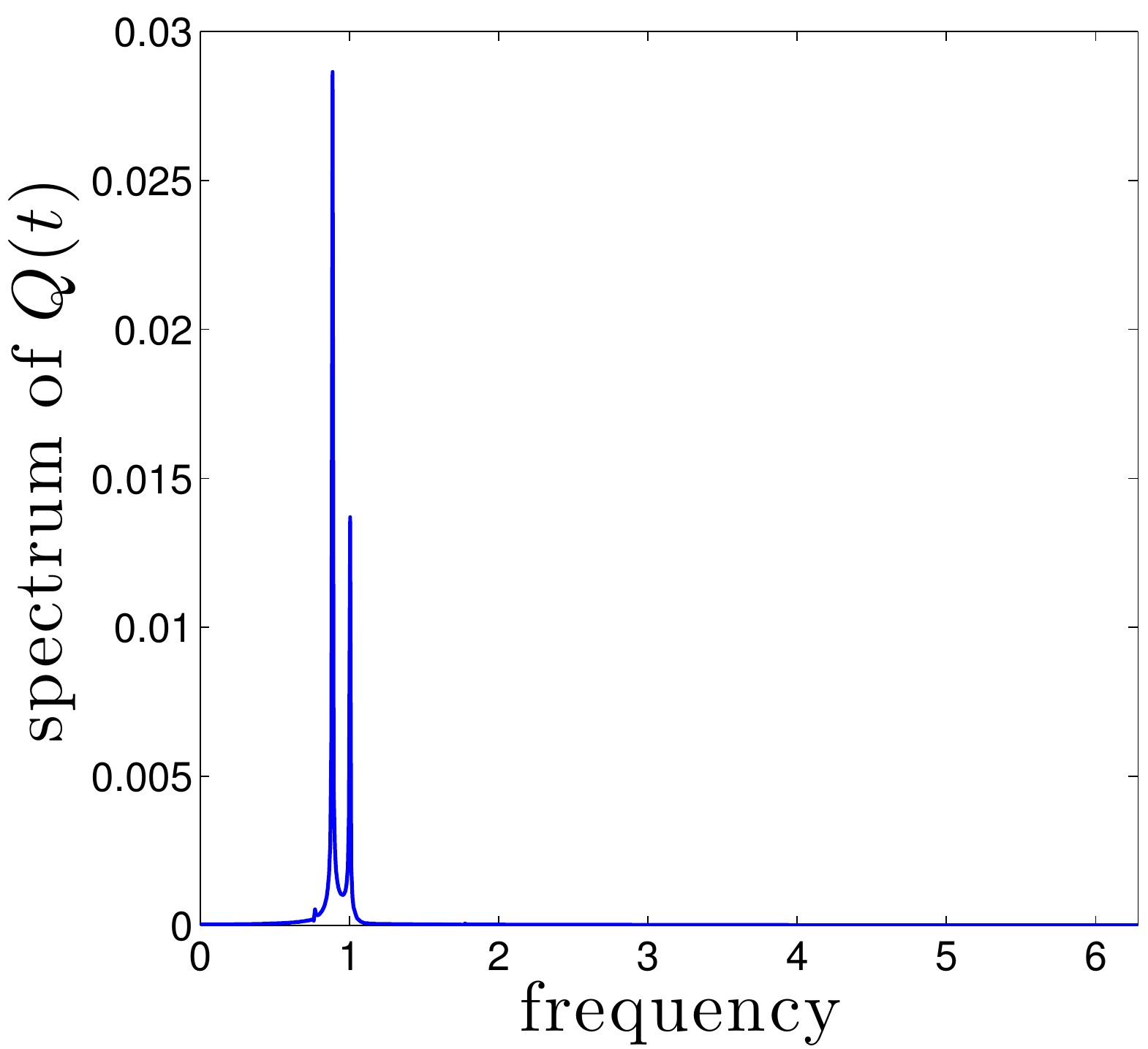}}
\subfigure[]{\includegraphics[width=0.49\textwidth,height=0.39\textwidth]{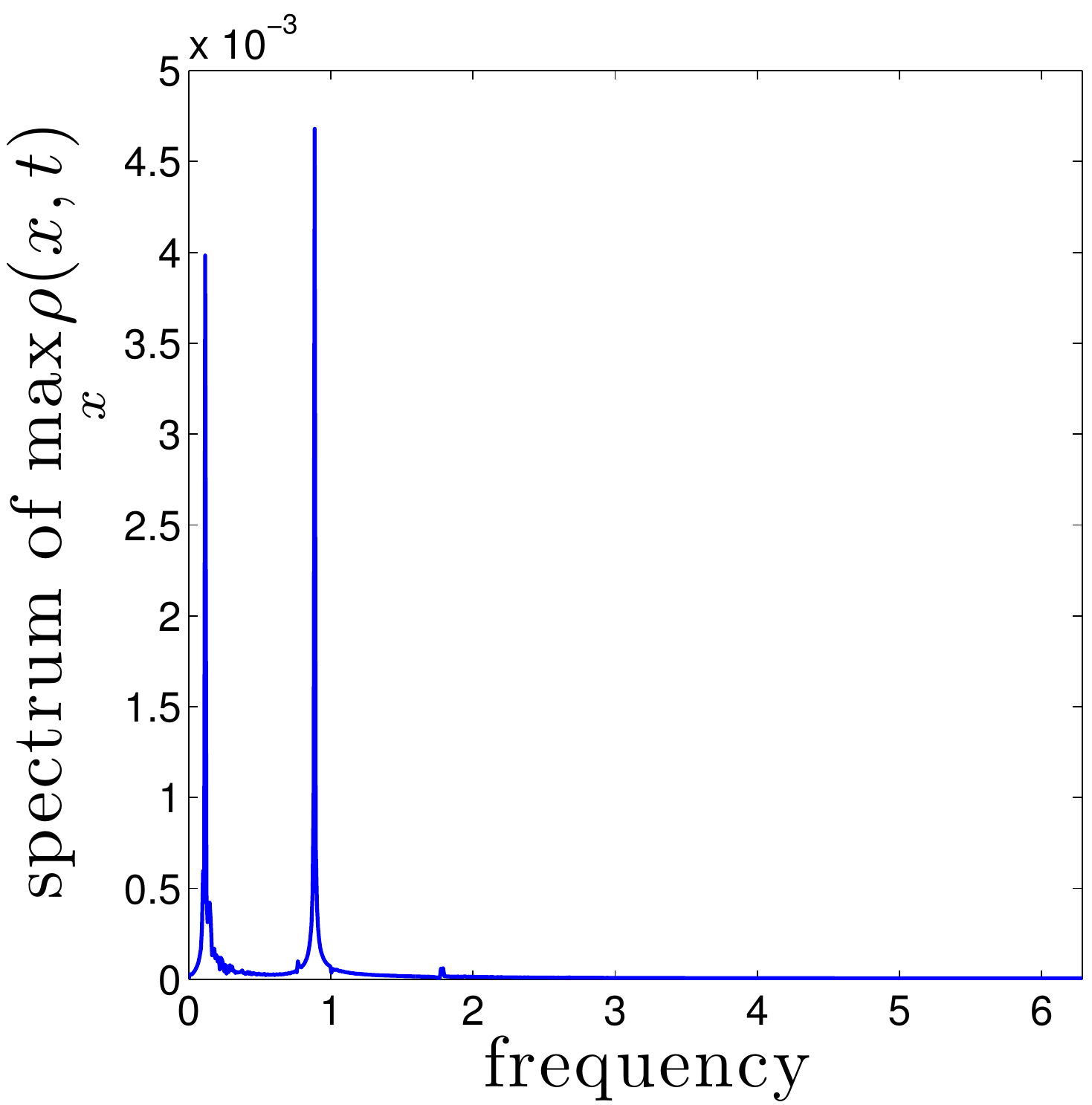}}
\caption{Intrinsic soliton oscillations for the case of a harmonic inhomogeneous force ($K=3 \pi/100 \approx 0.094248$) and $v_0=0.1$. Other parameters and initial conditions as in Fig.\ \ref{fig1}.  
Panels (a) and (b): charge from CC theory and simulation, respectively. 
Panels (c) and (d): amplitude $a=2 [m -\omega(t)]/g^2$ and $\max_{x} \rho(x,t)$, respectively.
Panel (e): DFT of $Q(t)$, soliton peak at $\omega_1=0.8875$ and phonon peak at $\omega_2=1.0053 \approx \sqrt{1+k^2}$ 
with $k=-K$. 
Panel (f): DFT of $\max_{x} \rho(x,t)$, peaks at $\omega_1=0.8875$, $\omega_3=\omega_2-\omega_1=0.11781$. 
}
\label{fig3}
\end{figure}

\begin{figure}
\subfigure[]{\includegraphics[width=0.49\textwidth,height=0.39\textwidth]{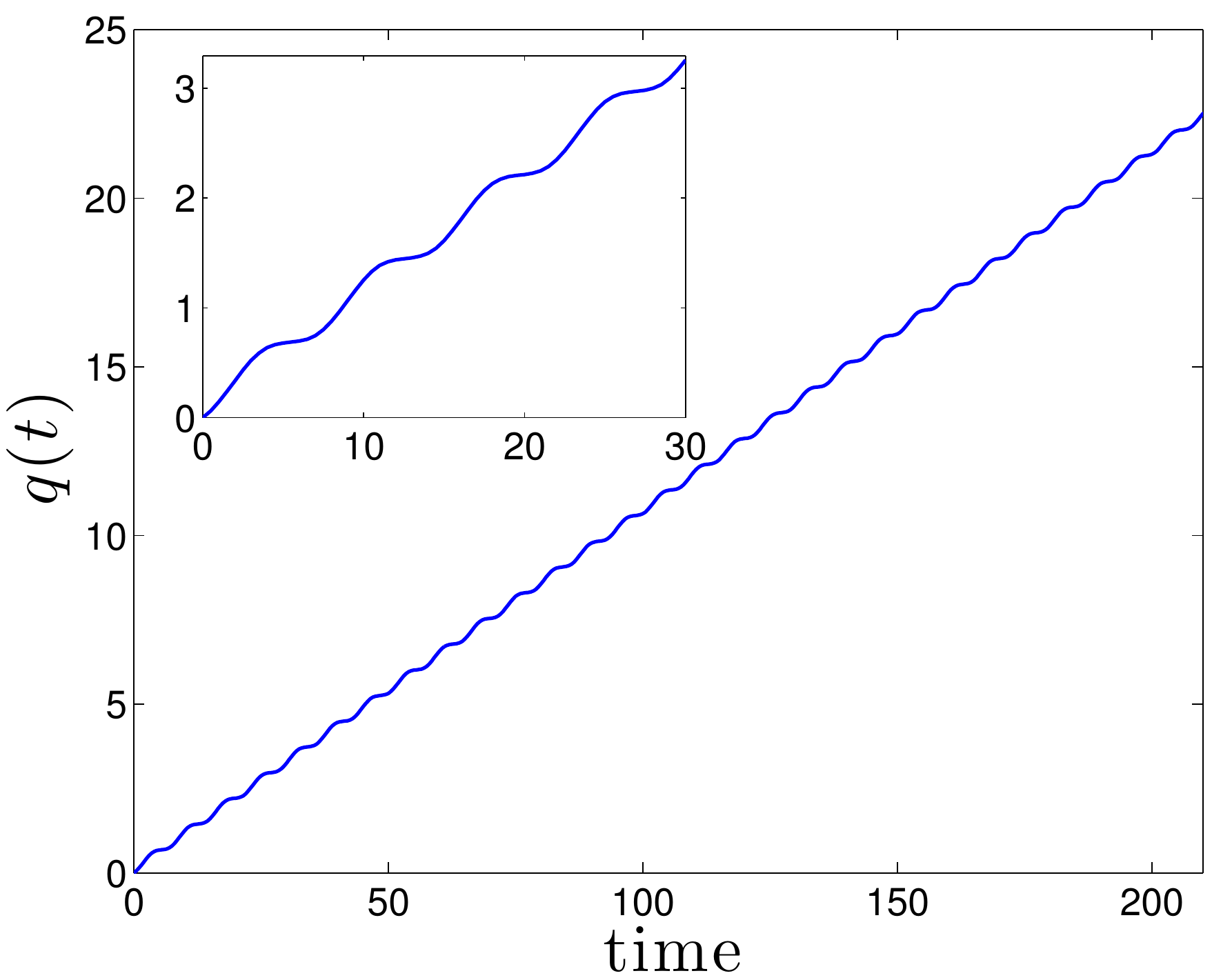}}
\subfigure[]{\includegraphics[width=0.49\textwidth,height=0.39\textwidth]{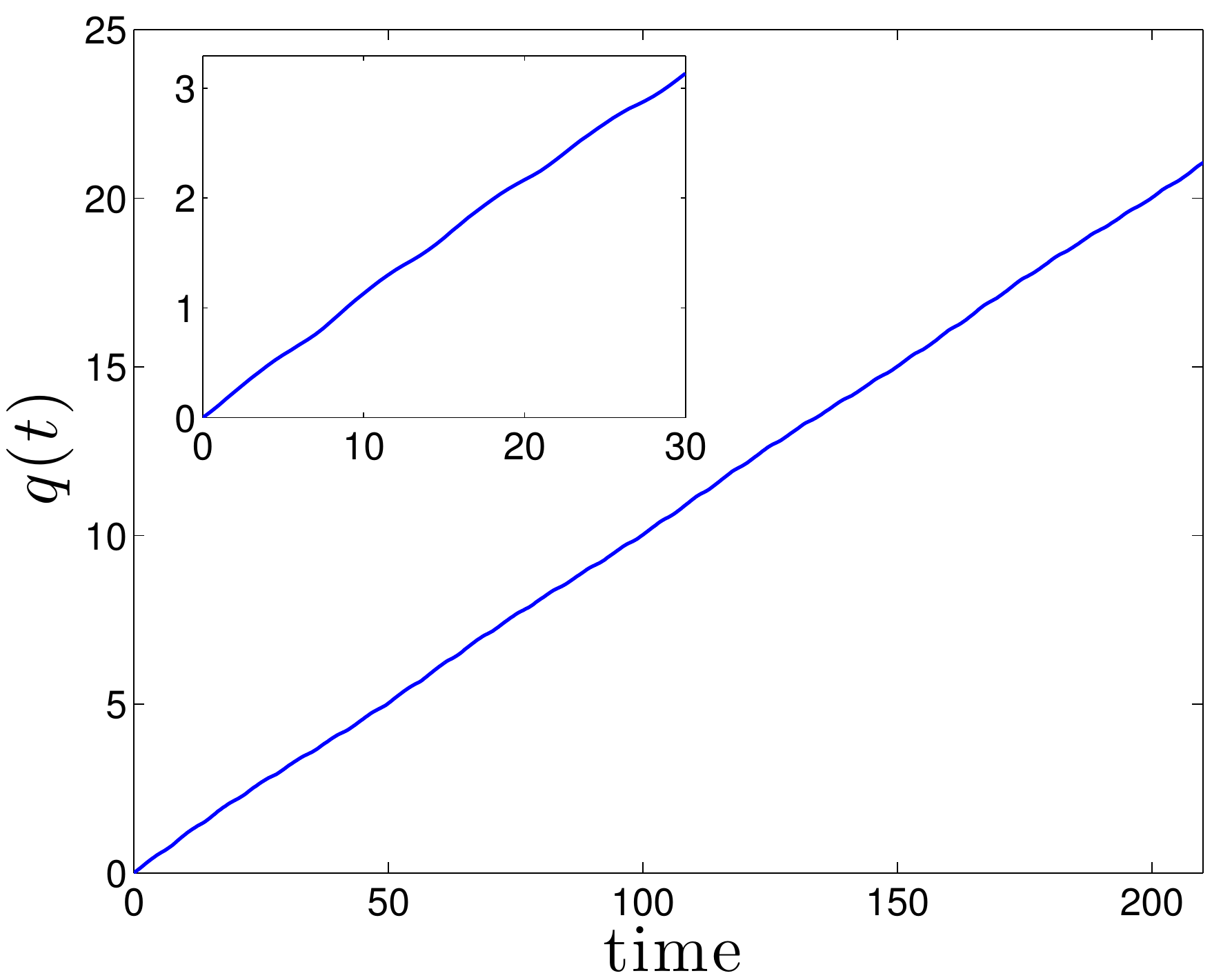}}
\subfigure[]{\includegraphics[width=0.49\textwidth,height=0.39\textwidth]{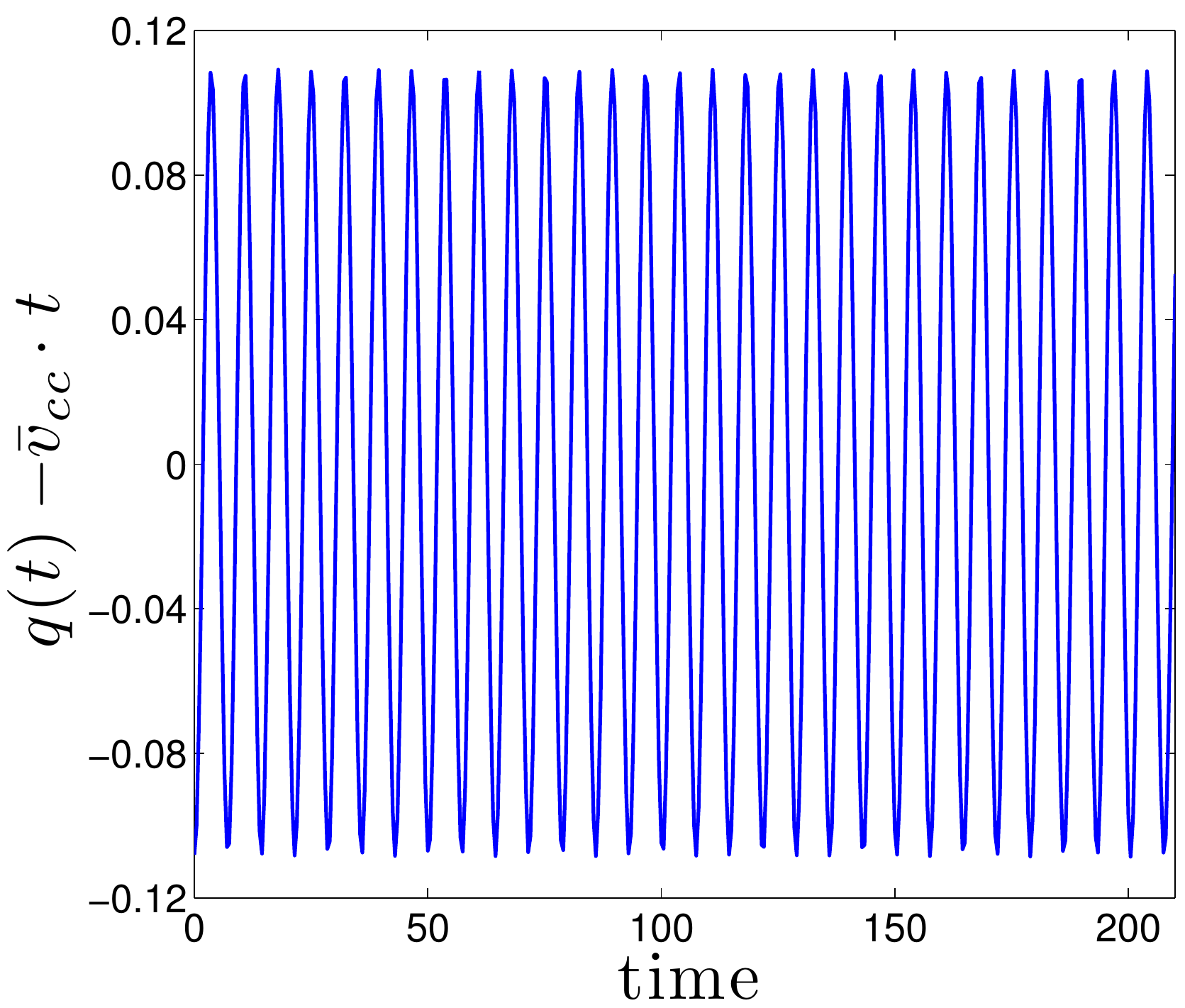}}
\subfigure[]{\includegraphics[width=0.49\textwidth,height=0.39\textwidth]{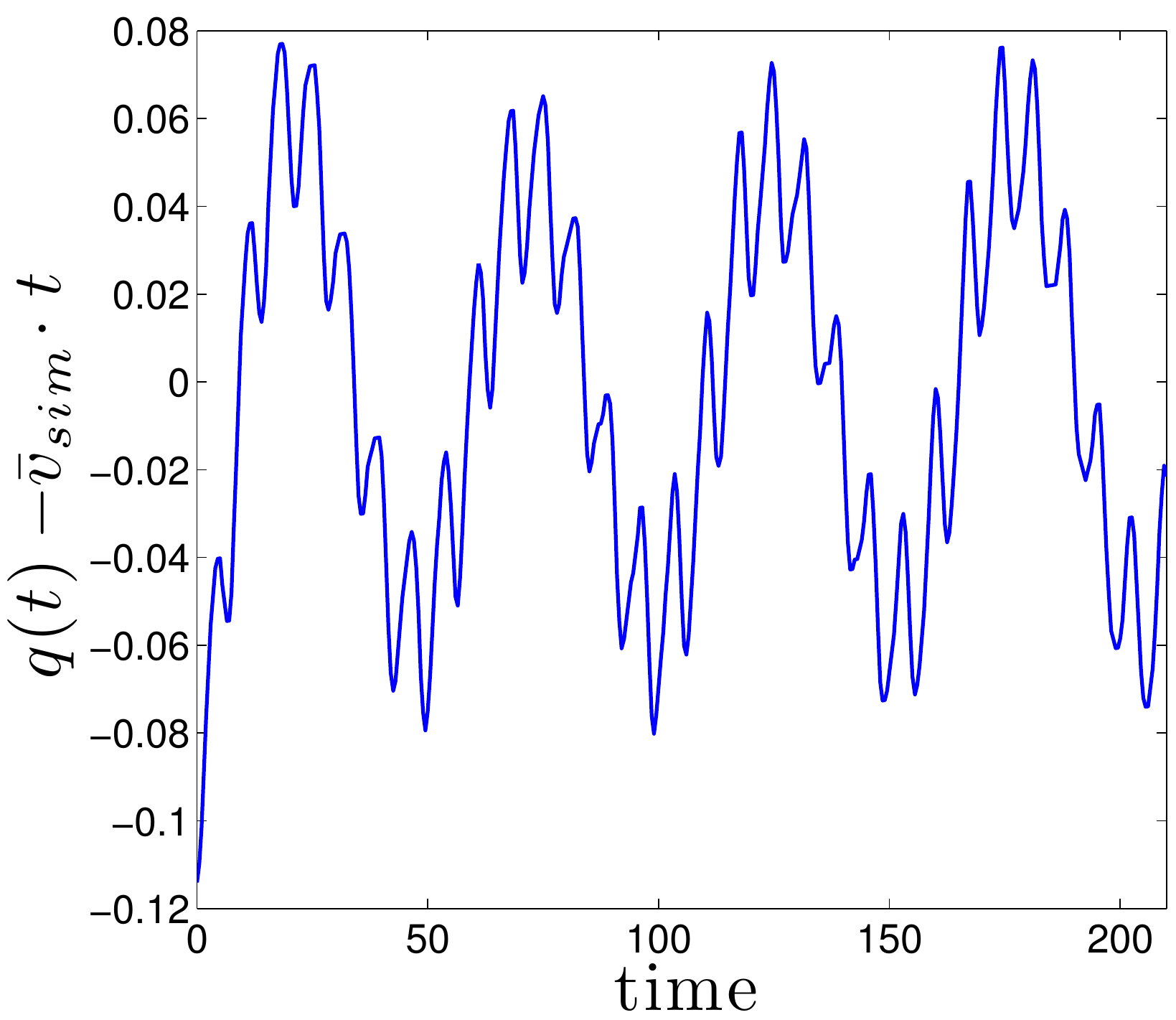}}
\subfigure[]{\includegraphics[width=0.49\textwidth,height=0.39\textwidth]{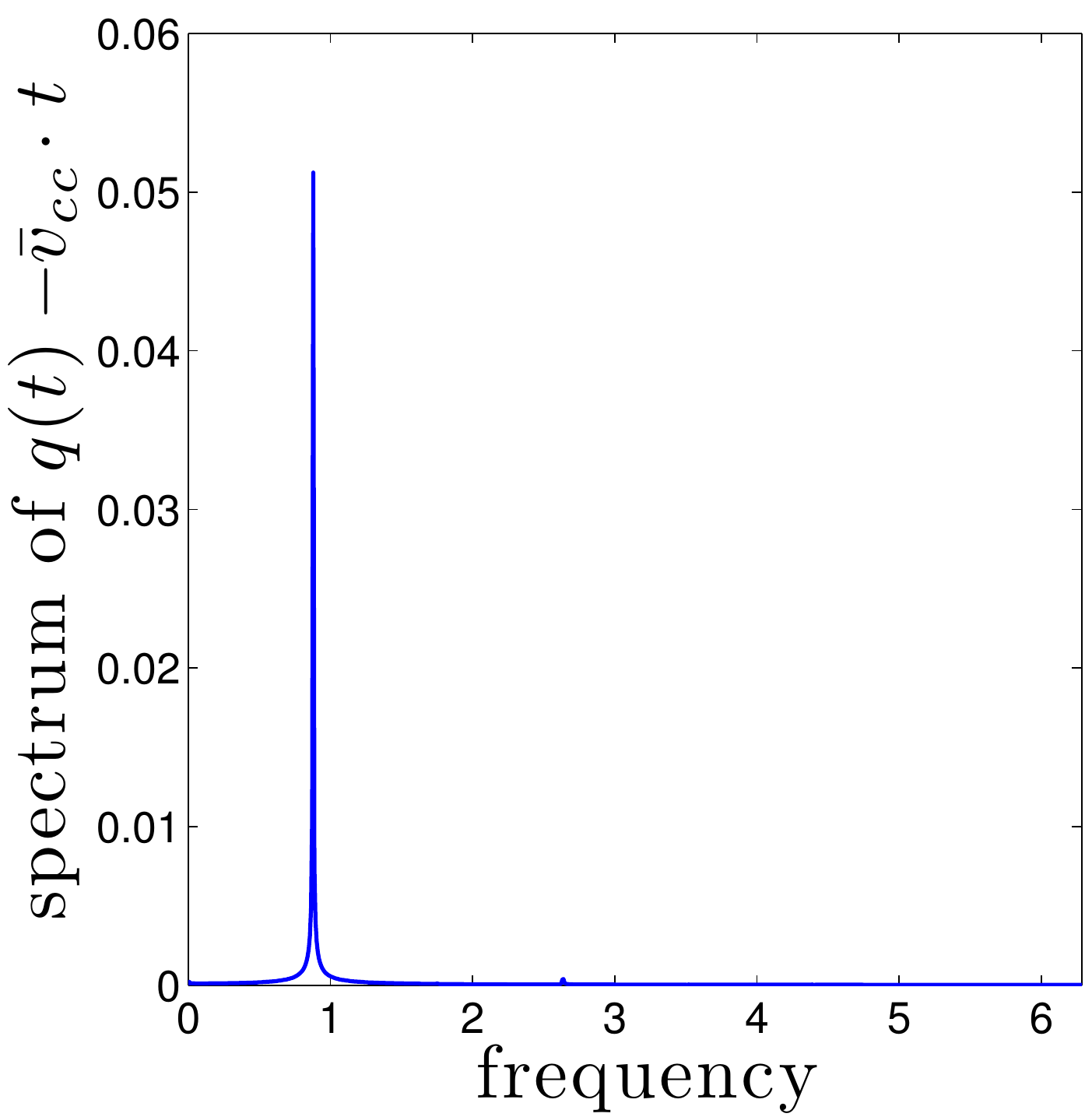}}
\subfigure[]{\includegraphics[width=0.49\textwidth,height=0.39\textwidth]{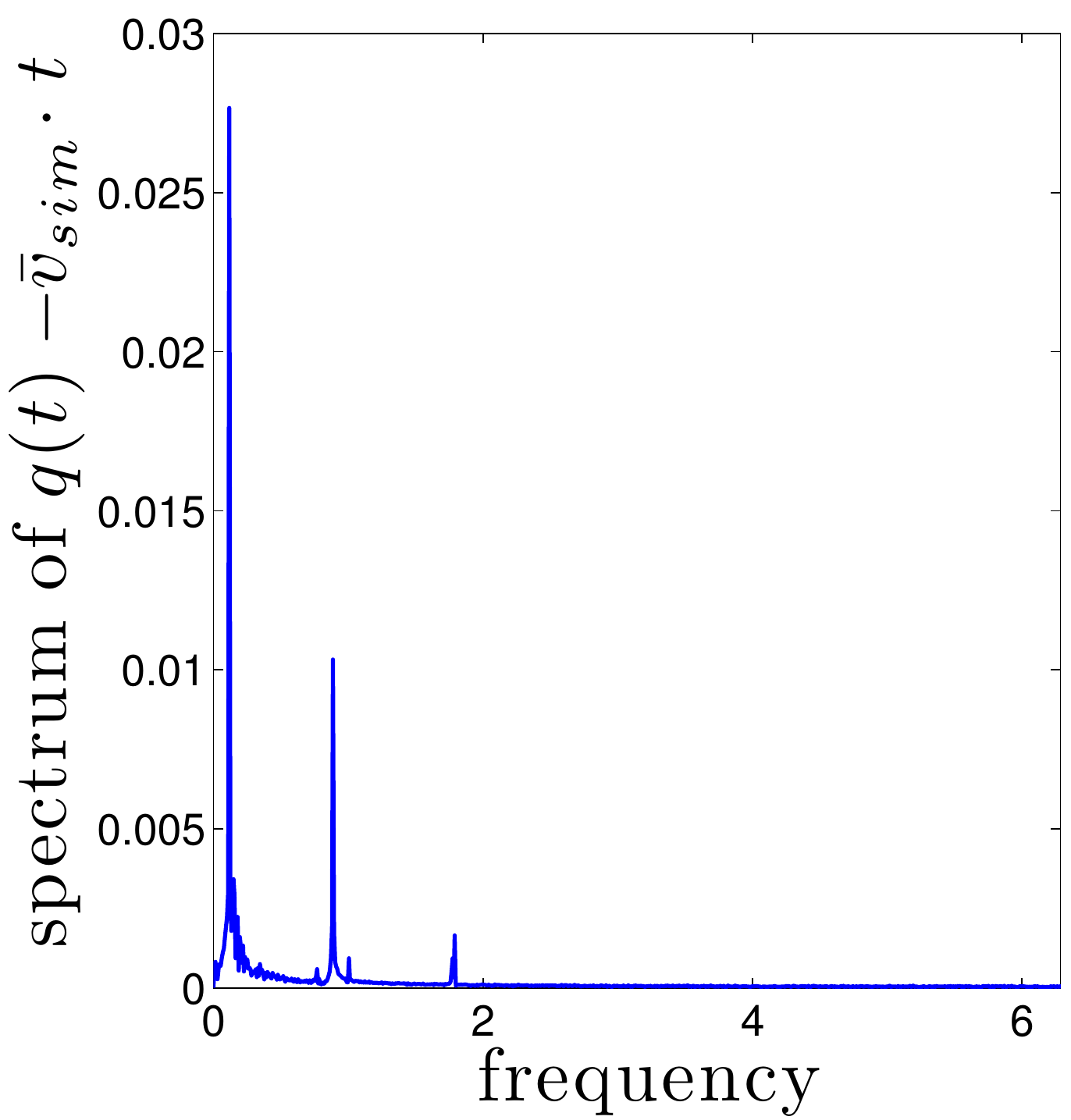}}
\caption{Oscillations of the translational motion of the soliton. $K=3 \pi/100 \approx 0.094248$ and $v_0=0.1$. 
Other parameters and initial conditions as in Fig.\ \ref{fig1}.  
Panels (a) and (b): $q(t)$ from CC theory and simulation, respectively. 
Panels (c) and (d): $q(t)-\bar{v} t$ from theory ($\bar{v}=\bar{v}_{cc}=0.1065$) 
 and simulation ($\bar{v}=\bar{v}_{sim}=0.099805$), respectively.
Panel (e): DFT of $q(t)-\bar{v}_{cc} t$, peak at $\omega_1=0.87965$.  
Panel (f): DFT of $q(t)-\bar{v}_{sim} t$,  peaks at $\omega_1=0.8875$, $2\omega_1=1.7750$ and at $\omega_3=\omega_2-\omega_1=0.11781$.  
}
\label{fig4}
\end{figure}

\begin{figure}
\subfigure[]{\includegraphics[width=0.49\textwidth,height=0.39\textwidth]{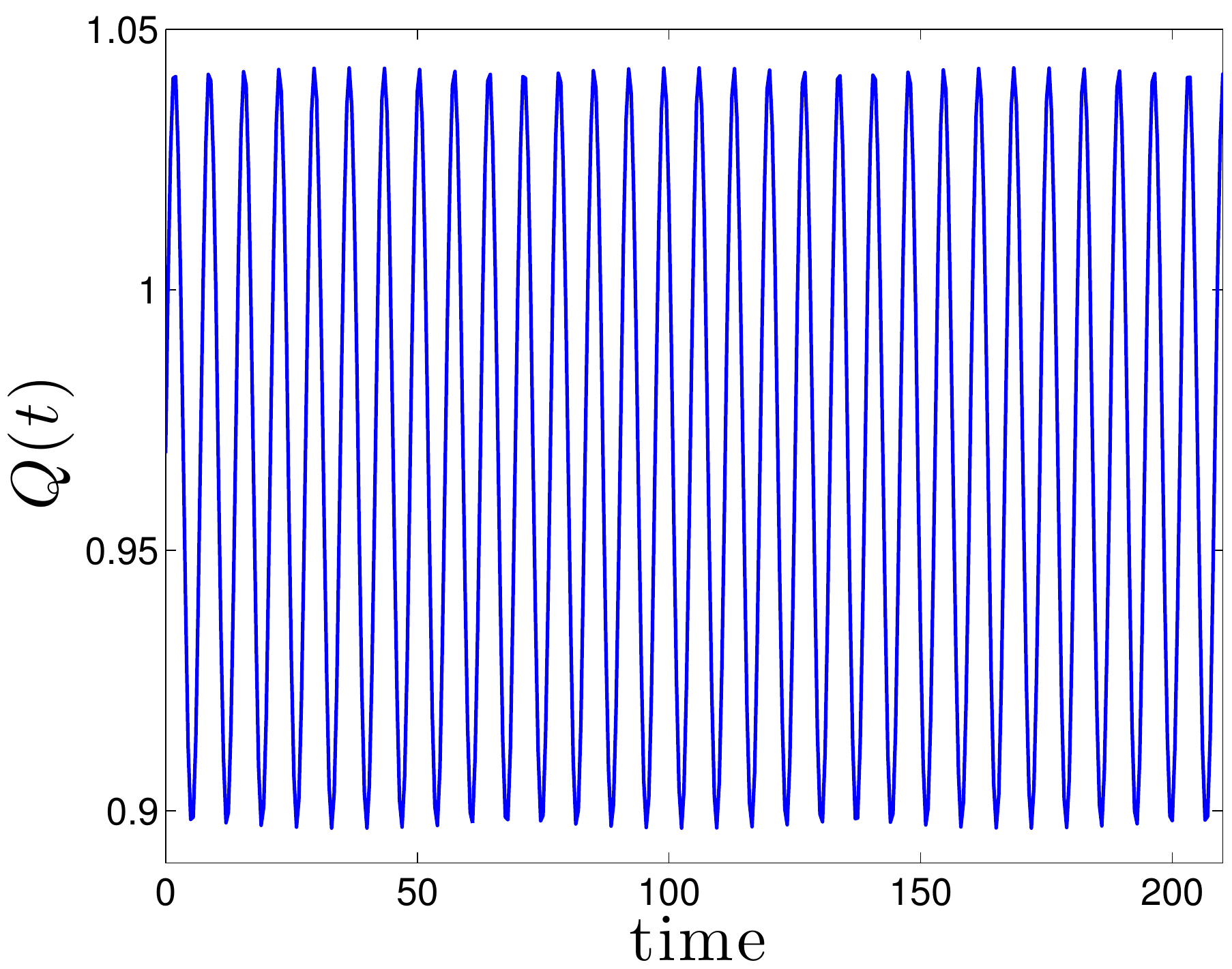}}
\subfigure[]{\includegraphics[width=0.49\textwidth,height=0.39\textwidth]{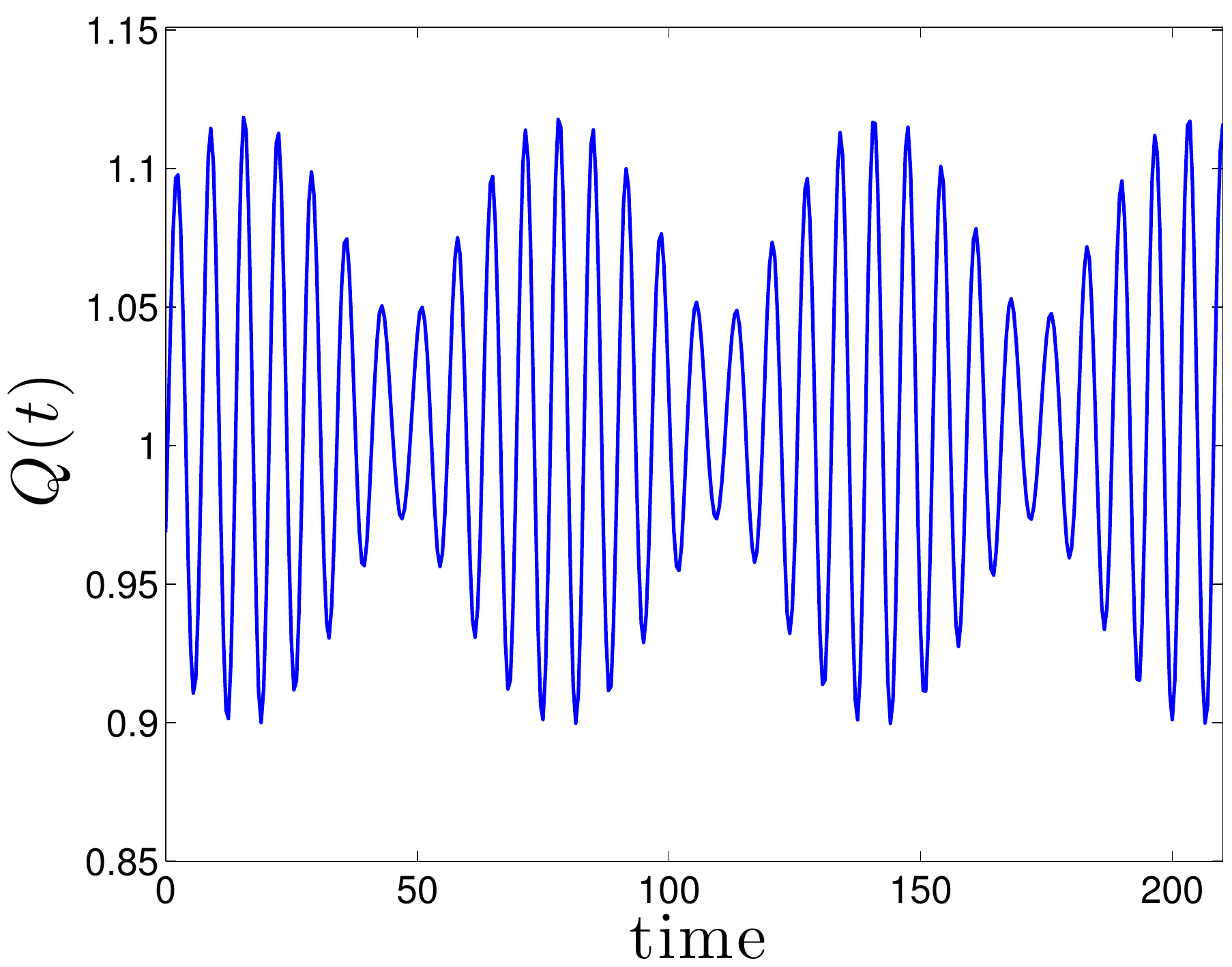}}
\subfigure[]{\includegraphics[width=0.49\textwidth,height=0.39\textwidth]{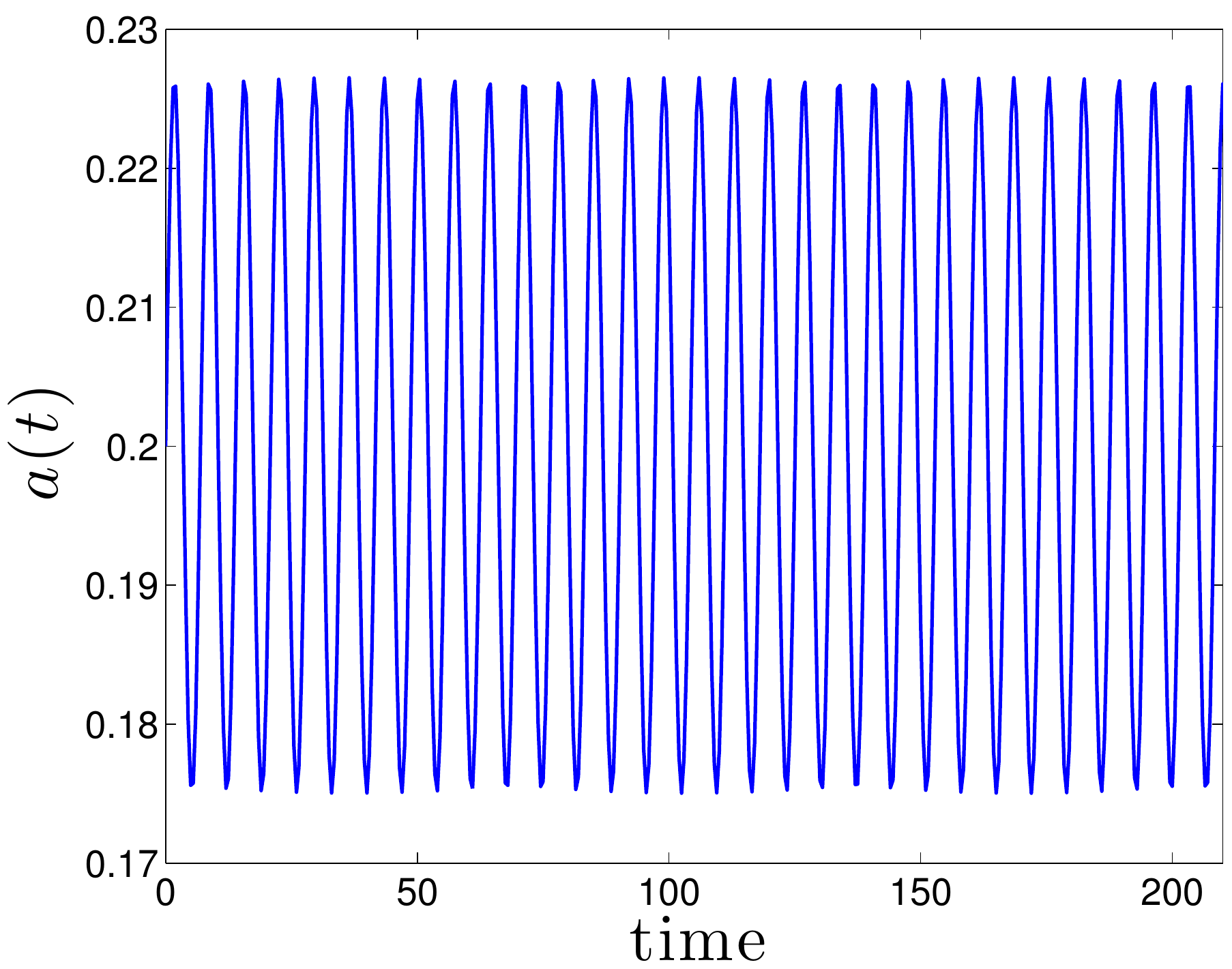}}
\subfigure[]{\includegraphics[width=0.49\textwidth,height=0.39\textwidth]{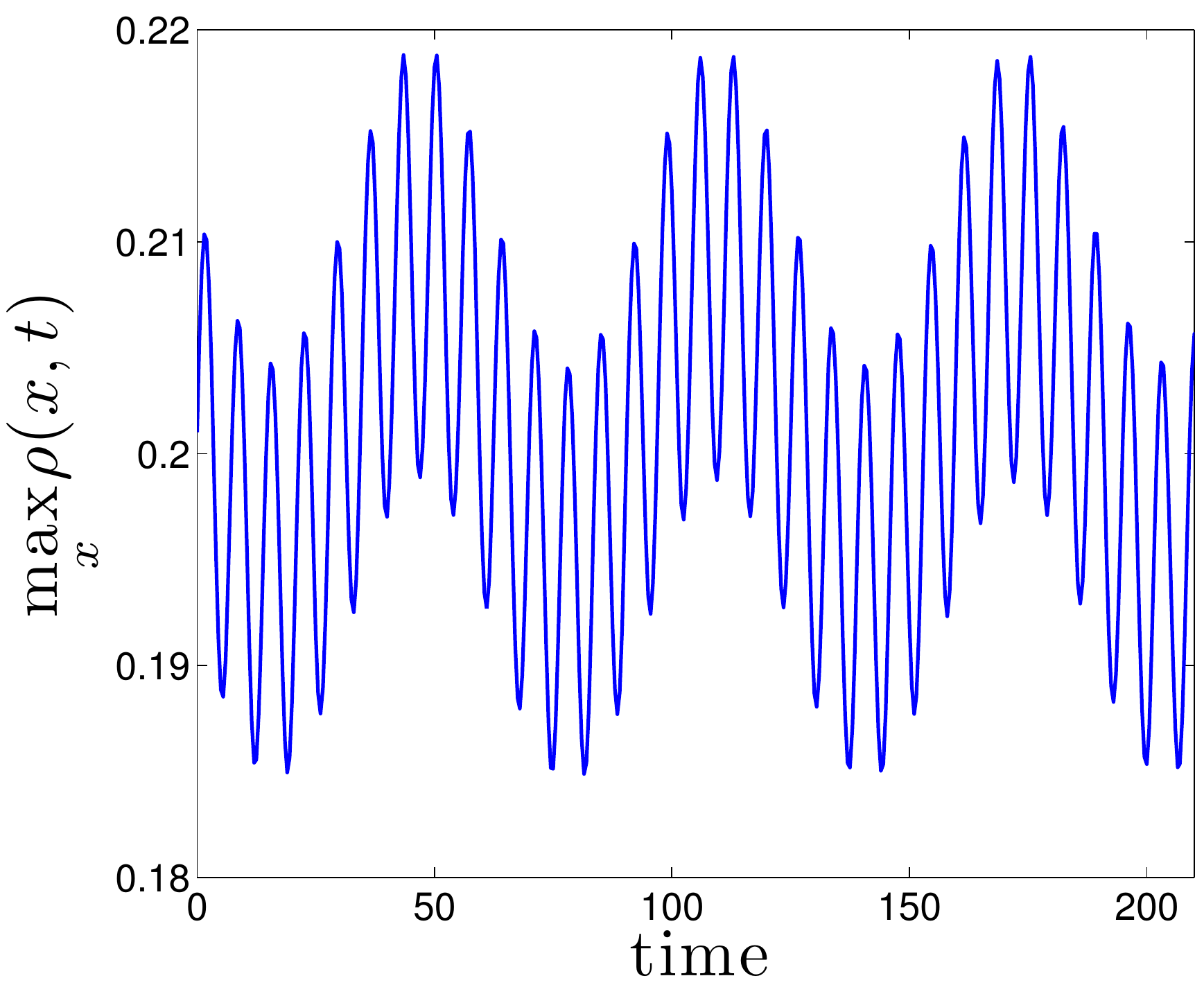}}
\subfigure[]{\includegraphics[width=0.49\textwidth,height=0.39\textwidth]{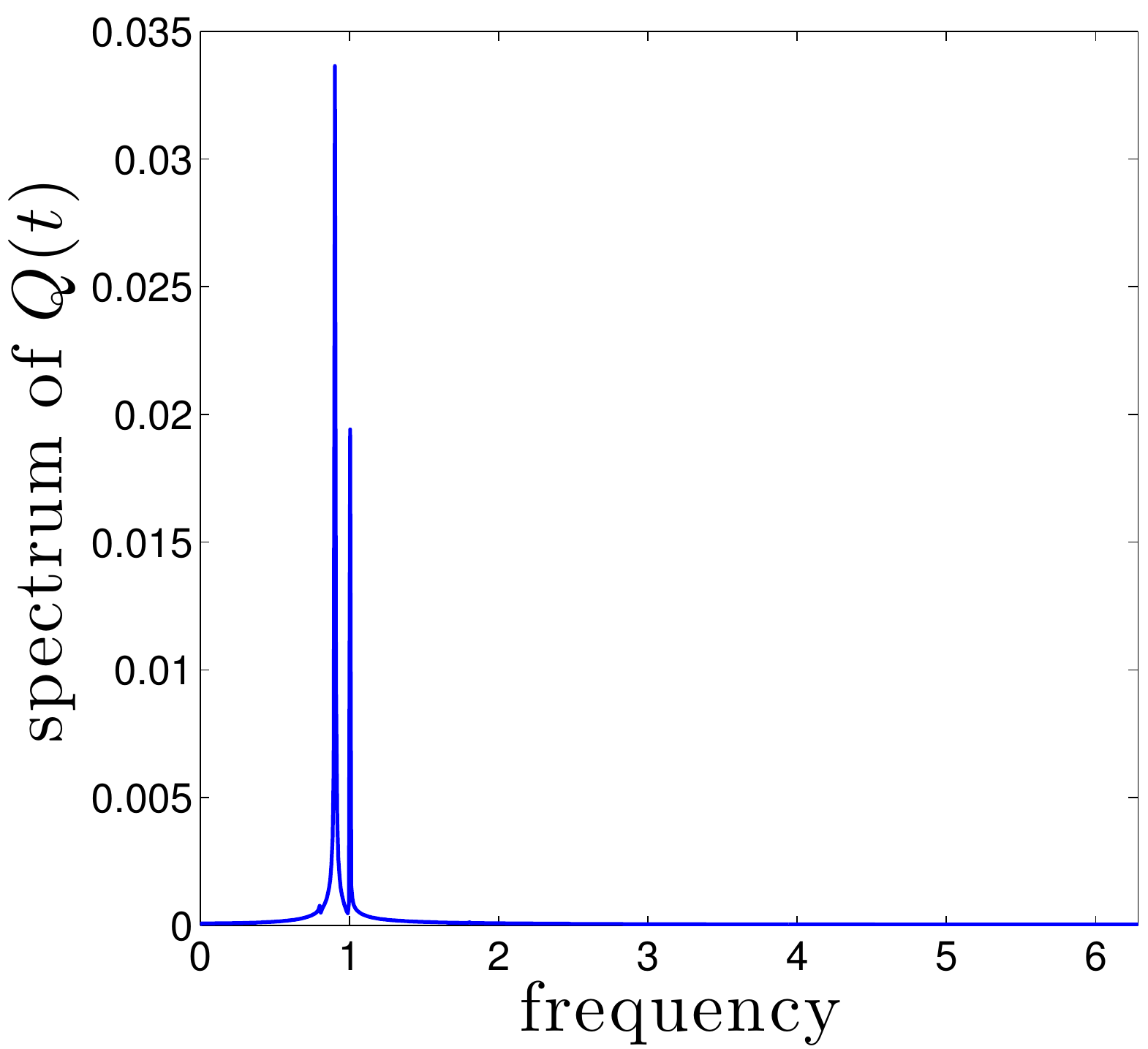}}
\subfigure[]{\includegraphics[width=0.49\textwidth,height=0.39\textwidth]{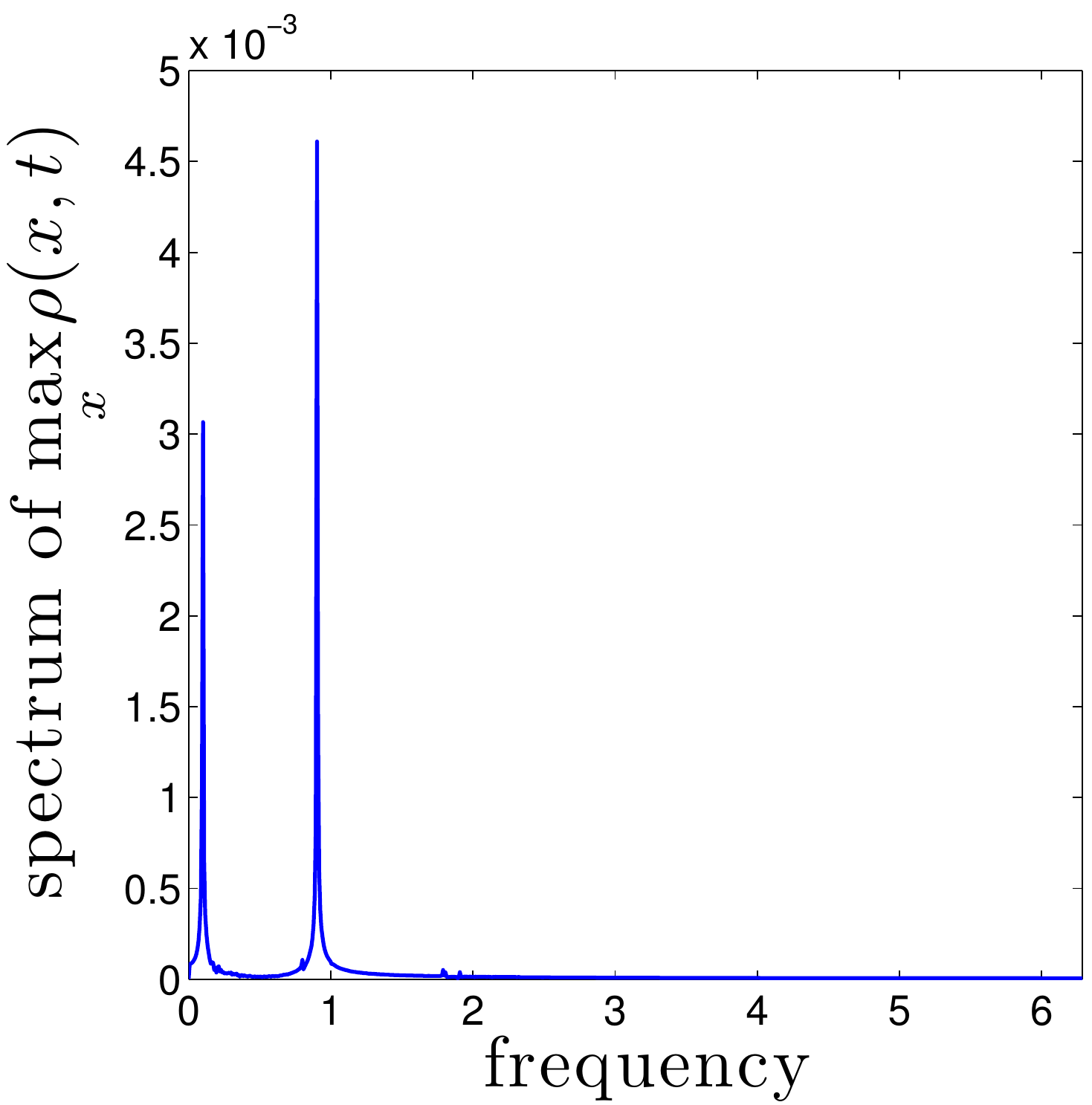}}
\caption{Intrinsic soliton oscillations for the case of a harmonic inhomogeneous force ($K=-3 \pi/100 \approx -0.094248$) and $v_0=0.1$. Other parameters and initial conditions as in Fig.\ \ref{fig1}.  
Panels (a) and (b): charge from CC theory and simulation, respectively. 
Panels (c) and (d): amplitude $a=2 [m -\omega(t)]/g^2$ and $\max_{x} \rho(x,t)$, respectively.
Panel (e): DFT of $Q(t)$, soliton peak at $\omega_1=0.9032$ and phonon peak at $\omega_2=1.0053 \approx \sqrt{1+k^2}$ with $k=-K$. 
Panel (f): DFT of $\max_{x} \rho(x,t)$,  peaks at $\omega_1=0.9032$, $\omega_3=\omega_2-\omega_1=0.1021$. 
}
\label{fig5}
\end{figure}

\begin{figure}
\subfigure[]{\includegraphics[width=0.49\textwidth,height=0.39\textwidth]{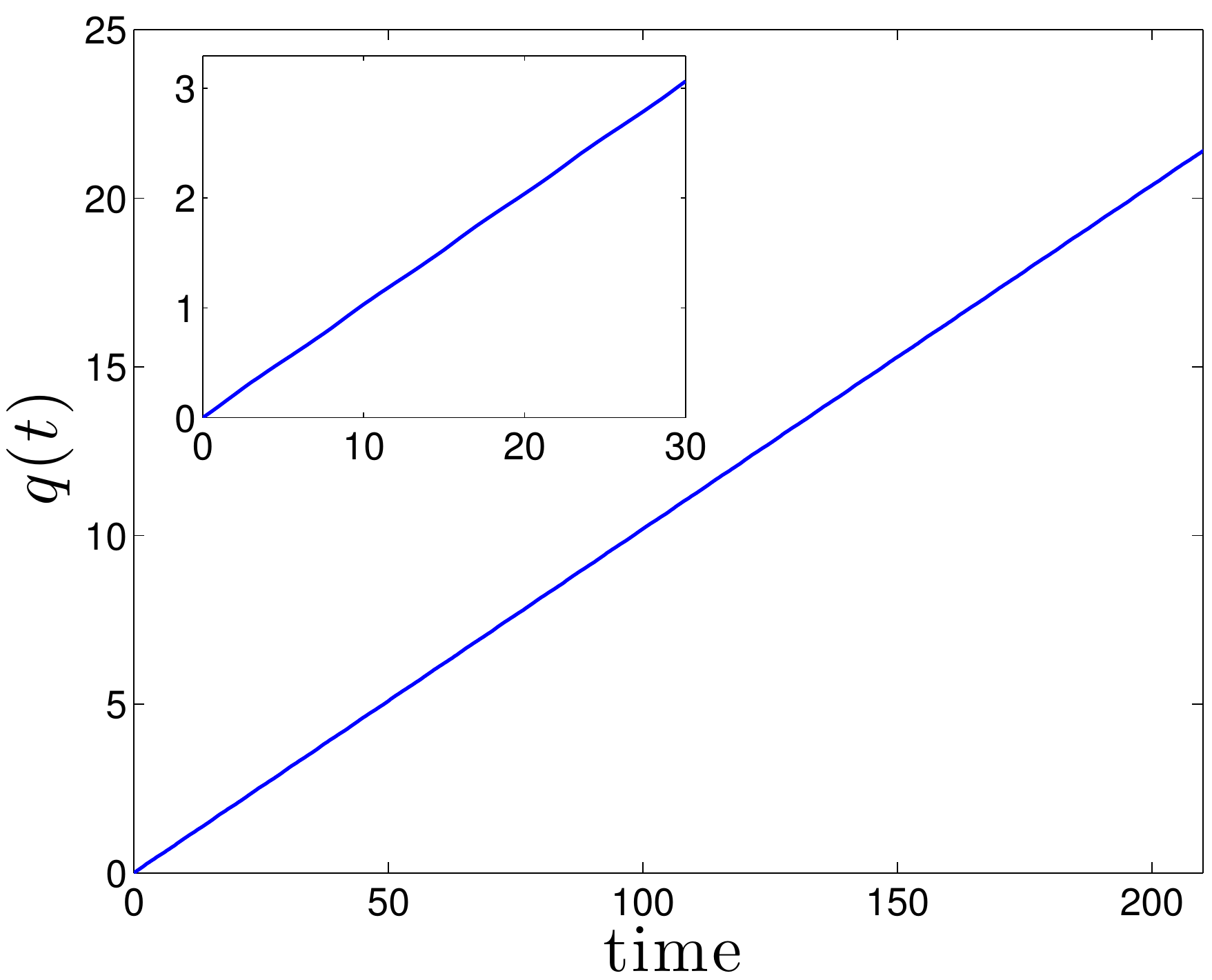}}
\subfigure[]{\includegraphics[width=0.49\textwidth,height=0.39\textwidth]{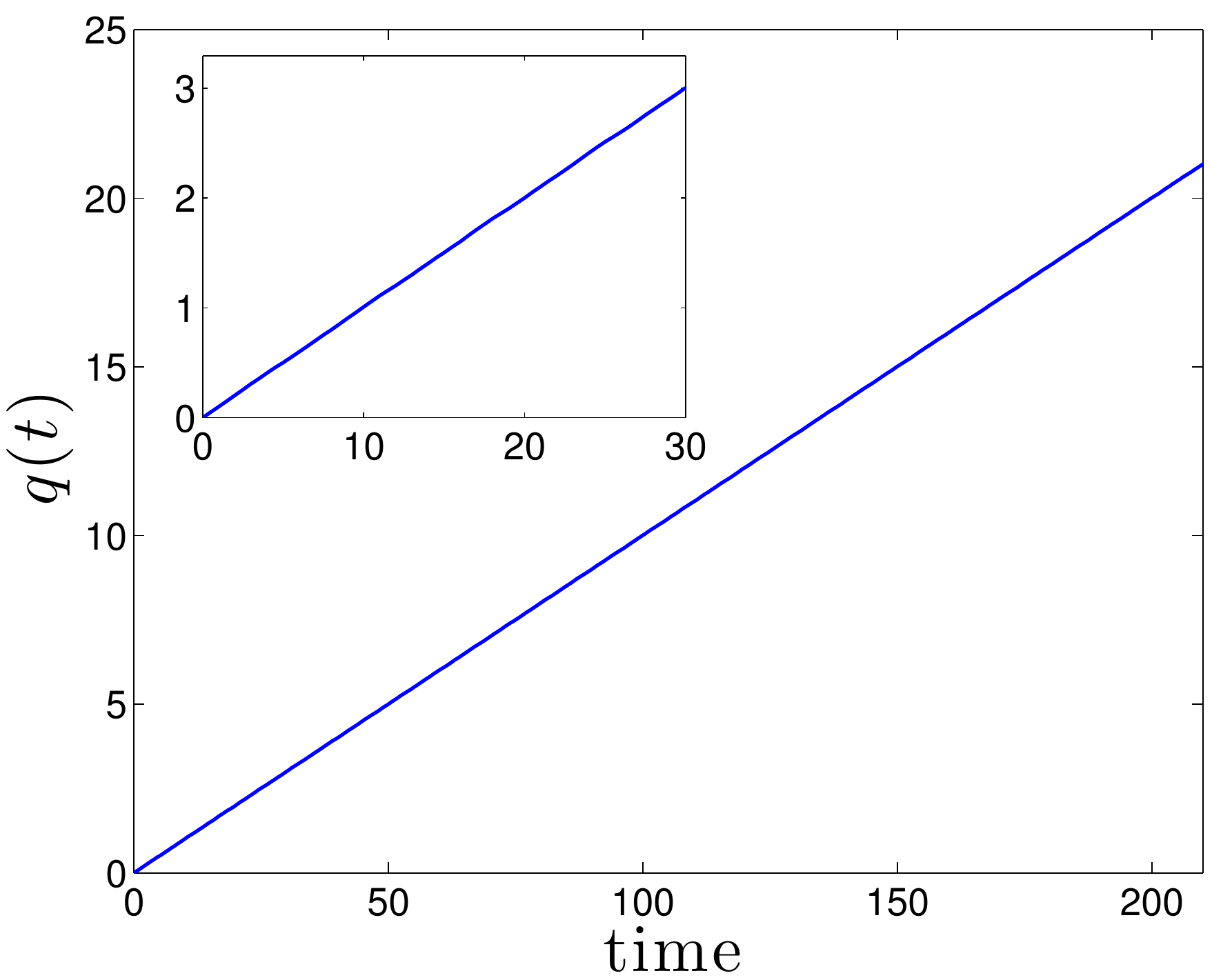}}
\subfigure[]{\includegraphics[width=0.49\textwidth,height=0.39\textwidth]{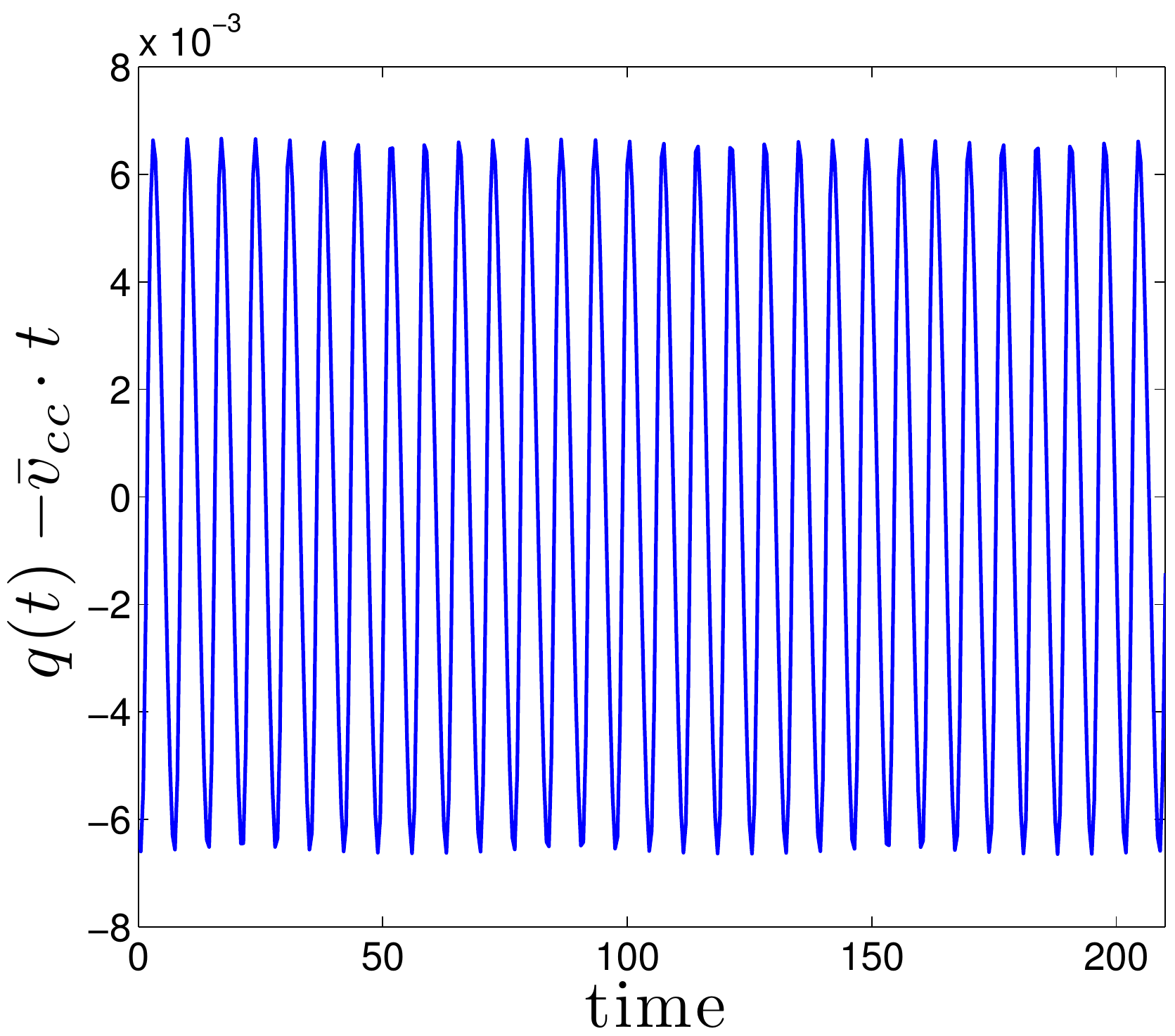}}
\subfigure[]{\includegraphics[width=0.49\textwidth,height=0.39\textwidth]{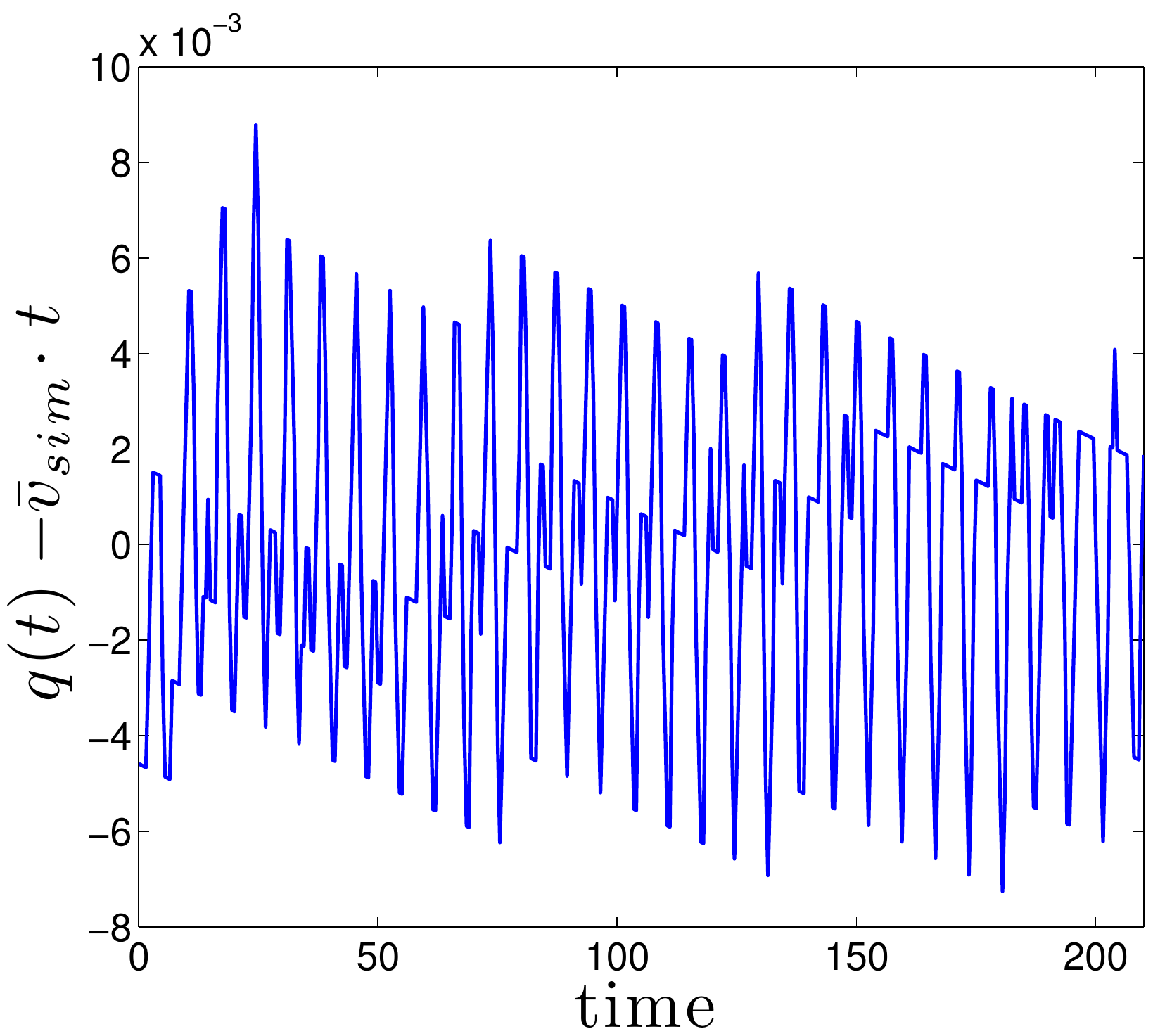}}
\subfigure[]{\includegraphics[width=0.49\textwidth,height=0.39\textwidth]{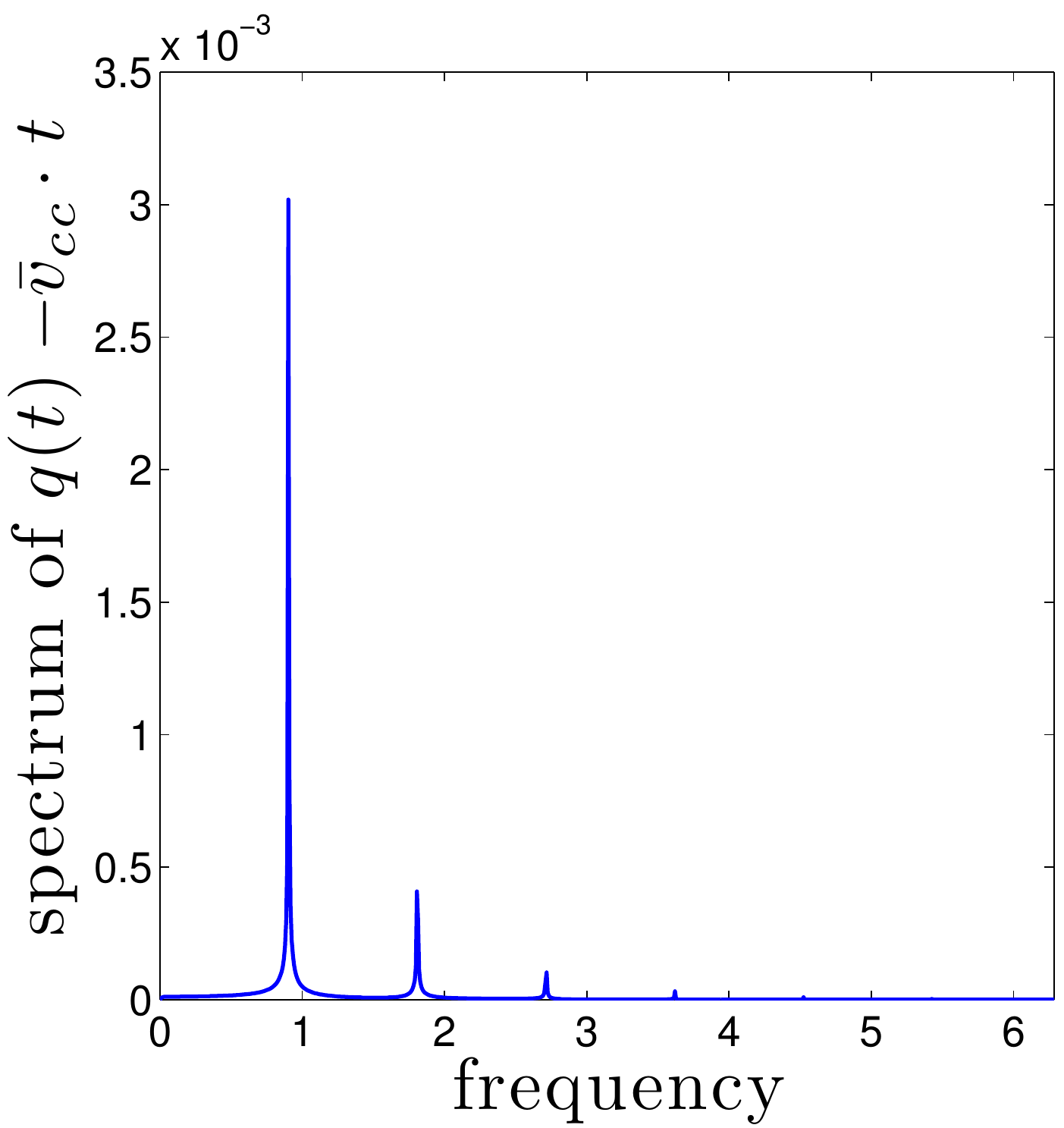}}
\subfigure[]{\includegraphics[width=0.49\textwidth,height=0.39\textwidth]{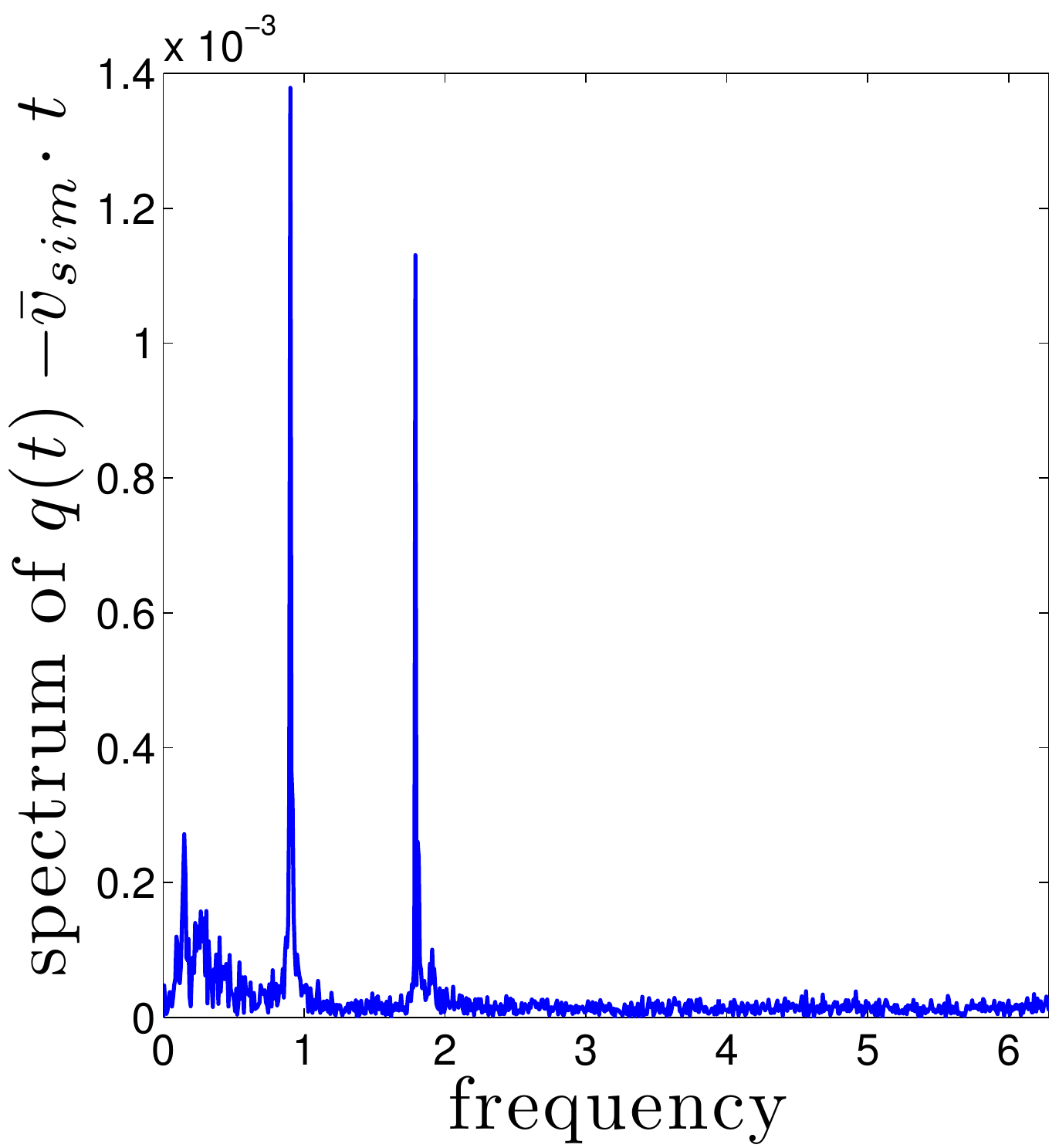}}
\caption{Oscillations of the translational motion of the soliton. $K=-3 \pi/100 \approx -0.094248$ and $v_0=0.1$. 
Other parameters and initial conditions as in Fig.\ \ref{fig1}.  
Panels (a) and (b): $q(t)$ from CC theory and simulation, respectively. 
Panels (c) and (d): $q(t)-\bar{v} t$ from theory ($\bar{v}=\bar{v}_{cc}=0.1019$) 
 and simulation ($\bar{v}=\bar{v}_{sim}=0.10005$), respectively.
Panel (e): DFT of $q(t)-\bar{v}_{cc} t$, soliton peaks at $\omega_1=0.9032$ and $2 \omega_1=1.8064$.  
Panel (f): DFT of $q(t)-\bar{v}_{sim} t$, peaks at $\omega_1=0.9032$ and $1.7907 \approx 2 \omega_1$.  
}
\label{fig6}
\end{figure}




\section{Summary} \label{sec8}

We investigated how a solitary wave solution of the nonlinear Dirac (NLD) equation evolves in time under an external force with the components $f_j = r_j \exp[-i(\nu_j t - K_j x)]$, $j =1,2$. As an ansatz for a collective coordinate (CC) theory we took the exact Lorentz boosted solitary wave solution of the unperturbed NLD equation. The collective variables are the soliton position $q(t)$, inverse width 
$\beta(t)$ and phase $\phi(t)$. The variable $\beta$ is related to the frequency $\omega(t) = \sqrt{m^2-\beta^2}$ that appears in the solitary wave solution and lies in the range $0 < \omega < m$. In the non-relativistic regime $\omega$ is close to the mass $m$. 
The forced NLD equation is obtained in a standard way from a Lagrangian density. We restricted ourselves to the case with $\nu_j = 0$, inserted our ansatz, integrated over space, and obtained the Lagrangian as a function of the collective coordinates. The  Lagrange equations are three coupled ODEs. In two special cases we obtained approximate analytical solutions, but in general the ODEs were solved numerically by a MATHEMATICA program. We chose parameters and initial conditions from the non-relativistic regime where solitary wave solutions are expected to be stable. The solutions are periodic in time, which means that the solitary waves exhibit intrinsic oscillations with frequency 
$\Omega_{cc}$. The translational motion is also affected, though weakly, because the position $q(t)$ oscillates around a mean trajectory $\bar{v}_{cc} t$. 
We compared our CC predictions with numerical simulations of the forced NLD equation: The solitary wave solutions are in fact stable and periodic. The observed frequency $\Omega_{sim} = \omega_1$ is nearly identical with $\Omega_{cc}$. However, $\bar{v}_{sim}$ agrees with $\bar{v}_{cc}$ with an error of about $14 \%$. The reason for this is that the CC theory does not include phonons (short for linear excitations). In fact, a specific plane wave phonon mode with wavenumber $k = -K$ is excited together with the intrinsic oscillations in order to conserve the total momentum. The predicted frequency $\Omega_K = \sqrt{m^2+K^2}$ agrees perfectly with the frequency $\omega_2$ in the spectra of all variables.

For the future work we plan to take initial conditions away from the non-relativistic regime, i.e. initial $\omega$ not close to $m$, and initial velocity not much smaller than the speed of light. Moreover, it will be very interesting to see what is the influence of time dependent external forces, i.e. non-vanishing $\nu_j$.

\section{Acknowledgments}  
This work was performed in part under the auspices of the United States Department of Energy. The authors would like to thank the Santa Fe Institute for its hospitality during the completion of this work. 
S.S. acknowledges financial support from the National Natural Science
Foundation of China (Nos.~11471025, 91330110, 11421101).
N.R.Q. 
acknowledges financial support from the Alexander von Humboldt Foundation (Germany) through Research Fellowship for Experienced Researchers SPA 1146358 STP and by the MICINN (Spain) through 
FIS2011-24540, and by Junta de Andalucia (Spain) under Projects No. FQM207, No. 
P06-FQM-01735, and No. P09-FQM-4643.  F.G.M. acknowledges the 
hospitality of the Mathematical Institute of the University of Seville (IMUS) and of the Theoretical 
Division and Center for Nonlinear Studies at Los Alamos National Laboratory, financial 
support by the Plan Propio of the University of Seville, and by the MICINN (Spain) through FIS2011-24540. 
A.K. acknowledges financial support from Department of Atomic Energy,
Government of India through a Raja Ramanna Fellowship. 

\appendix
\section{Relevant Integrals} \label{sec9}

For our ansatz in the rest frame, we have that  for $\kappa=1$
%
the charge $Q$
is
\bq \label{A1}
 Q=\int dx \Psi^{\dag} \Psi= \int dx (A^2 +B^2)=\frac{2 \beta^2}{ g^2 (m+ \omega)}    \int_{- \infty}^\infty dx  \frac{1+ \alpha^2 \tanh^2 \beta x}{(1-\alpha^2 \tanh^2 \beta x)^2} \sech^2 \beta x
=  \frac{2 \beta}{g^2 \omega} 
\eq
For Sec. V we need explicit expressions for the following  integrals  (in what follows,  $y= \tanh  \beta x$):
\ba
H_1 &&= -\frac{i}{2} \int dx \left[\bar \Psi \gamma^1 \partial_x \Psi -\partial_x \bar \Psi \gamma^1 \Psi \right] = \int dx  (B' A -A'B) = \frac{2 (m-\omega)}{g^2} \alpha  \int_{-1}^1 dy  \frac { 1-y^2} { ( 1- \alpha^2 y^2)^2} \nonumber \\
&&= \frac{2}{g^2} \left(2 \tanh ^{-1}\left(\sqrt{\frac{m-\omega }{\omega
   +m}}\right)-\beta \right) = I_0, \label{A2}
\ea
\ba
H_2 &=& m\int dx \bar \Psi \Psi =m I_1 = m \int dx (A^2 -B^2) =\frac{ 2 m \beta}{ g^2 (m+ \omega)}   \int_{-1}^1 dy  \frac { 1} { ( 1- \alpha^2 y^2)}= \frac{4 m \beta}{g^2 (m+ \omega)}  \frac{ \tanh ^{-1}(\alpha )}{\alpha } \nonumber \\
&& = \frac{4 m}{g^2} \tanh ^{-1}(\alpha )=M_0, \label{A3}
\ea
where $M_0$ is the mass in the rest frame. 
Note that $M_0$ has the property of going to zero as $\omega \to 1$.
\ba
I_2 && = \int dx (A^2 -B^2)^2 =\frac{ 4 \beta^3}{ g^4 (m+ \omega)^2}   \int_{-1}^1 dy  \frac { 1-y^2} { ( 1- \alpha^2 y^2)^2}= \frac{ 4 \beta^3}{ g^4 (m+ \omega)^2}  \left(\frac{\left(\alpha ^2+1\right) \tanh ^{-1}(\alpha )-\alpha }{\alpha ^3}\right) \nonumber \\
&& = \frac{2} {g^2} I_0=\frac{2}{g^2} H_1.  \label{A4}
 \ea
 To calculate the integral $J_j$ defined in (\ref{4.15}), first we rewrite it as 
\ba
J_j(\omega, \dot{q}) &=& \int_{-\infty}^{+\infty} dz A(z)  \cos(2 \beta a_j z) = 
 \frac{\sqrt{2 (m+\omega)} \beta}{g \omega}  \int_{-\infty}^{+\infty} dz   \frac{\cosh(\beta z) 
 \cosh(i 2 \beta a_j z)}{\frac{m}{\omega}+\cosh(2 \beta z)}  \nonumber \\
&=& \frac{\sqrt{2 (m+\omega)} \beta}{g \omega}  \int_{0}^{+\infty} dz   
\frac{\cosh[(1+i 2 a_j) \beta z]+ 
 \cosh[(1-i 2 a_j) \beta z]}{\frac{m}{\omega}+\cosh(2 \beta z)}.\label{A5}
 \ea
Now using expression (6) on page 357 of \cite{bk:PrudnikovBrychkov1986}, after some straightforward calculations we obtain
\ba  
J_j(\omega,\dot{q})&=&\frac{\pi \cos b_j}{g \sqrt{\omega} \cosh a_j \pi},  \label{A6}
\ea
where $a_j$ and $b_j$ are defined in Eq.\ (\ref{4.15}). The integral $N_j$ can be calculated in a similar way. 




%

\end{document}